\documentclass{article}
\usepackage{graphicx}
\DeclareSymbolFont{AMSa}{U}{msa}{m}{n}
\DeclareMathDelimiter\ulcorner{\mathopen} {AMSa}{"70}{AMSa}{"70}
\DeclareMathDelimiter\urcorner{\mathclose}{AMSa}{"71}{AMSa}{"71}
\makeatletter
\def\uufill{$\m@th\mathopen\ulcorner\mkern-7mu%
  \cleaders\hbox{\rule[6pt]{1dd}{1dd}}\hfill
  \mkern-7mu\mathclose\urcorner$}
\def\overbrack#1{\vbox{\m@th\ialign{##\crcr
      \uufill\crcr\noalign{\kern-\p@\nointerlineskip}%
      $\hfil\displaystyle{#1}\hfil$\crcr}}}
\makeatother

\title{Magnetic Monopole is Photon}

\author{Sze Kui Ng
\\ \small \it Department of Mathematics,
City University of Hong Kong, Hong Kong
\\ \small szekuing@yahoo.com.hk
}
\begin{document}
\date{}
\maketitle
\begin{abstract}

It is shown that a photon with a specific frequency can be
identified with the Dirac magnetic monopole. When a Dirac-Wilson
line forms a Dirac-Wilson loop, it is a photon. This loop model of
photon is exactly solvable. From the winding numbers of this
loop-form of photon, we derive the quantization properties of
energy and electric charge.

A new QED theory is presented that is free of ultraviolet
divergences. The Dirac-Wilson line is as the quantum photon
propagator of the new QED theory from which we can derive known
QED effects such as the anomalous magnetic moment and the Lamb
shift. The one-loop computation of these effects is simpler and is
more accurate than that in the conventional QED theory.
Furthermore, from the new QED theory, we have derived a new QED
effect. A new formulation of the Bethe-Salpeter (BS) equation
solves the difficulties of the BS equation and gives a modified
ground state of the positronium. By the mentioned new QED effect
and by the new formulation of the BS equation, a term in the
orthopositronium decay rate that is missing in the conventional
QED is found, resolving the orthopositronium lifetime puzzle
completely.

It is also shown that graviton can be constructed from 
photon, yielding a theory of quantum gravity that unifies
gravitation and electromagnetism.  We show that dark energies are the  gravitons and dark matters are matters constructed from gravitons, as electrons  constructed from photons. From  this idenification of  gravitons with dark energy  and the quantum  graviton propagator  we can compute the  cosmological constant and  show that the  cosmological constant is a very small positive constant  (and tends to 0 very slowly). This  resolves the cosmological constant problem.


\end{abstract}

\section{Introduction}\label{sec00}

It is well known that the quantum era of physics began with the
quantization of energy of electromagnetic field, from which Planck
derived the radiation formula. Einstein then introduced the
light-quantum to explain the photoelectric effects. This
light-quantum was regarded as a particle called photon
\cite{Pai}\cite{Pla}\cite{Ein}. Quantum mechanics was then
developed, ushering in the modern quantum physics era.
Subsequently, the quantization of the electromagnetic field and
the theory of quantum electrodynamics(QED) were established.

In this development of quantum theory of physics, the photon plays
a special role. While it is the beginning of quantum physics, it
is not easy to understand as is the quantum mechanics of other
particles described by the Schroedinger equation. In fact,
Einstein was careful in regarding the light-quantum as a particle,
and the acceptance of the light-quantum as a particle called
photon did not come about until much later \cite{Pai}. The quantum
field theory of electromagnetic field was developed for the
photon. However, such difficulties of the quantum field theory as
the ultraviolet divergences are well known. Because of the
difficulty of understanding the photon, Einstein once asked:
\textquotedblright What is the photon? \textquotedblright
\cite{Pai}.

On the other hand, based on the symmetry of the electric and
magnetic field described by the Maxwell equation and on the
complex wave function of quantum mechanics, Dirac derived the
concept of the magnetic monopole, which is hypothetically
considered as a particle with magnetic charge, in analogy to the
electron with electric charge. An important feature of this
magnetic monopole is that it gives the quantization of electric
charge. Thus it is interesting and important to find such
particles. However, in spite of much effort, no such particles
have been found \cite{Dir}\cite{Dir2}.

In this paper we shall establish a mathematical model of photon to
show that the magnetic monopole can be identified as a photon.
Before giving the detailed model, let us discuss some thoughts for
this identification in the following.

First, if the photon and the magnetic monopole are different types
of elementary quantum particles in the electromagnetic field, it
is odd that one type can be derived from the other. A natural
resolution of this oddity is the identification of the magnetic
monopole as a photon.

The quantum field theory of the free Maxwell equation is the basic
quantum theory of photon \cite{Zub}. This free field theory is a
linear theory and the models of the quantum particles obtained
from this theory are linear. However, a stable particle should be
a soliton, which is of the nonlinear nature. Secondly, the quantum
particles of the quantum theory of Maxwell equation are collective
quantum effects in the same way the phonons which are elementary
excitations in a statistical model. These phonons are usually
considered as quasi-particles and are not regarded as real
particles. Regarding the Maxwell equation as a statistical wave
equation of electromagnetic field, we have that the quantum
particles in the quantum theory of Maxwell equation are analogous
to the phonons. Thus they should be regarded as quasi-photons and
have properties of photons but not a complete description of
photons.

In this paper, a nonlinear model of photon is established. In the
model, we show that the Dirac magnetic monopole can be identified
with the photon with some frequencies. We provide a $U(1)$ gauge
theory of quantum electrodynamics (QED), from which we derive
photon as a quantum Dirac-Wilson loop $W(z,z)$ of this model. This
nonlinear loop model of the photon is exactly solvable and thus
may be regarded as a quantum soliton. From the winding numbers of
this loop model of the photon, we derive the quantization property
of energy in Planck's formula of radiation and the quantization
property of charge. We show that the quantization property of
charge is derived from the quantization property of energy (in
Planck's formula of radiation), when the magnetic monopole is
identified with photon with certain frequencies. This explains why
we cannot physically find a magnetic monopole. It is simply a
photon with a specific frequency.

From this nonlinear model of the photon, we also construct a model
of the electron which has a mass mechanism for generating mass of
the electron. This mechanism of generating mass supersedes the
conventional mechanism of generating mass (through the Higgs
particles) and makes hypothesizing the existence of the Higgs
particles unnecessary. This explains why we cannot physically find
such Higgs particles.

The new quantum gauge theory is similar to the conventional QED
theory
except that the former is not based on the four dimensional space-time $%
(x,t) $ but is based on the proper time $s$ in the theory of
relativity. Only in a later stage in the new quantum gauge theory,
the space-time variable $(x,t)$ is derived from the proper time
$s$ through the Lorentz metric $ds^{2}=dt^{2}-dx^{2}$ to obtain
space-time statistics and explain the observable QED effects.

The derived space variable $x$ is a random variable in this
quantum gauge theory. Recall that the conventional quantum
mechanics is based on the space-time. Since the space variable $x$
is actually a random variable as shown in the new quantum gauge
theory, the conventional quantum mechanics needs probabilistic
interpretation and thus has a most mysterious measurement problem,
on which Albert Einstein once remarked: "God does not play dice
with the universe." In contrast, the new quantum gauge theory does
not involve the mentioned measurement problem because it is not
based on the space-time and is deterministic. Thus this quantum
gauge theory resolves the mysterious measurement problem of
quantum mechanics.

Using the space-time statistics, we employ Feynman diagrams and
Feynman rules to compute the basic QED effects such as the vertex
correction, the photon self-energy and the electron self-energy.
In this computation of the Feynman integrals, the dimensional
regularization method in the conventional QED theory is also used.
Nevertheless, while the conventional QED theory uses it to reduce
the dimension 4 of space-time to a (fractional) number $n$ to
avoid the ultraviolet divergences in the Feynman integrals, the
new QED theory uses it to increase the dimension 1 of the proper
time to a number $n$ less than $4$, which is the dimension of the
space-time, to derive the space-time statistics. In the new QED
theory, there are no ultraviolet divergences, and the dimensional
regularization method is not used for {\it regularization}.

After this dimension increase, the renormalization method is used
to derive the well-known QED effects. Unlike the conventional QED
theory, the renormalization method is used in the new QED theory
to compute the space-time statistics, but not to remove the
ultraviolet divergences, since the ultraviolet divergences do not
occur in the new QED theory. From these QED effects, we compute
the anomalous magnetic moment and the Lamb shift \cite{Zub}. The
computation does not involve numerical approximation as does that
of the conventional QED and is simpler and more accurate.

For getting these QED effects, the quantum photon propagator
$W(z,z^{\prime })$, which is like a line segment connecting two
electrons, is used to derive the electrodynamic interaction (When
the quantum photon propagator $ W(z,z^{\prime })$ forms a closed
circle with $z=z^{\prime }$, it then becomes a photon $W(z,z)$).
From this quantum photon propagator, a photon propagator is
derived that is similar to the Feynman photon propagator in the
conventional QED theory.

The photon-loop $W(z,z)$ leads to the renormalized electric charge
$e$ and the mass $m$ of electron. In the conventional QED theory,
the bare charge $ e_{0}$ is of less importance than the
renormalized charge $e$, in the sense that it is unobservable. In
contrast, in this new theory of QED, the bare charge $e_{0}$ and
the renormalized charge $e$ are of equal importance. While the
renormalized charge $e$ leads to the physical results of QED, the
bare charge $e_{0}$ leads to the universal gravitation constant
$G$. It is shown that $e=n_{e}e_{0}$, where $n_{e}$ is a very
large winding number and thus $e_{0}$ is a very small number. It
is further shown that the gravitational constant $G=2e_{0}^{2}$
which is thus an extremely small number. This agrees with the fact
that the experimental gravitational constant $G$ is a very small
number. The relationships, $e=n_{e}e_{0}$ and $G=2e_{0}^{2}$, are
a part of a theory unifying gravitation and electromagnetism. In
this unified theory, the graviton propagator and the graviton are
constructed from the quantum photon propagator. This construction
leads to a theory of quantum gravity. In short, a new theory of
quantum gravity is developed from the new QED theory in this
paper, and unification of gravitation and electromagnetism is
achieved.

In this theory of  quantum gravity
 we show that dark energies are the  gavitons and dark matters are matters constructed from gravitons, as electrons constructed from photons. 
From the  gravitons identified as dark energy and the quantum  graviton propagator  we can compute the  cosmological constant and  show that the  cosmological constant is a very small positive constant (and tends to 0 very slowly). This then resolves the cosmological constant problem.

In this paper, we also derive a new QED effect from the seagull
vertex of the new QED theory. The conventional Bethe-Salpeter (BS)
equation is reformulated to resolve its difficulties (such as the
existence of abnormal solutions \cite{Bet}-\cite{Nie}) and to give
a modified ground state wave function of the positronium. By the
new QED effect and the reformulated BS equation, another new QED
effect, a term in the orthopositronium decay rate that is missing
in the conventional QED is discovered. Including the discovered
term, the computed orthopositronium decay rate now agrees with the
experimental rate, resolving the {\it orthopositronium lifetime
puzzle} completely \cite{Wes}-\cite{Kni2}. We note that the recent
resolution of this orthopositronium lifetime puzzle resolves the
puzzle only partially due to a special statistical nature of this
new term in the orthopositronium decay rate.

This paper is organized as follows. In Section 2 we give a brief
description of a new QED theory. With this theory, we introduce
the classical Dirac-Wilson loop in Section 3. We show that the
quantum version of this loop is a nonlinear exactly solvable model
and thus can be regarded as a soliton. We identify this quantum
Dirac-Wilson loop as a photon with the $U(1)$ group as the gauge
group. To investigate the properties of this Dirac-Wilson loop, we
derive a chiral symmetry from the gauge symmetry of this quantum
model. From this chiral symmetry, we derive, in Section 4, a
conformal field theory, which includes an affine Kac-Moody algebra
and a quantum Knizhnik-Zamolodchikov (KZ) equation. A main point
of our model on the quantum KZ equation is that we can derive two
KZ equations which are dual to each other. This duality is the
main point for the Dirac-Wilson loop to be exactly solvable and to
have a winding property which explains properties of photon. This
quantum KZ equation can be regarded as a quantum Yang-Mills
equation.

In Sections 5 to 8, we solve the Dirac-Wilson loop in a form with
a winding property, starting with the KZ equations. From the
winding property of the Dirac-Wilson loop, we derive, in Section 9
and Section 10, the quantization of energy and the quantization of
electric charge which are properties of photon and magnetic
monopole. We then show that the quantization property of charge is
derived from the quantization property of energy of Planck's
formula of radiation, when we identify photon with the magnetic
monopole for some frequencies. From this nonlinear model of
photon, we also derive a model of the electron in Section 11. In
this model of electron, we provide a mass mechanism for generating
mass to electron. In Section 12, we show that the photon with a
specific frequency can carry electric charge and magnetic charge,
since an electron is formed from a photon with a specific
frequency for giving the electric charge and magnetic charge. In
Section 13, we derive the statistics of photons and electrons from
the loop models of photons and electrons.  In Section 14 we show that the photon loops are the origin of magnetic properties.

In Sections 15 to 23, we derive a new theory of QED, wherein we
perform the computation of the known basic QED effects such as the
photon self-energy, the electron self-energy and the vertex
correction. In particular, we provide simpler and more accurate
computation of the anomalous magnetic moment and the Lamb shift.
Then in Section 24, we compute a new QED effect. Then from Section
25 to Section 26, we reformulate the Bethe-Salpeter (BS) equation.
With this new version of the BS equation and the new QED effect, a
modified ground state wave function of the positronium is derived.
Then by this modified ground state of the positronium, we derive
in Section 27 another new QED effect, a term missing in the
theoretic orthopositronium decay rate of the conventional QED
theory, and show that this new theoretical orthopositronium decay
rate agrees with the experimental decay rate, completely resolving
the orthopositronium life time puzzle \cite{Wes}-\cite{Kni2}.

In Section 28,  graviton is constructed from  photon. This leads
to a  theory of quantum gravity and a unification of
gravitation and electromagnetism.
In Section 29, we derive the quantum graviton propagator and graviton propagator, as similar to the photon case. 
Then in Section 30, we show that
the quantized energies of gravitons can be identified as dark
energy. Then in a way similar to the construction of electrons by
photons, we use gravitons to construct particles which can be
regarded as dark matter. We show that the gravitational force among gravitons
can be repulsive. This gives the determination of the cosmological constant and  the phenomenon of the accelerating expansion of the universe.
\cite{Rie}-\cite{Per2}.

 \section{New gauge model of QED}\label{sec1}

Let us construct a quantum gauge model, as follows. In probability
theory we have the Wiener measure $\nu$ which is a measure on the
space $C[t_0,t_1]$ of continuous functions \cite{Jaf}. This
measure is a well defined mathematical theory for the Brownian
motion and it may be symbolically written in the following form:
\begin{equation}
d\nu =e^{-L_0}dx \label{wiener}
\end{equation}
where $L_0 :=
\frac12\int_{t_0}^{t_1}\left(\frac{dx}{dt}\right)^2dt$ is the
energy integral of the Brownian particle and $dx =
\frac{1}{N}\prod_{t}dx(t)$ is symbolically a product of Lebesgue
measures $dx(t)$ and $N$ is a normalized constant.

Once the Wiener measure is defined we may then define other
measures on $C[t_0,t_1]$,  as follows \cite{Jaf}. Let a potential
term $\frac12\int_{t_0}^{t_1}Vdt$ be added to $L_0$. Then we have
a measure $\nu_1$ on $C[t_0,t_1]$ defined by:
\begin{equation}
d\nu_1 =e^{-\frac12\int_{t_0}^{t_1}Vdt}d \nu \label{wiener2}
\end{equation}
Under some condition on $V$ we have that $\nu_1$ is well defined
on $C[t_0,t_1]$. Let us call (\ref{wiener2}) as the Feymann-Kac
formula \cite{Jaf}.

Let us then follow this formula to construct a quantum  model of
electrodynamics, as follows. Then similar to the formula
(\ref{wiener2}) we construct a quantum model of electrodynamics
from the following energy integral:
\begin{equation}
\begin{array}{rl}
&-\int_{s_0}^{s_1}D ds:= -\int_{s_0}^{s_1}[\frac12\left(\frac{\partial A_1}{\partial
x^2}-\frac{\partial A_2}{\partial x^1}\right)^*
\left(\frac{\partial A_1}{\partial x^2}-\frac{\partial A_2}{\partial x^1}\right)+ \\
&\\
  &
 \left(
\frac{dZ^*}{ds} +ie_0(\sum_{j=1}^2A_j\frac{dx^j}{ds})Z^*\right)
\left( \frac{dZ}{ds}
-ie_0(\sum_{j=1}^2A_j\frac{dx^j}{ds})Z\right)]ds \\
\end{array}
\label{1.1}
\end{equation}
where the complex variable $Z=Z(z(s))$ and the real variables
$A_1=A_1(z(s))$ and $A_2=A_2(z(s))$ are continuous functions in a
form that they are in terms of a (continuously differentiable)
curve $z(s)=C(s)=(x^1(s),x^2(s)), s_0\leq s\leq s_1,
z(s_0)=z(s_1)$ in the complex plane where $s$ is a parameter
representing the proper time in relativity (We shall also write
$z(s)$ in the complex variable form
$C(s)=z(s)=x^1(s)+ix^2(s),s_0\leq s\leq s_1$). The complex
variable $Z=Z(z(s))$ represents a field of matter( such as the
electron) ($Z^*$ denotes its complex conjugate) and the real
variables $A_1=A_1(z(s))$ and $A_2=A_2(z(s))$ represent a
connection (or the gauge field of the photon) and $e_0$ denotes
the (bare) electric charge.

The integral (\ref{1.1}) has the following gauge symmetry:
\begin{equation}
\begin{array}{rl}
Z^{\prime}(z(s)) & := Z(z(s))e^{ie_0a(z(s))} \\
A'_j(z(s)) & := A_j(z(s))+\frac{\partial a}{\partial x^j} \quad
j=1,2
\end{array}
\label{1.2}
\end{equation}
where $a=a(z)$ is a continuously differentiable real-valued
function of $z$.

We remark that this QED theory is similar to the conventional
Yang-Mills gauge theories. A  feature of (\ref{1.1}) is that it is
not formulated with the four-dimensional space-time but is
formulated with the one dimensional proper time. This one
dimensional nature let this QED theory avoid the usual ultraviolet
divergence difficulty of quantum fields. As most of the theories
in physics are formulated with the space-time let us give reasons
of this formulation. We know that with the concept of space-time
we have a convenient way to understand physical phenomena and to
formulate theories such as the Newton equation, the Schroedinger
equation , e.t.c. to describe these physical phenomena. However we
also know that there are fundamental difficulties related to
space-time such as the ultraviolet divergence difficulty of
quantum field theory. To resolve these difficulties let us
reexamine the concept of space-time. We propose that the
space-time is a statistical concept which is not as basic as the
proper time in relativity. Because a statistical theory is usually
a convenient but incomplete description of a more basic theory
this means that some difficulties may appear if we formulate a
physical theory with the space-time. This also means that a way to
formulate a basic theory of physics is to formulate it not with
the space-time but with the proper time only as the parameter for
evolution. This is a reason that we use (\ref{1.1}) to formulate a
QED theory. In this formulation we regard the proper time as an
independent parameter for evolution. From (\ref{1.1}) we may
obtain the conventional results in terms of space-time by
introducing the space-time as a statistical method.

Let us explain in more detail how the space-time comes out as a
statistics. For statistical purpose when many electrons (or many
photons) present we introduce space-time $(t,x)$ as a statistical
method to write $ds^2$ in the form
\begin{equation}
ds^2=dt^2-dx^2 \label{lorentz}
\end{equation}
We notice that for a given $ds$ there may have many $dt$ and $dx$
which correspond to many electrons (or photons) such that
(\ref{lorentz}) holds. In this way the space-time is introduced as
a statistics.
 By (\ref{lorentz}) we shall derive statistical formulas
 for many electrons (or photons) from
 formulas obtained from (\ref{1.1}). In this way we obtain the Dirac equation as
 a statistical equation for electrons and the Maxwell equation as a statistical equation for photons.
  In this way we may regard the conventional QED theory as a
  statistical theory
  extended from the proper-time formulation of
  this QED theory (From the proper-time formulation of this QED theory we also have
  a theory of space-time statistics
  which give the results of the conventional QED theory). This statistical interpretation of
  the conventional QED theory is thus
  an explanation of the mystery that the conventional QED theory is successful in
  the computation of quantum effects of electromagnetic interaction while it has
  the difficulty of ultraviolet divergence.

 We notice that the relation (\ref{lorentz}) is the famous Lorentz metric (We may generalize it
 to other metric in general relativity). Here our understanding of the Lorentz metric is
 that it is a statistical formula
 where the proper time $s$ is more fundamental than the space-time $(t,x)$ in the sense
 that we first have the proper time and the space-time is introduced via the Lorentz metric only for
 the purpose of statistics. This reverses the order of appearance of
the proper time and the
 space-time in the history of relativity in which we first have the concept of space-time
 and then we have the concept of proper time which is introduced via the Lorentz metric.
  Once we understand that the space-time is a statistical concept from (\ref{1.1}) we can give
  a solution
  to the quantum measurement problem in the debate about quantum mechanics between Bohr and Einstein. In this debate Bohr insisted
  that with the probability interpretation quantum mechanics is very successful. On the other
  hand Einstein insisted that quantum mechanics is incomplete because of probability interpretation.
  Here we resolve this debate by constructing the above QED theory which is a quantum theory as
  the quantum mechanics
  and unlike quantum mechanics which needs probability interpretation we have that
  this QED theory is deterministic since it is not formulated with the space-time.

Similar to the usual Yang-Mills gauge theory we can generalize
this gauge theory with $U(1)$ gauge symmetry to nonabelian gauge
theories. As an illustration let us consider $SU(2)\otimes U(1)$
gauge symmetry where $SU(2)\otimes U(1)$ denotes the  direct
product of the groups $SU(2)$ and $U(1)$.

Similar to (\ref{1.1}) we consider the following energy integral:
\begin{equation}
L := 
\int_{s_0}^{s_1} [\frac12 Tr (D_1A_2-D_2A_1)^{*}(D_1A_2-D_2A_1) +
(D_0^*Z^*)(D_0Z)]ds \label{n1}
\end{equation}
where $Z= (z_1, z_2)^{T}$ is a two dimensional complex vector;
$A_j =\sum_{k=0}^{3}A_j^k t^k $ $(j=1,2)$ where $A_j^k$ denotes a
component of a gauge field $A^k$; $t^k=i T^k$ denotes a
generator of $SU(2)\otimes U(1)$ where $T^k$ denotes a
self-adjoint generator of $SU(2)\otimes U(1)$ 
 (here for simplicity we choose a
convention that the complex $i$ is absorbed by $t^k$ and $t^k$ is
absorbed by $A_j$; and  the notation $A_j$ is with a little
confusion with the notation $A_j$ in the above formulation of
(\ref{1.1}) where $A_j, j=1,2$ are real valued); and
$D_i=\frac{\partial}{\partial x_i}-e_0(\sum_{j=1}^2
A_j\frac{dx^j}{ds})$ for $i=1,2$; and
$D_0=\frac{d}{ds}-e_0(\sum_{j=1}^2 A_j\frac{dx^j}{ds})$
where $e_0$ is the bare electric charge for general interactions
including the strong and weak interactions.

From (\ref{n1}) we can develop a nonabelian gauge theory as
similar to that for the above abelian gauge theory. We have that
(\ref{n1}) is invariant under the following gauge transformation:
\begin{equation}
\begin{array}{rl}
Z^{\prime}(z(s)) & :=U(a(z(s)))Z(z(s)) \\
A_j^{\prime}(z(s)) & := U(a(z(s)))A_j(z(s))U^{-1}(a(z(s)))+
 U(a(z(s)))\frac{\partial U^{-1}}{\partial x^j}(a(z(s))),
j =1,2
\end{array}
\label{n2}
\end{equation}
where $U(a(z(s)))=e^{a(z(s))}$; $a(z(s))=\sum_k e_0 a^k (z(s))t^k$  for some functions $a^k$.
We shall mainly consider the case that $a$ is a function of the form $a(z(s))
=\sum_k \mbox{Re}\, \omega^k(z(s))t^k$ where $\omega^k$ are
analytic functions of $z$ (We let
$\omega(z(s)):=\sum_k\omega^k(z(s))t^k$ and we write
$a(z)=\mbox{Re}\,\omega(z)$).

The above gauge theory is based on the Banach space $X$ of
continuous functions $Z(z(s))$, $A_j(z(s))$, $j=1,2, s_0\leq s\leq
s_1$ on the one dimensional interval $\lbrack s_0, s_1 \rbrack$.

Since $L$ is positive and the theory is one dimensional (and thus
is simpler than the usual two dimensional Yang-Mills gauge theory)
we have that this gauge theory is similar to the Wiener measure
except that this gauge theory has a gauge symmetry. This gauge
symmetry gives a degenerate degree of freedom. In the physics
literature the usual way to treat the degenerate degree of freedom
of gauge symmetry is to introduce a gauge fixing condition to
eliminate the degenerate degree of freedom where each gauge fixing
will give equivalent physical results \cite{Fad}. There are
various gauge fixing conditions such as the Lorentz gauge
condition, the Feynman gauge condition, etc. We shall later in the
section on the Kac-Moody algebra adopt a gauge fixing condition
for the above gauge theory. This gauge fixing condition will also
be used to derive the quantum KZ equation in dual form which will
be regarded as a quantum Yang-Mill equation since its role will be
similar to the classical Yang-Mill equation derived from the
classical Yang-Mill gauge theory.

\section{Classical Dirac-Wilson loop } \label{sec4}

Similar to the Wilson loop in quantum field theory \cite{Wit} from
our quantum theory we introduce an analogue of Wilson loop, as
follows (We shall also call a Wilson loop as a Dirac-Wilson loop).

{\bf Definition}.
 A classical Wilson loop $W_R(C)$ is defined by :
\begin{equation}
W_R(C):= W(z_0, z_1):= Pe^{e_0\int_C A_jdx^j} \label{n4}
\end{equation}
where $R$ denotes a representation of $SU(2)$; $C(\cdot)=z(\cdot)$
is a fixed closed curve where the quantum gauge theories are based
on it as specific in the above section. As usual the notation $P$
in the definition of $W_R(C)$ denotes a path-ordered product
\cite{Wit}\cite{Kau}\cite{Baez}.

Let us give some remarks on the above definition of Wilson loop,
as follows.

1) We use the notation $W(z_0, z_1)$ to mean the Wilson loop
$W_R(C)$ which is based on the whole closed curve $z(\cdot)$. Here for
convenience we  use only the end points $z_0$ and $z_1$ of the
curve $z(\cdot)$ to denote this Wilson loop
(We keep in mind that the definition of $W(z_0, z_1)$ depends on
the whole curve $z(\cdot)$ connecting $z_0$ and $z_1$).

Then we  extend the definition of $W_R(C)$ to the case that
$z(\cdot)$ is not a closed curve with $z_0\neq z_1$. When
$z(\cdot)$ is not a closed curve we shall call $W(z_0, z_1)$ as a
Wilson line.

2) In constructing the Wilson loop we need to choose a
representation $R$ of the $SU(2)$ group. We shall see that because a Wilson line
$W(z_0, z_1)$ is with two variables $z_0$ and $z_1$ a natural representation of a Wilson line or
a Wilson loop is the tensor product of the usual two dimensional representation of
the $SU(2)$ for constructing the Wilson loop. $\diamond$

A basic property of a Wilson line $W(z_0,z_1)$ is that for a given continuous path $A_i, i=1,2$
on $[s_0, s_1]$
the Wilson line $W(z_0,z_1)$ exists on this path and has the
following transition property:
\begin{equation}
W(z_0,z_1)=W(z_0,z)W(z,z_1)
 \label{df2}
\end{equation}
where $W(z_0,z_1)$ denotes the Wilson line of a
curve $z(\cdot)$ which is with $z_0$ as the starting
point and $z_1$ as the ending point and $z$ is a
point on $z(\cdot)$ between $z_0$ and $z_1$.

This property can be prove as follows. We have that $W(z_0,z_1)$ is a limit
(whenever exists)
of ordered product of $e^{A_i\triangle x^i}$ and thus can be
written in the following form:
\begin{equation}
\begin{array}{rl}
W(z_0,z_1)= & I +
\int_{s^{\prime}}^{s^{\prime\prime}}
e_0 A_i(z(s))\frac{dx^i(s)}{ds}ds \\
&\\
 & + \int_{s^{\prime}}^{s^{\prime\prime}}
[\int_{s^{\prime}}^{s_2} e_0 A_i(z(s_1))\frac{dx^i(s_1)}{ds}ds_1]
e_0 A_i(z(s_2))\frac{dx^i(s_2)}{ds}ds_2 +\cdot\cdot\cdot
\end{array}
\label{df3}
\end{equation}
where $z(s^{\prime})=z_0$ and $z(s^{\prime\prime})=z_1$. Then
since $A_i$ are continuous on $[s^{\prime}, s^{\prime\prime}]$ and
$x^i(z(\cdot))$ are continuously differentiable on $[s^{\prime},
s^{\prime\prime}]$ we have that the series in (\ref{df3}) is
absolutely convergent. Thus the Wilson line $W(z_0,z_1)$ exists.
Then since $W(z_0,z_1)$ is the limit of ordered
product  we can write $W(z_0,z_1)$ in the form $W(z_0,z)W(z,z_1)$
by dividing $z(\cdot)$ into two parts at $z$. This proves the basic property
of Wilson line. $\diamond$

{\bf Remark (Classical version and quantum version of Wilsonloop)}. 
From the above property we have that the Wilson line
$W(z_0,z_1)$ exists in the classical pathwise sense where $A_i$
are as classical paths on $[s_0, s_1]$. This pathwise version of
the Wilson line $W(z_0,z_1)$; from the Feymann path integral point
of view; is as a partial description of the quantum version of the
Wilson line $W(z_0,z_1)$ which is as an operator when $A_i$ are as
operators. We shall in the next section derive and define a
quantum generator $J$ of $W(z_0,z_1)$ from the quantum gauge
theory. Then by using this generator $J$ we shall compute the
quantum version of the Wilson line $W(z_0,z_1)$.

We shall denote both the classical version and quantum version of Wilson line by
the same notation  $W(z_0,z_1)$ when there is no confusion.
$\diamond$

By following the usual approach of deriving a chiral symmetry from a gauge transformation of
a gauge field
we have the following chiral symmetry
which is derived by applying an analytic gauge transformation with an analytic function $\omega$
for the transformation:
\begin{equation}
W(z_0, z_1) \mapsto W^{\prime}(z_0, z_1)=U(\omega(z_1))
W(z_0, z_1)U^{-1}(\omega(z_0))
\label{n5}
\end{equation}
where $W^{\prime}(z_0, z_1)$ is a Wilson line with gauge field:
\begin{equation}
A_{\mu}^{\prime} =  \frac{\partial U(z)}{\partial x^{\mu}}U^{-1}(z) + U(z)A_{\mu}U^{-1}(z)
 \label{an5}
\end{equation}

This chiral symmetry is analogous to the chiral symmetry
of the usual guage theory where $U$ denotes an element of the gauge group \cite{Kau}.
Let us derive (\ref{n5}) as follows.
Let $U(z):=
U(\omega(z(s)))$ and $U(z+dz)\approx U(z)+\frac{\partial
U(z)}{\partial x^{\mu}}dx^{\mu}$ where $dz=(dx^1,dx^2)$. Following
 \cite{Kau} we have
\begin{equation}
\begin{array}{rl}
& U(z+ dz)(1+ e_0dx^{\mu}A_{\mu})U^{-1}(z)\\
&\\
 =& U(z+ dz)U^{-1}(z)
+ e_0 dx^{\mu}U(z+dz)A_{\mu}U^{-1}(s) \\
&\\
 \approx & 1+ \frac{\partial U(z)}{\partial
x^{\mu}}U^{-1}(z)dx^{\mu}
  +e_0 dx^{\mu}U(z+ dz)A_{\mu}U^{-1}(s) \\
  &\\
\approx & 1+ \frac{\partial U(z)}{\partial
x^{\mu}}U^{-1}(z)dx^{\mu}
+e_0 dx^{\mu}U(z)A_{\mu}U^{-1}(z) \\
&\\
 =: & 1+  \frac{\partial U(z)}{\partial
x^{\mu}}U^{-1}(z)dx^{\mu}
+e_0 dx^{\mu}U(z)A_{\mu}U^{-1}(z)\\
&\\
 =:& 1 +e_0 dx^{\mu}A_{\mu}^{\prime}
\end{array}
\label{n5b}
\end{equation}
From (\ref{n5b}) we have that (\ref{n5}) holds.

As analogous to
the WZW model in
conformal field theory \cite{Fra}\cite{Fuc}
from the above symmetry  we have the following formulas for the
variations $\delta_{\omega}W$ and $\delta_{\omega^{\prime}}W$ with
respect to this symmetry (\cite{Fra} p.621):
\begin{equation}
\delta_{\omega}W(z,z')=W(z,z')\omega(z)
\label{k1}
\end{equation}
and
\begin{equation}
\delta_{\omega^{\prime}}W(z,z')=-\omega^{\prime}(z')W(z,z')
\label{k2}
\end{equation}
where $z$ and $z'$ are independent variables and
$\omega^{\prime}(z')=\omega(z)$ when $z'=z$. In (\ref{k1}) the
variation is with respect to the $z$ variable while in (\ref{k2})
the variation is with respect to the $z'$ variable. This
two-side-variations when $z\neq z'$ can be derived as follows. For
the left variation we may let $\omega$ be analytic in a
neighborhood of $z$ and  extended as a continuously differentiable
function to a neighborhood of $z'$ such that $\omega(z')=0$ in
this neighborhood of $z'$. Then from (\ref{n5}) we have that
(\ref{k1}) holds. Similarly we may let $\omega^{\prime}$ be
analytic in a neighborhood of $z'$ and extended as a continuously
differentiable function to a neighborhood of $z$ such that
$\omega^{\prime}(z)=0$ in this neighborhood of $z$. Then we have
that (\ref{k2}) holds.

\section{A gauge fixing condition and affine Kac-Moody algebra} \label{sec6}

This section has two related purposes. One purpose is to find a
gauge fixing condition for eliminating the degenerate degree of
freedom from the gauge invariance of the above quantum gauge
theory in section 2. Then another purpose is to find an equation
for defining a generator $J$ of the Wilson line $W(z,z')$. This
defining equation of $J$ can then be used as a gauge fixing
condition. Thus with this defining equation of $J$ the
construction of the quantum gauge theory in section 2 is then
completed.

 We shall derive a quantum loop algebra (or the
affine Kac-Moody algebra) structure from the Wilson line $W(z,z')$
for the generator $J$ of $W(z,z')$. To this end let us first
consider the classical case. Since $W(z,z')$ is constructed from $
SU(2)$ we have that the mapping $z \to W(z,z')$ (We consider
$W(z,z')$ as a function of $z$ with $z'$ being fixed) has a loop
group structure \cite{Lus}\cite{Seg}. For a loop group we have the
following generators:
\begin{equation}
J_n^a = t^a z^n \qquad n=0, \pm 1, \pm 2, ...
\label{km1}
\end{equation}
These generators satisfy the following algebra:
\begin{equation}
[J_m^a, J_n^b] =
if_{abc}J_{m+n}^c
\label{km2}
\end{equation}
This is  the so called loop algebra \cite{Lus}\cite{Seg}. Let us
then introduce the following generating function $J$:
\begin{equation}
J(w) = \sum_a J^a(w)=\sum_a j^a(w) t^a
\label{km3}
\end{equation}
where we define
\begin{equation}
J^a(w)= j^a(w) t^a :=
\sum_{n=-\infty}^{\infty}J_n^a(z) (w-z)^{-n-1}
\label{km3a}
\end{equation}

From $J$ we have
\begin{equation}
J_n^a=  \frac{1}{2\pi i}\oint_z dw (w-z)^{n}J^a(w)
\label{km4}
\end{equation}
where $\oint_z$ denotes a closed contour integral  with center $z$. This formula
can be interpreted as that
$J$ is the generator of the loop group and that
$J_n^a$ is the directional generator in the direction
$\omega^a(w)= (w-z)^n$. We may generalize $(\ref{km4})$
to the following  directional generator:
\begin{equation}
  \frac{1}{2\pi i}\oint_z dw \omega(w)J(w)
\label{km5}
\end{equation}
where the analytic function
$\omega(w)=\sum_a \omega^a(w)t^a$ is regarded
as a direction and we define
\begin{equation}
 \omega(w)J(w):= \sum_a \omega^a(w)J^a
\label{km5a}
\end{equation}

Then since $W(z,z')\in SU(2)$, from the variational formula
(\ref{km5}) for the loop algebra of the loop group of $SU(2)$ we
have that the variation of $W(z,z')$ in the direction $\omega(w)$
is given by
\begin{equation}
W(z,z')
  \frac{1}{2\pi i}\oint_z dw \omega(w)J(w)
\label{km6}
\end{equation}

Now let us consider the quantum case which is based on the quantum
gauge theory in section 2. For this quantum case we shall define a
quantum generator $J$ which is analogous to the $J$ in
(\ref{km3}). We shall choose the equations (\ref{n8b}) and
(\ref{n6}) as  the equations for defining the quantum generator
$J$. Let us first give a formal derivation of the equation
(\ref{n8b}), as follows.
 Let us consider the
following formal functional integration:
\begin{equation}
\langle W(z,z')A(z) \rangle := \int dA_1dA_2dZ^{*}dZ  e^{-L}
W(z,z')A(z) \label{n8a}
\end{equation}
where $A(z)$ denotes a field from the quantum gauge theory (We
first let $z'$ be fixed as a parameter).

Let us  do a calculus of variation on this integral to derive a variational
equation by applying a gauge transformation on (\ref{n8a}) as follows
(We remark that such variational equations are usually called the
Ward identity in the  physics literature).

Let $(A_1,A_2,Z)$ be regarded as a coordinate system of the integral
(\ref{n8a}).
Under a gauge transformation (regarded as
a change of coordinate) with gauge function $a(z(s))$ this coordinate
is changed to another coordinate denoted by
$(A_1^{\prime}, A_2^{\prime}, Z^{\prime})$.
As similar to the usual change of variable for integration we have that
the integral  (\ref{n8a}) is unchanged
under a change of variable and we have the following
equality:
\begin{equation}
\begin{array}{rl}
& \int dA_1^{\prime}
 dA_2^{\prime}dZ^{\prime *}dZ^{\prime}
 e^{-L^{\prime}} W^{\prime}(z,z')A^{\prime}(z)
= \int dA_1dA_2dZ^{*}dZ  e^{-L} W(z,z')A(z)
\end{array}
\label{int}
\end{equation}
where $W^{\prime}(z,z')$ denotes the Wilson line based on
$A_1^{\prime}$ and $A_2^{\prime}$ and similarly $A^{\prime}(z)$
denotes  the field obtained from $A(z)$ with $(A_1, A_2,Z)$
replaced by $(A_1^{\prime}, A_2^{\prime},Z^{\prime})$.

Then it can be shown that the differential is unchanged under a
gauge transformation \cite{Fad}:
\begin{equation}
dA_1^{\prime}
dA_2^{\prime}dZ^{\prime *}dZ^{\prime}
= dA_1dA_2dZ^{*}dZ
\label{int2}
\end{equation}
Also by the gauge invariance property the factor $e^{-L}$ is
unchanged under a gauge transformation. Thus from (\ref{int}) we
have
\begin{equation}
0 = \langle W^{\prime}(z,z')A^{\prime}(z)\rangle -
  \langle W(z,z')A(z)\rangle
\label{w1}
\end{equation}
where the correlation notation
$\langle \cdot\rangle$ denotes the integral with
respect to the differential
\begin{equation}
e^{-L}dA_1dA_2dZ^{*}dZ
\label{w1a}
\end{equation}

We can now carry out the calculus of variation. From the gauge
transformation we have the formula:
\begin{equation}
W^{\prime}(z,z')=U(a(z))W(z,z')U^{-1}(a(z'))\,, \label{aw1a}
\end{equation}
where $a(z)=\mbox{Re}\,\omega(z)$. This
gauge transformation gives a variation of $W(z,z')$ with the
gauge function $a(z)$
as the variational direction $a$
in the variational formulas (\ref{km5}) and  (\ref{km6}). Thus analogous
to the variational formula (\ref{km6}) we have that the variation
of $W(z,z')$ under this gauge transformation is given by
\begin{equation}
W(z,z')
  \frac{1}{2\pi i}\oint_z dw a(w)J(w)
\label{int3}
\end{equation}
where the generator $J$ for this variation is to
be specific. This $J$ will be a quantum generator
which generalizes the classical generator $J$ in
(\ref{km6}).
Thus under a gauge transformation with gauge function $a(z)$ from (\ref{w1}) we have the
following variational equation:
\begin{equation}
0= \langle W(z,z')[\delta_{a}A(z)+\frac{1}{2\pi i}\oint_z
dw a(w)J(w)A(z)]\rangle
\label{w2}
\end{equation}
where $\delta_{a}A(z) $
denotes the variation of the field
$A(z)$ in the direction $a(z)$.
From this equation an ansatz of
$J$ is that $J$ satisfies the following equation:
\begin{equation}
W(z,z')[\delta_{a}A(z)+\frac{1}{2\pi i}\oint_z
dw a(w)J(w)A(z)] =0 \label{n8bb}
\end{equation}
From this equation we have the following variational equation:
\begin{equation}
\delta_{a}A(z)=\frac{-1}{2\pi i}\oint_z dw a(w)J(w)A(z)
\label{n8bre}
\end{equation}
This completes the formal calculus of variation. Now (with the
above derivation as a guide) we choose the following equation (\ref{n8b}) as one of the
equation for defining the generator $J$:
\begin{equation}
\delta_{\omega}A(z)=\frac{-1}{2\pi i}\oint_z dw\omega(w)J(w)A(z)
\label{n8b}
\end{equation}
where we generalize the direction $a(z)=\mbox{Re}\,\omega(z)$ to
the analytic direction $\omega(z)$ (This generalization has the
effect of extending the real measure of the pure gauge part of the
gauge theory to include the complex Feynman path integral since it
gives the transformation $ds\to i ds$ for the integral of the
Wilson line $W(z, z')$).

Let us now choose one more equation for determine the generator
$J$ in (\ref{n8b}). This choice will be as
a gauge fixing
condition. As analogous to the WZW model in conformal field
theory \cite{Fra}\cite{Fuc} \cite{Kni}  let us consider a $J$
given by
\begin{equation}
J(z) := -k_0 W^{-1}(z, z')\partial_z W(z, z') \label{n6}
\end{equation}
where we define $\partial_z=\partial_{x^1} +i\partial_{x^2} $ and we set $z'=z$ after
the differentiation with respect to $z$; $ k_0>0 $ is a constant which is fixed when
the $J$ is determined to be of the form (\ref{n6}) and the minus sign is chosen by
convention. In the WZW model \cite{Fra}\cite{Kni}
 the $J$ of the form (\ref{n6})
is the  generator  of the chiral symmetry of the WZW model. We can
write the $J$ in (\ref{n6}) in the following form:
\begin{equation}
 J(w) = \sum_a J^a(w) =
\sum_a j^a(w) t^a
\label{km8}
\end{equation}
We see that the generators $t^a$ of $SU(2)$ appear in this form of
$J$ and  this form is analogous to the classical $J$ in
(\ref{km3}). This shows that
 this $J$ is a possible candidate for the generator
$J$ in (\ref{n8b}).

Since $W(z,z')$ is constructed by gauge field we need to have a
gauge fixing for the computations related to $W(z,z')$. Then since
the $J$ in (\ref{n8b}) and (\ref{n6}) is constructed from
$W(z,z')$ we have that in defining this $J$ as the generator $J$
of $W(z,z')$ we have chosen a condition for the gauge fixing. In
this paper we shall always choose this defining equations
(\ref{n8b}) and (\ref{n6}) for $J$ as the gauge fixing condition.

In summary we introduce the following definition.

{\bf Definition}. The generator $J$ of the quantum Wilson line $W(z,z')$ whose classical version
is defined by (\ref{n4}), is an operator defined by the two conditions (\ref{n8b}) and (\ref{n6}).
$\diamond$

{\bf Remark}. We remark that the condition (\ref{n6}) first defines $J$ classically. Then
the condition (\ref{n8b}) raises this classical $J$ to the quantum generator $J$. $\diamond$

Now we want to show that this generator $J$ in (\ref{n8b}) and
(\ref{n6}) can be uniquely solved (This means that the gauge
fixing condition has already fixed the gauge that the degenerate
degree of freedom of gauge invariance has been eliminated so that
we can carry out computation).

Let us now solve $J$.
From (\ref{n5}) and (\ref{n6}) we
have that the variation $\delta_{\omega}J$ of the generator $J$ in
(\ref{n6}) is given by \cite{Fra}(p.622) \cite{Kni}:
\begin{equation}
\delta_{\omega}J= \lbrack J, \omega\rbrack -k_0\partial_z \omega
\label{n8c}
\end{equation}

From (\ref{n8b}) and (\ref{n8c}) we have that $J$ satisfies the
following relation of current algebra
\cite{Fra}\cite{Fuc}\cite{Kni}:
\begin{equation}
J^a(w)J^b(z)=\frac{k_0\delta_{ab}}{(w-z)^2}
+\sum_{c}if_{abc}\frac{J^c(z)}{(w-z)} \label{n8d}
\end{equation}
where as a convention the regular term of the product
$J^a(w)J^b(z)$ is omitted. Then by following
\cite{Fra}\cite{Fuc}\cite{Kni} from (\ref{n8d}) and (\ref{km8})
we can show that the $J_n^a$ in (\ref{km3})  for the corresponding Laurent series of
the quantum generator $J$
satisfy the
following  Kac-Moody algebra:
\begin{equation}
[J_m^a, J_n^b] = if_{abc}J_{m+n}^c + k_0 m\delta_{ab}\delta_{m+n, 0}
\label{n8}
\end{equation}
where $k_0$ is  usually called the central extension or the level of
the Kac-Moody algebra.

{\bf Remark}. Let us also consider the other side of the chiral
symmetry.
Similar to the $J$ in (\ref{n6}) we define a generator
$J^{\prime}$ by:
\begin{equation}
J^{\prime}(z')= k_0 \partial_{z'}W(z, z')W^{-1}(z, z') \label{d1}
\end{equation}
where after differentiation with respect to $z'$ we set $z=z'$.
Let us then consider
 the following formal correlation:
\begin{equation}
\langle A(z')W(z,z') \rangle := \int
dA_1dA_2dZ^{*}dZ
  A(z')W(z,z')e^{-L}
\label{n8aa}
\end{equation}
where $z$ is fixed. By an approach similar to the above derivation
of (\ref{n8b}) we have the following  variational equation:
\begin{equation}
\delta_{\omega^{\prime}}A(z') =\frac{-1}{2\pi i}\oint_{z^{\prime}}
dwA(z')J^{\prime}(w) \omega^{\prime}(w) \label{n8b1}
\end{equation}
where as a gauge fixing we choose the $J^{\prime}$ in (\ref{n8b1})
be the $J^{\prime}$ in (\ref{d1}). Then similar to (\ref{n8c}) we
also have
\begin{equation}
\delta_{\omega^{\prime}}J^{\prime}= \lbrack  J^{\prime},
\omega^{\prime}\rbrack -k_0 \partial_{z'} \omega^{\prime}
\label{n8c1}
\end{equation}
Then from (\ref{n8b1}) and (\ref{n8c1}) we can derive the current
algebra and the Kac-Moody algebra for $J^{\prime}$ which are of
the same form of (\ref{n8d}) and (\ref{n8}). From this we  have
$J^{\prime}=J$. $\diamond$

Now with the above current algebra $J$ and the formula (\ref{n8b}) we can
follow the usual approach
in conformal field theory to derive a quantum
Knizhnik-Zamolodchikov (KZ) equation for the product of
primary fields in a conformal field theory \cite{Fra}\cite{Fuc}\cite{Kni}.
We derive the KZ equation for the product of $n$ Wilson lines $W(z, z')$.
Here an important point is that from the two sides of
$W(z, z')$  we can derive two quantum KZ equations which are
dual to each other. These two quantum KZ equations are different from the usual KZ equation in that
they are equations for the quantum operators $W(z, z')$ while the usual KZ equation is for
the correlations of quantum operators.
With this difference we can follow the usual approach
in conformal field theory to derive the following quantum
Knizhnik-Zamolodchikov equation \cite{Fra} \cite{Fuc}\cite{Ng}:
\begin{equation}
\partial_{z_i}
 W(z_1, z_1^{\prime})\cdot\cdot\cdot
W(z_n, z_n^{\prime})
=\frac{-e_0^2}{k_0+g_0}
\sum_{j\neq i}^{n}
\frac{\sum_a t_i^a \otimes t_j^a}{z_i-z_j}
 W(z_1, z_1^{\prime})\cdot\cdot\cdot
W(z_n, z_n^{\prime})
\label{n9}
\end{equation}
for $i=1, ..., n$ where $g_0$ denotes the dual Coxeter number of a group multiplying with $e_0^2$
and we have $g_0=2e_0^2$ for the group $SU(2)$ (When the gauge group is $U(1)$ we have $g_0=0$).
We remark that in (\ref{n9}) we have
defined $t_i^a:= t^a$ and:
\begin{equation}
\begin{array}{rl}
  t_i^a \otimes t_j^a W(z_1, z_1^{\prime})\cdot\cdot\cdot
W(z_n, z_n^{\prime})
 := W(z_1, z_1^{\prime})\cdot\cdot\cdot
 [t^aW(z_i, z_i^{\prime})]\cdot\cdot\cdot
[t^aW(z_j, z_j^{\prime})]\cdot\cdot\cdot
W(z_n, z_n^{\prime})
\end{array}
\label{n9a}
\end{equation}

It is interesting and important that we also have
the following quantum Knizhnik-Zamolodchikov equation with repect to
the $z_i^{\prime}$ variables which is dual to (\ref{n9}):
\begin{equation}
\partial_{z_i^{\prime}}
 W(z_1,z_1^{\prime})\cdot\cdot\cdot W(z_n,z_n^{\prime})
= \frac{-e_0^2}{k_0+g_0}\sum_{j\neq i}^{n}
 W(z_1, z_1^{\prime})\cdot\cdot\cdot
W(z_n, z_n^{\prime})
\frac{\sum_a t_i^a\otimes t_j^a}{z_j^{\prime}-z_i^{\prime}}
\label{d8}
\end{equation}
for $i=1, ..., n$
where we have defined:
\begin{equation}
\begin{array}{rl}
   W(z_1, z_1^{\prime})\cdot\cdot\cdot
W(z_n, z_n^{\prime})t_i^a \otimes t_j^a 
 := W(z_1, z_1^{\prime})\cdot\cdot\cdot
 [W(z_i, z_i^{\prime})t^a]\cdot\cdot\cdot
[W(z_j, z_j^{\prime})t^a]\cdot\cdot\cdot
W(z_n, z_n^{\prime})
\end{array}
\label{d8a}
\end{equation}

{\bf Remark}. From the quantum gauge theory we derive the above
quantum KZ equation in dual form by calculus of variation. This
quantum KZ equation in dual form may be considered as a quantum
Euler-Lagrange equation or as a quantum Yang-Mills equation since
it is analogous to the classical Yang-Mills equation which is
derived from the classical Yang-Mills gauge theory by calculus of
variation.
 $\diamond$

\section{Solving quantum KZ equation in dual form}\label{sec8a}

Let us consider the following product of two quantum
Wilson lines:
\begin{equation}
G(z_1,z_2, z_3, z_4):=
 W(z_1, z_2)W(z_3, z_4)
\label{m1}
\end{equation}
where the two quantum Wilson lines $W(z_1, z_2)$ and
$W(z_3, z_4)$ represent two pieces
of curves starting at $z_1$ and $z_3$ and ending at
$z_2$ and $z_4$ respectively.
We have that this product $G$ satisfies the KZ equation for the
variables $z_1$, $z_3$ and satisfies the dual KZ equation
for the variables $z_2$ and $z_4$.
Then
by solving the two-variables-KZ equation in (\ref{n9}) we have that a form of $G$ is
given by \cite{Chari}\cite{Koh}\cite{Dri}:
\begin{equation}
e^{-\hat{t}\log [\pm (z_1-z_3)]}C_1
\label{m2}
\end{equation}
where $\hat{t}:=\frac{e_0^2}{k_0+g_0}\sum_a t^a \otimes t^a$
and $C_1$ denotes a constant matrix which is independent
of the variable $z_1-z_3$.

We see that $G$ is a multi-valued analytic function where the
determination of the $\pm$ sign depended on the choice of the
branch.

Similarly by solving the dual two-variable-KZ equation
 in (\ref{d8}) we have that
$G$ is of the form
\begin{equation}
C_2e^{\hat{t}\log [\pm (z_4-z_2)]}
\label{m3}
\end{equation}
where $C_2$ denotes a constant matrix which is independent
of the variable $z_4-z_2$.

From (\ref{m2}), (\ref{m3}) and letting:
\begin{equation}
C_1=Ae^{\hat{t}\log[\pm (z_4-z_2)]}, \quad
C_2=e^{-\hat{t}\log[\pm(z_1-z_3)]}A
 \label{am3}
\end{equation}
 where $A$ is a constant matrix  we have that
$G$ is given by
\begin{equation}
G(z_1, z_2, z_3, z_4)=
e^{-\hat{t}\log [\pm (z_1-z_3)]}Ae^{t\log [\pm (z_4-z_2)]}
\label{m4}
\end{equation}
where at the singular case that $z_1=z_3$ we simply define $\log [\pm (z_1-z_3)]=0$. Similarly
for $z_2=z_4$.

Let us find a form of the initial operator $A$. We notice that
there are two operators 
$\Phi_{\pm}(z_1-z_3):=e^{-\hat{t}\log [\pm (z_1-z_3)]}$ and
$\Psi_{\pm}(z_4-z_2) := e^{\hat{t}\log [\pm (z_4-z_2)]}$
acting on
the two sides of $A$ respectively where  the two independent
variables  $z_1, z_3$ of $\Phi_{\pm}$ are mixed from the two
quantum Wilson lines $W(z_1, z_2)$ and $W(z_3, z_4)$ respectively
and the the two independent variables  $z_2, z_4$ of $\Psi_{\pm}$
are mixed from the two quantum Wilson lines $W(z_1, z_2)$ and
$W(z_3, z_4)$ respectively. From this we determine the form of $A$
as follows.

Let $D$ denote a representation of $SU(2)$. Let $D(g)$ represent an element $g$ of  $SU(2)$
and let $D(g)\otimes D(g)$ denote the tensor product representation of $SU(2)$. Then
in the KZ equation we define
\begin{equation}
[t^a\otimes t^a] [D(g_1)\otimes D(g_1)]\otimes
[D(g_2)\otimes D(g_2)]
:=[t^aD(g_1)\otimes D(g_1)]\otimes
[t^aD(g_2)\otimes D(g_2)]
\label{tensorproduct}
\end{equation}
and
\begin{equation}
[D(g_1)\otimes D(g_1)]\otimes
[D(g_2)\otimes D(g_2)][t^a\otimes t^a]
:=[D(g_1)\otimes D(g_1)t^a]\otimes
[D(g_2)\otimes D(g_2)t^a]
\label{tensorproduct2}
\end{equation}

Then we let $U({\bf a})$
denote the universal
enveloping algebra
where ${\bf a}$ denotes an algebra which is formed by the Lie
algebra $su(2)$ and the identity matrix.

Now let the initial operator $A$ be of the form $A_1\otimes A_2\otimes A_3\otimes A_4$
with $A_i,i=1,...,4$
taking values in $U({\bf a})$.
In
this case we have that in (\ref{m4}) the operator
$\Phi_{\pm}(z_1-z_2):=e^{-\hat{t}\log [\pm (z_1-z_3)]}$ acts on $A$ from
the left via the following formula:
\begin{equation}
t^a\otimes t^a A=
[t^a A_1]\otimes A_2\otimes [t^a A_3]\otimes A_4
\label{ini2}
\end{equation}

Similarly the operator
$\Psi_{\pm}(z_1-z_2):=e^{\hat{t}\log [\pm (z_1-z_3)]}$
in (\ref{m4}) acts on $A$ from the right via the following formula:
\begin{equation}
A t^a\otimes t^a =
A_1\otimes [A_2 t^a]\otimes A_3\otimes[A_4 t^a]
\label{ini3}
\end{equation}

We may generalize the above tensor product of
two quantum Wilson lines as follows.
Let us consider a tensor product of $n$ quantum Wilson lines:
$W(z_1, z_1^{\prime})\cdot\cdot\cdot W(z_n, z_n^{\prime})$
where the variables $z_i$, $z_i^{\prime}$
are all independent. By solving the two KZ equations
we have that this tensor product is given by:
\begin{equation}
W(z_1, z_1^{\prime})\cdot\cdot\cdot W(z_n, z_n^{\prime})
=\prod_{ij} \Phi_{\pm}(z_i-z_j)
A\prod_{ij}
\Psi_{\pm}(z_i^{\prime}-z_j^{\prime})
\label{tensor}
\end{equation}
where $\prod_{ij}$ denotes a product of
$\Phi_{\pm}(z_i-z_j)$ or
$\Psi_{\pm}(z_i^{\prime}-z_j^{\prime})$
for $i,j=1,...,n$ where $i\neq j$.
In (\ref{tensor}) the initial operator
$A$ is represented as a tensor product of
operators $A_{iji^{\prime}j^{\prime}}, i,j,i^{\prime}, j^{\prime}=1,...,n$ where
each $A_{iji^{\prime}j^{\prime}}$ is of the form of the initial operator $A$ in
the above tensor product of two-Wilson-lines case and is acted by $\Phi_{\pm}(z_i-z_j)$ or
$\Psi_{\pm}(z_i^{\prime}-z_j^{\prime})$ on its two sides respectively.

\section{Computation of quantum Wilson lines}\label{sec 8aa}

Let us consider the following product of two quantum
Wilson lines:
\begin{equation}
G(z_1,z_2, z_3, z_4):=
W(z_1, z_2)W(z_3, z_4)
\label{h1}
\end{equation}
where the two quantum Wilson lines $W(z_1, z_2)$ and
$W(z_3, z_4)$ represent two pieces
of curves starting at $z_1$ and $z_3$ and ending at
$z_2$ and $z_4$ respectively.
As shown in the above section we have that $G$
is given by the following formula:
\begin{equation}
G(z_1, z_2, z_3, z_4)=
e^{-\hat{t}\log [\pm (z_1-z_3)]}Ae^{\hat{t}\log [\pm (z_4-z_2)]}
\label{m4a}
\end{equation}
where the product is
a 4-tensor.
Let us set $z_2=z_3$. Then
the 4-tensor $W(z_1, z_2)W(z_3, z_4)$ is reduced to the 2-tensor
$W(z_1, z_2)W(z_2, z_4)$.
By using (\ref{m4a}) the 2-tensor
$W(z_1, z_2)W(z_2, z_4)$
is given by:
\begin{equation}
W(z_1, z_2)W(z_2, z_4)
=e^{-\hat{t}\log [\pm (z_1-z_2)]}A_{14}e^{\hat{t}\log [\pm (z_4-z_2)]}
\label{closed1}
\end{equation}
where $A_{14}=A_1\otimes A_4$ is a 2-tensor reduced from the 4-tensor
$A=A_1\otimes A_2\otimes A_3\otimes A_4$ in (\ref{m4a}). In this reduction the $\hat{t}$ operator
of $\Phi=e^{-\hat{t}\log [\pm (z_1-z_2)]}$ acting on the left side of $A_1$ and $A_3$ in $A$ is reduced
to acting on the left side of $A_1$ and $A_4$ in $A_{14}$. Similarly
the $\hat{t}$ operator of $\Psi=e^{-\hat{t}\log [\pm (z_4-z_2)]}$ acting on the right side of $A_2$
and $A_4$ in $A$ is reduced to acting on the right side of $A_1$ and $A_4$ in $A_{14}$.

Then since $t$ is a 2-tensor operator we have that $t$ is as a matrix acting on the two sides of
the 2-tensor $A_{14}$ which is also as a matrix with the same dimension as $t$.
Thus $\Phi$ and $\Psi$ are as matrices of the same dimension as the matrix
$A_{14}$  acting on $A_{14}$ by the usual matrix operation.
Then since $t$ is a Casimir operator for the 2-tensor group representation of $SU(2)$ we have that
$\Phi $  and $\Psi $ commute
with $A_{14}$ since  $\Phi $  and $\Psi$ are exponentials
of $t$ (We remark that $\Phi $  and $\Psi $ are in general not commute with
the 4-tensor initial operator $A$).
Thus we have
\begin{equation}
e^{-\hat{t}\log [\pm (z_1-z_2)]}A_{14}e^{\hat{t}\log[\pm (z_4-z_2)]}
=e^{-\hat{t}\log [\pm (z_1-z_2)]}e^{\hat{t}\log[\pm (z_4-z_2)]}A_{14}
\label{closed1a}
\end{equation}

We let $W(z_1, z_2)W(z_2, z_4)$ be as a representation of the quantum Wilson line $W(z_1,z_4)$
and we write $W(z_1,z_4)=W(z_1, z_2)W(z_2, z_4)$.
Then we have the following representation of
$W(z_1,z_4)$:
\begin{equation}
W(z_1,z_4)=W(z_1,w_1)W(w_1,z_4)
=e^{-\hat{t}\log [\pm (z_1-w_1)]}e^{\hat{t}\log[\pm (z_4-w_1)]}A_{14}
\label{closed1a1}
\end{equation}
This representation of the quantum Wilson line $W(z_1,z_4)$ means that the line (or path)
with end points $z_1$ and $z_4$ is specific that it passes the intermediate point $w_1=z_2$.
This representation shows the quantum nature that the path is  not specific
at other intermediate points except the intermediate point $w_1=z_2$. This unspecification of
the path is of the same quantum nature of the Feymann path description of quantum mechanics.

Then let us consider another representation of the quantum Wilson line $W(z_1,z_4)$.
We consider $W(z_1,w_1)W(w_1,w_2)W(w_2,z_4)$ which is obtained from
the tensor $W(z_1,w_1)W(u_1,w_2)W(u_2,z_4)$ by two reductions where $z_j$, $w_j$, $u_j$, $j=1,2$
are independent variables. For this representation we have:
\begin{equation}
W(z_1,w_1)W(w_1,w_2)W(w_2,z_4)
=e^{-\hat{t}\log [\pm (z_1-w_1)]}e^{-\hat{t}\log [\pm (z_1-w_2)]}
e^{\hat{t}\log[\pm (z_4-w_1)]}e^{\hat{t}\log[\pm (z_4-w_2)]}A_{14}
\label{closed1a2}
\end{equation}
This representation of the quantum Wilson line $W(z_1,z_4)$ means that the line (or path)
with end points $z_1$ and $z_4$ is specific that it passes the intermediate points $w_1$ and $w_2$.
This representation shows the quantum nature that the path is  not specific
at other intermediate points except the intermediate points $w_1$ and $w_2$.
This unspecification of the path is of the same quantum nature of
the Feymann path description of quantum mechanics.
Similarly we may represent the quantum Wilson line $W(z_1,z_4)$ by path with end points $z_1$
and $z_4$ and is specific only to pass at finitely many intermediate points.
Then we let the quantum Wilson line $W(z_1,z_4)$ as an equivalent class of all these representations. Thus we may write
\begin{equation}
\begin{array}{ll}
W(z_1,z_4) =W(z_1,w_1)W(w_1,z_4)=
W(z_1,w_1)W(w_1,w_2)W(w_2,z_4)=\cdots
\end{array}
\label{aclosed1a2}
\end{equation}
where the equalities are in the sense of in the same equivalent class.

{\bf Remark}. Since $A_{14}$ is a 2-tensor
we have that a natural group representation for the Wilson line $W(z_1,z_4)$ is
the 2-tensor group representation of the group $SU(2)$.

\section{Representing braiding of curves by quantum Wilson lines}\label{sec 9aa}

Consider again the product $G(z_1, z_2, z_3,
z_4)=W(z_1,z_2)W(z_3,z_4)$. We have that $G$ is a multi-valued
analytic function where the determination of the $\pm$ sign
depended on the choice of the branch.

Let the two pieces of curves be crossing at $w$. Then we have $W(z_1,z_2)=W(z_1,w)W(w,z_2)$ and
 $W(z_3,z_4)=W(z_3,w)W(w,z_4)$. Thus we have
\begin{equation}
W(z_1,z_2)W(z_3,z_4)=
W(z_1,w)W(w,z_2)W(z_3,w)W(w,z_4)
\label{h2}
\end{equation}

If we interchange $z_1$ and $z_3$, then from
(\ref{h2}) we have the following ordering:
\begin{equation}
 W(z_3,w)W(w, z_2)W(z_1,w)W(w,z_4)
\label{h3}
\end{equation}

Now let us choose a  branch. Suppose that
these two curves are cut from a knot and that
following the orientation of a knot the
curve represented by  $W(z_1,z_2)$ is before the
curve represented by  $W(z_3,z_4)$. Then we fix a branch such that the  product in (\ref{m4a}) is
with two positive signs :
\begin{equation}
W(z_1,z_2)W(z_3,z_4)=
e^{-\hat{t}\log(z_1-z_3)}Ae^{\hat{t}\log(z_4-z_2)}
\label{h4}
\end{equation}

Then if we interchange $z_1$ and $z_3$ we have
\begin{equation}
W(z_3,w)W(w, z_2)W(z_1,w)W(w,z_4) =
e^{-\hat{t}\log[-(z_1-z_3)]}Ae^{\hat{t}\log(z_4-z_2)}
\label{h5}
\end{equation}
From (\ref{h4}) and (\ref{h5}) as a choice of branch we have
\begin{equation}
W(z_3,w)W(w, z_2)W(z_1,w)W(w,z_4) =
R W(z_1,w)W(w,z_2)W(z_3,w)W(w,z_4)
\label{m7a}
\end{equation}
where $R=e^{-i\pi \hat{t}}$ is the monodromy of the KZ equation.
In (\ref{m7a}) $z_1$ and $z_3$ denote two points on a closed curve
such that along the direction of the curve the point
$z_1$ is before the point $z_3$ and in this case we choose
a branch such that the angle of $z_3-z_1$ minus the angle
of $z_1-z_3$ is equal to $\pi$.

{\bf Remark}. We may use other representations of $W(z_1,z_2)W(z_3,z_4)$. For example we may use
the following representation:
\begin{equation}
\begin{array}{rl}
 &W(z_1,w)W(w, z_2)W(z_3,w)W(w,z_4)\\
 &\\
= &e^{-\hat{t}\log(z_1-z_3)}e^{-2\hat{t}\log(z_1-w)}e^{-2\hat{t}\log(z_3-w)}Ae^{t\log(z_4-z_2)}
e^{2\hat{t}\log(z_4-w)}e^{2\hat{t}\log(z_2-w)}
\end{array}
\label{h4a}
\end{equation}
Then the interchange of $z_1$ and $z_3$ changes only $z_1-z_3$ to $z_3-z_1$.
Thus the formula (\ref{m7a}) holds. Similarly
all other representations of $W(z_1,z_2)W(z_3,z_4)$ will give the same result. $\diamond$

Now from (\ref{m7a}) we can take a convention that the ordering
(\ref{h3}) represents that the curve represented by  $W(z_1,z_2)$
is up-crossing the curve represented by  $W(z_3,z_4)$ while
(\ref{h2}) represents zero crossing of these two curves.

Similarly from the dual KZ equation as a choice of branch which
is consistent with the above formula we have
\begin{equation}
W(z_1,w)W(w,z_4)W(z_3,w)W(w,z_2)=
W(z_1,w)W(w,z_2)W(z_3,w)W(w,z_4)R^{-1}
\label{m8a}
\end{equation}
where $z_2$ is before $z_4$. We take a convention that the
ordering in (\ref{m8a}) represents that the curve represented by
$W(z_1,z_2)$ is under-crossing the curve represented by
$W(z_3,z_4)$. Here along the orientation of a closed curve the
piece of curve represented by $W(z_1,z_2)$ is before the piece of
curve represented by $W(z_3,z_4)$. In this case since the angle of
$z_3-z_1$ minus the angle of $z_1-z_3$ is equal to $\pi$ we have
that the angle of $z_4-z_2$ minus the angle of $z_2-z_4$ is also
equal to $\pi$ and this gives the $R^{-1}$ in this formula
(\ref{m8a}).

From (\ref{m7a}) and (\ref{m8a}) we have
\begin{equation}
 W(z_3,z_4)W(z_1,z_2)=
RW(z_1,z_2)W(z_3,z_4)R^{-1} \label{m9}
\end{equation}
where $z_1$ and $z_2$ denote the end points of a curve which is before a curve
with end points $z_3$ and $z_4$.
From (\ref{m9}) we see that the algebraic structure of these
quantum Wilson lines $W(z,z')$
is analogous to the quasi-triangular quantum
group \cite{Fuc}\cite{Chari}.

\section{Computation of quantum Dirac-Wilson loop}\label{sec10a}

Let us consider again the quantum Wilson line $W(z_1,z_4)=W(z_1,
z_2)W(z_2, z_4)$. Let us set $z_1=z_4$. In this case the quantum
Wilson line forms a closed loop. Now in (\ref{closed1a}) with
$z_1=z_4$ we have that $e^{-\hat{t}\log  \pm (z_1-z_2)}$ and $e^{\hat{t}\log
\pm (z_1-z_2)}$ which come from the two-side KZ equations cancel
each other and from the multi-valued property of the $\log$
function we have
\begin{equation}
W(z_1, z_1) =R^{N}A_{14} \quad\quad N=0, \pm 1, \pm 2, ...
\label{closed2}
\end{equation}
where $R=e^{-i\pi \hat{t}}$ is the monodromy of the KZ equation \cite{Chari}.

{\bf Remark}. It is clear that if we use other representation of
the quantum Wilson loop $W(z_1,z_1)$ (such as
the representation $W(z_1,z_1)=W(z_1,w_1)W(w_1,w_2)W(w_2,z_1)$) then we will get
the same result as (\ref{closed2}).

{\bf Remark}. For simplicity we shall drop the subscript of $A_{14}$ in (\ref{closed2})
and simply write $A_{14}=A$.

\section{Winding number of Dirac-Wilson loop as quantization}\label{sec8b}

We have the equation (\ref{closed2})
 where the integer $N$ is as a winding number.
 Then when the gauge group is $U(1)$ we have
\begin{equation}
W(z_1, z_1)
=R_{U(1)}^{N} A
\label{closed21a}
\end{equation}
where  $R_{U(1)}$ denotes the monodromy of the KZ equation for $U(1)$. We have
\begin{equation}
R_{U(1)}^{N}=e^{iN\frac{\pi e_0^2}{k_0+g_0}} \qquad N=0, \pm 1, \pm 2, ...
\label{closed21}
\end{equation}
 where the constant $e_0$  denotes the bare electric charge (and $g_0=0$ for $U(1)$ group).
 The winding number $N$ is as  the quantization property of photon.
We show in the following section that the Dirac-Wilson loop $W(z_1, z_1)$ with
the abelian $U(1)$ group is a model of the photon.

\section{Magnetic monopole is photon with specific frequency}

We see that the Dirac-Wilson loop is an exactly solvable nonlinear
observable. Thus we may regard it as a quantum soliton of the
above gauge theory. In particular for the abelian gauge theory
with $U(1)$ as gauge group we regard the Dirac-Wilson loop as a
quantum soliton of the electromagnetic field. We now want to show
that this soliton has all the properties of photon and thus we may
identify it with the photon. First we see that from
(\ref{closed21}) it has discrete energy levels of light-quantum
given by
\begin{equation}
N h \nu :=N\frac{\pi e_0^2 }{k_0}, \qquad N=0,1,2,3, . . .
\label{planck}
\end{equation}
 where $h$ is the Planck's constant; $\nu$ denotes a frequency and the constant $k_0>0$
 is determined from this formula.
 This formula is from the monodromy $R_{U(1)}$ for the abelian gauge theory. We see that
 the Planck's constant $h$ comes out from this winding property of the Dirac-Wilson loop.
 Then since this Dirac-Wilson loop is a loop we have that it has the polarization property
 of light by the right hand rule along the loop and this polarization can also be regarded as
 the spin of photon. Now since this loop is a quantum soliton which behaves as a particle we have
 that this loop is a basic particle of the above abelian gauge theory where
 the abelian gauge property is considered as the fundamental property of electromagnetic field.
 This shows that the Dirac-Wilson loop has properties of photon. We shall later show that from
 this loop model of photon we can describe the absorption and emission of photon by an electron.
 This property of absorption and emission is considered as a basic principle of
 the light-quantum hypothesis of Einstein \cite{Pai}. From these properties of
 the Dirac-Wilson loop we may identify it with the photon.

On the other hand from Dirac's analysis of the magnetic monopole
\cite{Dir} we have that the property of magnetic monopole comes
from a closed line integral of vector potential of the
electromagnetic field which is similar to the Dirac-Wilson loop.
Now from this Dirac-Wilson loop we can define the magnetic charge
$q$ and the minimal magnetic charge $q_{min}$ which are given by:
\begin{equation}
e q:= e nq_{min}:= n_e e_0  n\frac{n_m e_0 \pi }{k_0}, \qquad n=0,1,2,3, . . .
\label{dirac}
\end{equation}
where $e:=n_e e_0 $ is as the observed electric charge for some positive integer $n_e$; and
$q_{min}:=\frac{n_m e_0 \pi }{k_0}$ for some positive integer $n_m$
and we write $N=nn_e n_m, n=0,1,2,3, . . . $ (By absorbing the constant $k_0$ to $e_0^2$ we may
let $k_0=1$).
This shows that the Dirac-Wilson loop gives the property of magnetic monopole for some frequencies.
Since this loop is a quantum soliton which behaves as a particle we have that
this Dirac-Wilson loop may be identified with the magnetic monopole for some frequencies.
Thus we have that photon may be identified with the magnetic monopole for some frequencies.

With this identification we have the following interesting
conclusion: Both the energy quantization of electromagnetic field
and the charge quantization property come from the same property
of photon. Indeed we have:
\begin{equation}
nh\nu_1 :=n \frac{n_e n_m e_0^2 \pi }{k_0}
=eq, \qquad n=0,1,2,3, . . .
\label{dirac2}
\end{equation}
where $\nu_1$ denotes a frequency. This formula shows that the
energy quantization gives the charge quantization and thus these
two quantizations are from the same property of the photon when
photon is modelled by the Dirac-Wilson loop and identified with
the magnetic monopole for some frequencies. We notice that between
two energy levels $neq_{min}$ and $(n+1)eq_{min}$ there are other
energy levels which may be regarded as the excited states of a
particle with charge $ne$.

\section{Nonlinear loop model of electron}

In this section let us also give a loop model to the electron.
This loop model of electron is based on the above loop model of
the photon. From the loop model of photon we also construct an
observable which gives mass to the electron and is thus a mass
mechanism for the electron.

Let W(z,z) denote a Dirac-Wilson loop which represents a photon. Let
$Z$ denotes the complex variable for electron in (\ref{1.1}). We then consider
the following observable:
\begin{equation}
 W(z,z)Z
\label{electron}
\end{equation}
Since $W(z,z)$ is solvable we have that this observable is also solvable where in solving $W(z,z)$
the variable $Z$ is fixed. We let this observable be identified with the electron.
Then we consider the following observable:
\begin{equation}
 Z^* W(z,z)Z
\label{electron1}
\end{equation}
This observable is with a scalar factor $Z^* Z$ where $Z^*$ denotes the complex conjugate of $Z$
and we regard it as the mass mechanism of the electron (\ref{electron}).
For this observable we model the energy levels with specific frequencies of $W(z,z)$ as
the mass levels of electron and the mass $m$ of electron is the lowest energy level $h\nu_1$
with specific frequency $\nu_1$ of $W(z,z)$ and is given by:
\begin{equation}
 mc^2=h\nu_1
\label{electron2}
\end{equation}
where $c$ denotes the constant of the speed of light and $\nu_1 :=\frac{n_m n_e e_0^2 \pi }{k}$.
From this model of the mass mechanism of electron we have that electron
is with mass $m$ while photon is with zero mass because there does not have such
a mass mechanism $Z^* W(z,z)Z$ for the photon. From this definition of mass we have
the following formula relating the observed electric charge $e$ of electron,
the magnetic charge $q_{min}$ of magnetic monopole and the mass $m$ of electron:
\begin{equation}
 mc^2= e q_{min}=h\nu_1
\label{electron3}
\end{equation}

By using the nonlinear model $W(z,z)Z$ to represent an electron we can then describe
the absorption and emission of a photon by an electron where photon is as a parcel
of energy described by the loop $W(z,z)$, as follows. Let $W(z,z)Z$ represents an electron
and let $W_1(z_1,z_1)$ represents a photon. Then we have the observable $(W(z,z)+W_1(z_1,z_1))Z$
which represents an electron having absorbed the photon $W_1(z_1,z_1)$.
This property of absorption and emission is as a basic principle of
the hypothesis of light-quantum stated by Einstein \cite{Pai}. Let us quote
the following paragraph from \cite{Pai}:

..., First, the light-quantum was conceived of as a parcel of energy as far as
the properties of pure radiation (no coupling to matter) are concerned. Second, Einstein made
the assumption--he call it the heuristic principle--that also in its coupling to matter
(that is, in emission and absorption), light is created or annihilated in similar discrete parcels
of energy. That, I believe, was Einstein's one revolutionary contribution to physics.
It upset all existing ideas about the interaction between light and matter. ....

\section{Photon with specific frequency carries electric and
magnetic charges}

In this loop model of photon we have that the observed electric charge $e:=n_e e_0 $ and
the magnetic charge $q_{min}$ are carried by the photon with some specific frequencies.
Let us here describe the physical effects from this property of photon that photon
with some specific frequency carries the electric and magnetic charge. From
the nonlinear model of electron we have that an electron $W(z,z)Z$ also carries
the electric charge when a photon $W(z,z)$ carrying the electric and magnetic charge
is absorbed to form the electron $W(z,z)Z$. This means that the electric charge of an electron
is from the electric charge carried by a photon.
Then an interaction (as the electric force) is formed between two electrons
(with the electric charges).
On the other hand  since photon carries the constant $e_0^2 $ of
the bare electric charge $e_0 $ we have that between two photons
there is an interaction which is similar to the electric force
between two electrons (with the electric charges). This
interaction however may not be of the same magnitude as the
electric force with the magnitude $e^2$ since the photons may not
carry the frequency for giving the electric and magnetic charge.
Then for stability such interaction between two photons tends to
give repulsive effect to give the diffusion phenomenon among
photons.

Similarly an electron $W(z,z)Z$ also carries the magnetic charge when a photon $W(z,z)$ carrying
the electric and magnetic charge is absorbed to form the electron $W(z,z)Z$. This means that
the magnetic charge of an electron is from the magnetic charge carried by a photon.
Then a closed-loop interaction (as the magnetic force) may be  formed between two electrons
(with the magnetic charges).
On the other hand  since photon carries the constant $e_0^2 $ of
the bare electric charge $e_0 $ we have that between two photons
there is an interaction which is similar to the magnetic force
between two electrons (with the magnetic charges). This
interaction however may not be of the same magnitude as the
magnetic force with the magnetic charge $q_{min}$ since the
photons may not carry the frequency for giving the electric and
magnetic charge. Then for stability such interaction between two
photons tends to give repulsive effect to give the diffusion
phenomenon among photons.

\section{Statistics of photons and electrons}

The nonlinear model $W(z,z)Z$ of an electron  gives a relation
between photon and electron where the photon is modelled by
$W(z,z)$ which is with a specific frequency for $W(z,z)Z$ to be an
electron, as described in the above Sections. We want to show that
from this nonlinear model we may also derive the required
statistics of photons and electrons that photons obey the
Bose-Einstein statistics and electrons obey the Fermi-Dirac
statistics. We have that $W(z,z)$ is as an operator acting on $Z$.
Let $W_1(z,z)$ be a photon. Then we have that the nonlinear model
$W_1(z,z)W(z,z)Z$ represents that the photon $W_1(z,z)$ is
absorbed by the electron $W(z,z)Z$ to form an electron
$W_1(z,z)W(z,z)Z$. Let $W_2(z,z)$ be another photon. The we have
that the model $W_1(z,z)W_2(z,z)W(z,z)Z$ again represents an
electron where we have:
\begin{equation}
\begin{array}{rl}
W_1(z,z)W_2(z,z)W(z,z)Z = W_2(z,z)W_1(z,z)W(z,z)Z\,.
\end{array}
 \label{aelectron}
\end{equation}

More generally the model:
\begin{equation}
\prod_{n=1}^N W_n(z,z)W(z,z)Z
\label{ad2}
\end{equation}
represents that the photons $W_n (z,z), n\,{=}\,1,2,\dots, N$ are
absorbed by the electron $W(z,z)Z$. This model shows that
identical (but different) photons can appear identically and it
shows that photons obey the Bose-Einstein statistics. From the
polarization of the Dirac-Wilson loop $W(z,z)$ we may assign spin
$1$ to a photon represented by $W(z,z)$.

Let us then consider statistics of electrons. The observable
$Z^*W(z,z)Z$ gives mass to the electron $W(z,z)Z$ and thus this
observable is as a scalar and thus is assigned with spin $0$. As
the observable $W(z,z)Z$ is between $W(z,z)$ and $Z^*W(z,z)Z$
which are with spin $1$ and $0$ respectively we thus assign spin
$\frac12$ to the observable $W(z,z)Z$ and thus electron
represented by this observable $W(z,z)Z$ is with  spin $\frac12$.
Then let $Z_1$ and $Z_2$ be two independent complex variables for
two electrons and let $W_1(z,z)Z_1$ and $W_2(z,z)Z_2$ represent
two electrons. Let $W_3(z,z)$ represents a photon. Then the model:
\begin{equation}
W_3(z,z)(W_1(z,z)Z_1+W_2(z,z)Z_2)
\label{ad3}
\end{equation}
means that two electrons
are in the same state that the operator $W_3(z,z)$ is acted on the
two electrons. However this model means that a photon $W(z,z)$ is
absorbed by two distinct electrons and this is impossible. Thus
the models $W_3(z,z)W_1(z,z)Z_1$ and $W_3(z,z)W_2(z,z)Z_2$ cannot
both exist and this means that electrons obey Fermi-Dirac
statistics.
Thus this nonlinear loop model of photon and electron gives the
required statistics of photons and electrons.

\section{Origin of magnetic properties}

Since the magnetic charge and magnetic monopole is derived from the photon loop $W(z,z)$ (or the Dirac-Wilson loop) we see that the photon loop is the origin of all the observable magnetic properties. In particular all the loop form of magnetic properties such as the magnetic flux and the magnetic vortex in the phenomena of superconductivity are from the photon loop $W(z,z)$.
When the Planck constant parameter $h=0$, the photon loop $W(z,z)$ or the electron loop $W(z,z)Z$ is in the ground state. In this case the photon loop gives the magnetic property of the ground state of photon or gives the magnetic property of the ground state of electron. Then since $h\neq 0$ gives dynamical effect we have the ground state of photon is the origin of static magnetic properties while the dynamical magnetic properties are from the photon loops with $h\neq 0$.
We shall show that this will give the ferromagnetic and antiferromagnetic properties of materials.

\section{Photon propagator and quantum photon propagator}

Let us then investigate the quantum Wilson line $W(z_0,z)$ with
$U(1)$ group where $z_0$ is fixed for the photon field. We want to
show that this quantum Wilson line $W(z_0,z)$ may be regarded as
the quantum photon propagator for a photon propagating from $z_0$
to $z$.
As we have shown in the above section on computation of quantum
Wilson line; to compute $W(z_0,z)$ we need to write $W(z_0,z)$ in
the form of two (connected) Wilson lines:
$W(z_0,z)=W(z_0,z_1)W(z_1,z)$ for some $z_1$ point.  Then we have:
\begin{equation}
W(z_0,z_1)W(z_1,z)=e^{-\hat{t}\log [\pm (z_1-z_0)]}Ae^{\hat{t}\log [\pm
(z-z_1)]} \label{graviton6b}
\end{equation}
where $t=-\frac{e_0^2}{k_0}$ for the $U(1)$ group ($k_0>0$ is a
constant and we may for simplicity let $k_0=1$) where the term
$e^{-\hat{t}\log [\pm (z-z_0)]}$ is obtained by solving the first form
of the dual form of the KZ equation and the term $e^{\hat{t}\log [\pm
(z_0-z)]}$ is obtained by solving the second form of the dual form
of the KZ equation.

Then we may write $W(z_0,z)$ in the following form:
\begin{equation}
W(z_0,z)=W(z_0,z_1)W(z_1,z)=e^{-\hat{t}\log \frac{(z_1-z_0)}{(z-z_1)}}A
\label{graviton6}
\end{equation}

Let us fix $z_1$ with $z$ such that:
\vspace*{-2pt}
\begin{equation}
\frac{|z_1-z_0|}{|z-z_1|}=\frac{r_1}{n_e^2} \label{egraviton6}
\end{equation}
for some positive integer $n_e $ such that $r_1\leq n_e^2$;
and we let $z$ be a point on a path of connecting $z_0$ and $z_1$
and then a closed loop is formed with $z$ as the starting and
ending point. (This loop can just be the photon-loop of the
electron in this electromagnetic interaction by this photon
propagator (\ref{graviton6}).) Then (\ref{graviton6}) has a factor
$e_0^2\log \frac{r_1}{n_e^2}$ which is the fundamental solution of
the two dimensional Laplace equation and is analogous to the
fundamental solution $\frac{e^2}{r}$ (where $e\,{:=}\,e_0 n_e$
denotes the observed (renormalized) electric charge and $r$
denotes the three dimensional distance) of the three dimensional
Laplace equation for the Coulomb's law. Thus the operator
$W(z_0,z)\,{=}\,W(z_0,z_1)W(z_1,z)$ in (\ref{graviton6}) can be
regarded as the quantum photon propagator propagating from $z_0$
to $z$.

We remark that when there are many photons we may introduce the
space variable $x$ via the Lorentz metric $ds^2=dt^2-dx^2$ as a
statistical variable to obtain the Coulomb's law $\frac{e^2}{r}$
from the fundamental solution $e_0^2\log \frac{r_1}{n_e^2}$ as a
statistical law for electricity (We shall give such a space-time
statistics later).

The quantum photon propagator (\ref{graviton6}) gives a repulsive
effect since it is analogous to the Coulomb's law $\frac{e^2}{r}$.
On the other hand we can reverse the sign of $\hat{t}$ such that
this photon operator can also give an attractive effect:
\begin{equation}
W(z_0,z)=W(z_0,z_1)W(z_1,z)=e^{\hat{t}\log
\frac{(z-z_1)}{(z_1-z_0)}}A\,, \label{graviton6a}
\end{equation}
where we fix $z_1$ with $z_0$ such that:
\begin{equation}
\frac{|z-z_1|}{|z_1-z_0|}=\frac{r_1}{n_e^2} \label{egraviton6a}
\end{equation}
for some positive integer $n_e $ such that $r_1\geq n_e^2$; and we
again let $z$ be a point on a path of connecting $z_0$ and $z_1$
and then a closed loop is formed with $z$ as the starting and
ending point. (This loop again can just be the photon-loop of the
electron in this electromagnetic interaction by this photon
propagator (\ref{graviton6}).) Then (\ref{graviton6a}) has a
factor $-e_0^2\log \frac{r_1}{n_e^2}$ which is the fundamental
solution of the two dimensional Laplace equation and is analogous
to the attractive fundamental solution $-\frac{e^2}{r}$ of the
three dimensional Laplace equation for the Coulomb's law.
Thus the quantum photon propagator in (\ref{graviton6}) (and in
(\ref{graviton6a})) can give repulsive or attractive effect
between two points $z_0$ and $z$ for all $z$ in the complex plane.
These repulsive or attractive effects of the quantum photon
propagator correspond to two charges of the same sign and of
different sign respectively.
On the other hand when $z=z_0$ the quantum Wilson line
$W(z_0,z_0)$  in (\ref{graviton6}) which is the quantum photon
propagator becomes a quantum Wilson loop $W(z_0,z_0)$ which is
identified as a photon, as shown in the above sections.

Let us then derive a usual form of photon propagator from the
quantum photon propagator $W(z_0,z)$. Let us choose a path
connecting $z_0$ and $z$. Let us consider the following path:
\begin{equation}
z=z(s)=z_1+ a_0[\theta(s_1-s)e^{-i\beta_1(s_1-s)}+
\theta(s-s_1)e^{i\beta_1(s_1-s)}] \label{g1}
\end{equation}
where $\beta_1>0 $ is a parameter and $z(s_0)=z_0$ for some proper
time $s_0$; and $a_0$ is some complex constant; and $\theta$ is a
step function given by $\theta(s)=0 $ for $s<0$, $\theta(s)=1 $
for $s\geq 0$. Then on this path we have:
\begin{equation}
\begin{array}{rl}
  & W(z_0,z)
= W(z_0,z_1)W(z_1,z) =e^{\hat{t}\log \frac{(z-z_1)}{(z_1-z_0)}}A
=  e^{\hat{t}\log \frac{a_0[\theta(s-s_1)e^{-i\beta_1(s_1-s)}+
\theta(s_1-s)e^{i\beta_1(s_1-s)}]}{(z_1-z_0)}}A \\
&\\
=&  e^{\hat{t}\log b[\theta(s-s_1)e^{-i\beta_1(s_1-s)}+
\theta(s_1-s)e^{i\beta_1(s_1-s)}]}A
=b_0 [\theta(s-s_1)e^{-i\hat{t}\beta_1(s_1-s)}+
\theta(s_1-s)e^{it\beta_1(s_1-s)}]A\\
\end{array}
\label{g2}
\end{equation}
for some complex constants $b$ and $b_0$. From this chosen of the
path (\ref{g1}) we have that the quantum photon propagator is
proportional to the following expression:
\begin{equation}
\frac{1}{2\lambda_1}[\theta(s-s_1)e^{-i\lambda_1(s-s_1)}+
\theta(s_1-s)e^{i\lambda_1(s-s_1)}] \label{g3}
\end{equation}
where we define $\lambda_1=-\hat{t}\beta_1=e_0^2\beta_1>0$.
 We see that this is the usual propagator of a
particle $x(s)$ of harmonic oscillator with mass-energy parameter
$\lambda_1 >0$ where $x(s)$ satisfies the following harmonic
oscillator equation:
\begin{equation}
\frac{d^2x}{ds^2} =-\lambda_1^2 x(s) \label{g4}
\end{equation}
We regard (\ref{g3}) as the propagator of a photon with
mass-energy parameter $\lambda_1$. Fourier transforming (\ref{g3})
we have the following form of photon propagator:
\begin{equation}
\frac{i}{k_E^2-\lambda_1} \label{g5}
\end{equation}
where we use the notation $k_E$ (instead of the notation $k$) to
denote the proper energy of photon. We shall show in the next
section that from this  photon propagator by space-time statistics
we can get a propagator with the $k_E$ replaced by the
energy-momemtum four-vector $k$ which is similar to the Feynman
propagator (with a mass-energy parameter $\lambda_1
>0$). We thus see that
the quantum photon propagator $W(z_0,z)$ gives a classical form of
photon propagator in the conventional QED theory.

Then we notice that while $\lambda_1 >0$ which may be think of as
the mass-energy parameter of a photon the original quantum photon
propagator $W(z_0,z)$ can give the Coulomb potential and thus give
the effect that the photon is massless. Thus the photon
mass-energy parameter $\lambda_1 >0$ is consistent with the
property that the photon is massless. Thus in the following
sections when we compute the vertex correction and the Lamb shift
we shall then be able to let $\lambda_1 >0$ without contradicting
the property that the photon is massless. This then can solve the
infrared-divergence problem of QED.

We remark that if we choose other form of paths for connecting
$z_0$ and  $z$ we can get other forms of  photon propagator
corresponding to a choice of gauge.  From the property of gauge
invariance the final result should not depend on the form of
propagators. We shall see that this is achieved by
renormalization. This property of renormalizable is as a property
related to the gauge invariance. Indeed we notice that the quantum
photon propagator with a photon-loop $W(z,z)$ attached to an
electron represented by $Z$ has already given the renormalized
charge $e$ (and the renormalized mass $m$ of the electron) for the
electromagnetic interaction.
 It is clear that this renormalization
by the quantum photon propagator with a photon-loop $W(z,z)$ is
independent of the chosen photon propagator (because it does not
need to choose a photon propagator). Thus the renormalization
method as that in the conventional QED theory for the chosen of a
photon propagator (corresponding to a choice of gauge) should give
the observable result which does not depend on the form of the
photon propagators since these two forms of renormalization must
give the same effect of renormalization. In the following section
and the sections from section \ref{QED1} to section \ref{QED8} on
quantum electrodynamics (QED) we shall investigate the
renormalization method which is analogous to that of the
conventional QED theory and the computation of QED effects by
using this renormalization method.

\section{Renormalization}

In this section and the following sections from section \ref{QED1}
to section \ref{QED8} on quantum electrodynamics (QED) we shall
use the density (\ref{1.1}) and the notations from this density
where $A_j, j=1,2$ are real components of the photon field.
Following the conventional QED theory let us consider the
following renormalization:
\begin{equation}
\begin{array}{rl}
& A_j=z_A^{\frac12}A_{jR}, \,\, j=1,2; \quad Z = z_Z^{\frac12}Z_R; \quad
 e_0=\frac{z_e}{z_Z z_A^{\frac12}}e=\frac{1}{n_e}e
\end{array}
\label{ren1}
\end{equation}
where $z_A$, $z_Z$, and $z_e$ are renormalization constants to be
determined and $A_{Rj}, j=1,2$, $Z_R$ are renormalized fields.
From this renormalization the density $D$ of QED in (\ref{1.1})
can be written in the following form:
\begin{equation}
\begin{array}{rl}
D &=\frac12 z_A\left(\frac{\partial A_{1R}}{\partial
x^2}-\frac{\partial A_{2R}}{\partial x^1}\right)^*
\left(\frac{\partial A_{1R}}{\partial x^2}-\frac{\partial A_{2R}}{\partial x^1}\right)+ \\
&\\
  &
\quad  z_Z \left( \frac{dZ_R^*}{ds}
+ie(\sum_{j=1}^2A_{jR}\frac{dx^j}{ds})Z_R^*\right) \left(
\frac{dZ_R}{ds}
-ie(\sum_{j=1}^2A_{jR}\frac{dx^j}{ds})Z_R\right)\\
&\\
&=\{\frac12 \left(\frac{\partial A_{1R}}{\partial
x^2}-\frac{\partial A_{2R}}{\partial x^1}\right)^*
\left(\frac{\partial A_{1R}}{\partial x^2}-\frac{\partial A_{2R}}{\partial x^1}\right) 
 + \frac{dZ_R^*}{ds}\frac{dZ_R}{ds}  +\mu^2Z_R^*Z_R-\mu^2Z_R^*Z_R\\
&\\
&\quad +ie(\sum_{j=1}^2A_{jR}\frac{dx^j}{ds})Z_R^*\frac{dZ_R}{ds}
-ie(\sum_{j=1}^2A_{jR}\frac{dx^j}{ds})\frac{dZ_R^*}{ds}Z_R
+e^2(\sum_{j=1}^2A_{Rj}\frac{dx^j}{ds})^2Z_R^*Z_R \} \\
&\\
&\quad  + \{ (z_A-1)[\frac12 \left(\frac{\partial A_{1R}}{\partial
x^2}-\frac{\partial A_{2R}}{\partial x^1}\right)^*
\left(\frac{\partial A_{1R}}{\partial x^2}-\frac{\partial A_{2R}}{\partial x^1}\right)] 
+ (z_Z-1)\frac{dZ_R^*}{ds}\frac{dZ_R}{ds}  \\
&\\
& \quad
+(z_e-1)[+ie(\sum_{j=1}^2A_{jR}\frac{dx^j}{ds})Z_R^*\frac{dZ_R}{ds}
-ie(\sum_{j=1}^2A_{jR}\frac{dx^j}{ds})\frac{dZ_R^*}{ds}Z_R] \\
&\\
& \quad
+(\frac{z_e^2}{z_Z}-1)e^2(\sum_{j=1}^2A_{jR}\frac{dx^j}{ds})^2Z_R^*Z_R
\} 
 :=D_{phy} + D_{cnt}
\end{array}
\label{2.1}
\end{equation}
where $D_{phy}$ is as the physical term and the $D_{cnt}$ is as
the counter term; and in $D_{phy}$ the positive parameter $\mu$ is
introduced for perturbation expansion and for renormalization.

As the Ward-Takahashi identities in the conventional QED theory
being derived by the gauge invariance property of the conventional
QED theory; by using the gauge invariance property of this QED
theory we can also derive the corresponding Ward-Takahashi
identities for this QED theory. From these Ward-Takahashi
identities we then show that there exists a renormalization
procedure such that $z_e=z_Z$; as similar to that in the
conventional QED theory. From this relation $z_e=z_Z$ we then
have:
\begin{equation}
e_0=\frac{e}{z_A^{\frac12}}=\frac{1}{n_e} e
 \label{ree}
\end{equation}
and that in (\ref{2.1}) we have $\frac{z_e^2}{z_Z}-1=z_e-1$.

\section{Feynman diagrams and Feynman rules for QED}\label{QED1}

Let us then transform $ds$ in (\ref{1.1}) to $\frac{1}{(\beta+i
h)}ds$ where $\beta, h>0$ are parameters and $h$ is as the Planck
constant. The parameter $h$ will give the dynamical effects of QED
(as similar to the conventional QED). Here for simplicity we only
consider the limiting case that $\beta \rightarrow 0$ and we let
$h=1$. From this transformation we get the Lagrangian
$\mathcal{L}$ from $-\int_{s_0}^{s_1}Dds$ changing to
$\int_{s_0}^{s_1}\mathcal{L}ds$. Then we write
$\mathcal{L}=\mathcal{L}_{phy} +\mathcal{L}_{cnt}$ where
$\mathcal{L}_{phy}$ corresponds to $D_{phy}$  and
$\mathcal{L}_{cnt}$ corresponds to $D_{cnt}$. Then from the
following term in $\mathcal{L}_{phy}$:
\begin{equation}
-i[(\frac{dZ_R}{ds})^{*}\frac{dZ_R}{ds}-\mu^2Z_R^*Z_R]
 \label{2.1e1}
\end{equation}
and by the perturbation expansion of
$e^{\int_{s_0}^{s_1}\mathcal{L}ds}$ we have the following
propagator:
\begin{equation}
\frac{i}{p_E^2-\mu^2}
 \label{2.1e2}
\end{equation}
which is as the (primitive) electron propagator where $p_E$
denotes the proper energy variable of electron.

Then from the pure gauge part of $\mathcal{L}_{phy}$ we get the
photon propagator (\ref{g5}), as done in the above sections and
the section on photon propagator.
Then from  $\mathcal{L}_{phy}$ we have the following seagull
vertex term:
\begin{equation}
ie^2(\sum_{j=1}^2A_{jR}\frac{dx^j}{ds})^2Z_R^* Z_R \label{v1a}
\end{equation}
This seagull vertex term gives the vertex factor $ie^2$ (We remark
that the $ds$ of the paths $\frac{dx^j}{ds}$ are not transformed
to $\frac{1}{(\beta+i h)}ds$ since these paths are given paths and
thus  are independent of the transformation $ds \to
\frac{1}{(\beta+i h)}ds$).

From this vertex by using the photon propagator (\ref{g5}) in the
above section we get the following term:
\begin{equation}
 \frac{ie^2}{2\pi}\int\frac{i dk_E}{k_E^2-\lambda_1^2}
= -\frac{ie^2}{2\lambda_1} =: -i\omega^2 \label{2.7}
\end{equation}
The parameter $\omega$ is regarded as the mass-energy parameter
 of electron.

Then from the perturbation expansion of
$e^{\int_{s_0}^{s_1}\mathcal{L}ds}$ we have the following
geometric series (which is similar to the Dyson series in the
conventional QED):
\begin{equation}
\frac{i}{p_E^2-\mu^2}+\frac{i}{p_E^2-\mu^2}(-i\omega^2+i\mu^2)\frac{i}{p_E^2-\mu^2}+...
=\frac{i}{p_E^2-\mu^2-\omega^2+\mu^2}=\frac{i}{p_E^2-\omega^2}
\label{g01}
\end{equation}
where the term $i\mu$ of $-i\omega^2+i\mu^2$ is the $i\mu$ term in
$\mathcal{L}_{phy}$ (The other term $-i\mu$ in $\mathcal{L}_{phy}$
has been used in deriving ((\ref{2.1e2}))). Thus we have the
following electron propagator:
\begin{equation}
\frac{i}{p_E^2-\omega^2} \label{g02}
\end{equation}
This is as the electron propagator with mass-energy parameter
$\omega$. From $\omega$ we shall get the mass $m$ of electron (We
shall later introduce a space-time statistics to get the usual
electron propagator of the Dirac equation. This usual electron
propagator is as the statistical electron propagator). As the
Feynman diagrams in the conventional QED we represent this
electron propagator by a straight line.

In the above sections and the section on the photon propagator we
see that the photon-loop $W(z,z)$ gives the renormalized charge
$e=n_e e_0$ and the renormalized mass $m$ of electron from the
bare charge $e_0$ by the winding numbers of the photon loop such
that $m$ is with the winding number factor $n_e$. Then we see that
the above one-loop energy integral of the photon gives the
mass-energy parameter $\omega$ of electron which gives the mass
$m$ of electron. Thus these two types of photon-loops are closely
related (from the relation of photon propagator and quantum photon
propagator) such that the mass $m$ obtained by the winding numbers
of the photon loop $W(z,z)$ reappears in the one-loop energy
integral (\ref{2.7}) of the photon.

Thus we see that even there is no mass term in the Lagrangian of
this gauge theory the mass $m$ of the electron can come out from
the gauge theory. This actually resolves the mass problem of
particle physics that particle can be with mass even without the
mass term. Thus we do not need the Higgs mechanism for generating
masses to particles.

On the other hand from the one-loop-electron  form of the seagull
vertex we have the following term:
\begin{equation}
 \frac{ie^2}{2\pi}\int\frac{idp_E}{p_E^2-\mu^2}= -\frac{ie^2}{2\mu}=: -i\lambda_2^2
\label{g03}
\end{equation}
Then for photon from the perturbation expansion of
$e^{\int_{s_0}^{s_1}\mathcal{L}ds}$ we have the following
geometric series:
\begin{equation}
\frac{i}{k_E^2-\lambda_1^2}+
\frac{i}{k_E^2-\lambda_1^2}(-i\lambda_2^2)\frac{i}{k_E^2-\lambda_1^2}+...
=\frac{i}{k_E^2-\lambda_1^2-\lambda_2^2}=:\frac{i}{k_E^2-\lambda_0^2}
\label{g04}
\end{equation}
where we define $\lambda_0^2=\lambda_1^2+\lambda_2^2$. Thus we
have the following photon propagator:
\begin{equation}
\frac{i}{k_E^2-\lambda_0^2} \label{g05}
\end{equation}
which is of the same form as (\ref{g5}) where we replace
$\lambda_1$ with $\lambda_0$.
 As the Feynman diagrams in
the conventional QED we represent this photon propagator by a wave
line.

Then the  following interaction term in $\mathcal{L}_{phy}$:
\begin{equation}
-ie\frac{dZ_R^*}{ds}(\sum_{j=1}^2A_{jR}\frac{dx^j}{ds})Z_R
+ie\frac{dZ_R}{ds}(\sum_{j=1}^2A_{jR}\frac{dx^j}{ds})Z_R^*
 \label{v1}
\end{equation}
gives the vertex factor $(-ie)(p_{E}+q_{E})$ which corresponds to
the usual vertex of Feynman diagram with two electron straight
lines (with energies $p_{E}$ and $q_{E}$) and one photon wave line
in the conventional QED.

Then as the Feynman rules in the conventional QED a sign factor
$(-1)^n$, where $n$ is the number of the electron loops in a
Feynman diagram, is to be included for the Feynman diagram.

\section{Statistics with space-time}\label{sec2}

Let us introduce space-time as a statistical method for a large
amount of basic variables $Z_R$ and $A_{1R}$, $A_{2R}$. As an
illustration let us consider the electron propagator
$\frac{i}{p_E^2-\omega^2}$ and the following Green's  function
corresponding to it:
\begin{equation}
\frac{i}{2\pi}\int\frac{e^{-ip_E(s-s')}dp_E}{p_E^2-\omega^2}
\label{3.1}
\end{equation}
where $s$ is the proper time. We imagine each electron (and
photon) occupies a space region (This is the creation of the
concept of space which is associated to the electron. Without the
electron this space region does not exist). Then we write
\begin{equation}
p_E (s-s')=p_E(t-t')- {\bf p}({\bf x} -{\bf x}') \label{3.2}
\end{equation}
where ${\bf p} ({\bf x} -{\bf x}')$ denotes the inner product of
the three dimensional vectors {\bf p} and ${\bf x} -{\bf x}'$ and
$(t,{\bf x})$ is the time-space coordinate where {\bf x} is in the
space region occupied by $Z_R(s)$ and that
\begin{equation}
\omega^2-{\bf p}^2=m^2>0 \label{3.3}
\end{equation}
where $m$ is the mass of electron. This mass $m$ is greater than 0
since each $Z_R$ occupies a space region which implies that when
$t-t'$ tends to 0 we can have that $|{\bf x} -{\bf x}'|$ does not
tend to 0 (${\bf x}$ and ${\bf x}'$ denote two coordinate points
in the regions occupied by $Z_R(s)$ and $Z_R(s')$ respectively)
and thus (\ref{3.3}) holds.
  Then by linear summing the effects of a large amount of basic variables
$Z_R$  and letting $\omega$ varies from $m$ to $\infty$ from
(\ref{3.1}), (\ref{3.2}) and (\ref{3.3}) we get the following
statistical expression:
\begin{equation}
\frac{i}{(2\pi)^4}\int \frac{e^{-ip(x-x')}dp}{p^2-m^2} \label{3.5}
\end{equation}
which is the usual Green's function of a free field with mass $m$
where $p$ is a four vector and $x=(t,{\bf x})$.

The result of the above statistics is that (\ref{3.5}) is induced
from (\ref{3.1}) with the scalar product $p^2$ of a scalar $p$
changed to an indefinite inner product $p^2$ of a four vector $p$
(For simplicity we have with a little confusion adopted the same
notation $p$ for both scalar and vector) and the parameter
$\omega$ is reduced to $m$.

Let us then introduce Fermi-Dirac statistics for electrons. As
done by Dirac for deriving the Dirac equation we factorize
$p^2-m^2$ into the following form:
\begin{equation}
p^2-m^2=(p_E-\omega)(p_E+\omega)=(\gamma_{\mu}p^{\mu}-m)(\gamma_{\mu}p^{\mu}+m)
\label{d01}
\end{equation}
where $\gamma_{\mu}$ are the Dirac matrices. Then from (\ref{3.5})
we get the following Green's function:
\begin{equation}
\frac{i}{(2\pi)^4}\int
e^{-ip(x-x')}\frac{\gamma_{\mu}p^{\mu}+m}{p^2-m^2}dp =
\frac{i}{(2\pi)^4}\int
\frac{e^{-ip(x-x')}dp}{\gamma_{\mu}p^{\mu}-m} \label{d02}
\end{equation}
Thus we have the Fermi-Dirac statistics that the statistical
electron propagator is of the form
$\frac{i}{\gamma_{\mu}p^{\mu}-m}$ which is the propagator of the
Dirac equation and is the electron propagator of the conventional
QED.

Let us then consider statistics of photons.
  Since the above quantum gauge theory of photons is a gauge theory which is gauge invariant
  we have that
  the space-time statistical equation for photons should be gauge invariant. Then since the Maxwell
  equation is the only gauge invariant equation for electromagnetism which is based on
  the space-time
  we have that the Maxwell equation must be a statistical equation for
  photons.

  Then let us consider the vertexes.
The tree vertex (\ref{v1}) with three lines (one for photon and
two for electron) gives the factor $-ie(p_{E}+q_E)$ where $p_E$
and $q_E$ are from the factor $\frac{dZ_R}{ds}$ for electron.

We notice that this vertex is with two electron lines (or electron
propagator) and one photon line (or photon propagator). In doing a
statistics on this photon line when it is as an external
electromagnetic field on the electron this photon line is of  the
statistical form $\gamma_{\mu}A^{\mu}$ where $A^{\mu}$ denotes the
four electromagnetic potential fields of the Maxwell equation.
Thus the vertex $-ie(p_E+q_E)$ after statistics is changed to the
form $-ie(p_E+q_E)\frac{\gamma^{\mu}}{2}$ where for each
$\gamma^{\mu}$ a factor $\frac12 $ is introduced for statistics.

Let us then introduce the on-mass-shell condition as in the
conventional QED theory (\cite{Zub}). As similar to the
on-mass-shell condition in the conventional QED theory our
on-mass-shell condition is that $p_E=m$ where $m$ is the
observable mass of the electron. In this case $-ie
(p_E+q_E)\frac{\gamma^{\mu}}{2}$ is changed to $-ie
m\gamma^{\mu}$. Then the $m$ is absorbed to the two external
spinors $\frac{1}{\sqrt{E}}u$ (where $E$ denotes the energy of the
electron satisfied the Dirac equation while the $E$ of $p_E$ is
only as a notation) of the two electrons lines attached to this
vertex such that the factor $\frac{1}{\sqrt{E}}$ of spin $0$ of
the Klein-Gordon equation is changed to the factor
$\sqrt{\frac{m}{E}}$ of spinors of the Dirac equation. In this
case we have the magnitude of $p_E$ and $q_E$ reappears in the two
external electron lines with the factor $\sqrt{m}$. The
statistical vertex then becomes $-ie\gamma^{\mu}$. This is exactly
the usual vertex in the conventional QED. Thus after a space-time
statistics on the original vertex $-ie(p_E+q_E)$ we get the
statistical vertex $-ie\gamma^{\mu}$ of the conventional QED.

\section{Basic effects of quantum electrodynamics}

  To illustrate this new theory of QED let us
compute the three basic effects of QED: the one-loop photon and
electron self-energies and the one-loop vertex correction.

As similar to the conventional QED we have the Feynman rules such
that the one-loop photon self-energy is given by the following
Feynman integral:
\begin{equation}
i\Pi_0(k_E):=i^2(-i)^2\frac{e^2}{2\pi} \int
dp_E\frac{(2p_E+k_E)(2p_E+k_E)}{(p_E^2-\omega^2)((k_E+p_E)^2-\omega^2)}
\label{2.16a}
\end{equation}
where $e$ is the renormalized electric charge.
Then as the Feynman rules in the conventional QED for the
space-time statistics a statistical sign factor $(-1)^j$, where
$j$ is the number of the electron loops in a Feynman diagram, will
be included for the Feynman diagram. Thus for the one-loop photon
self-energy (\ref{2.16a}) a statistical factor $(-1)^j$ will be
introduced to this one-loop photon self-energy integral.

Then similarly we have the Feynman rules such that the one-loop
electron self-energy is given by the following Feynman integral:
\begin{equation}
-i\Sigma_0(p_E):=i^2(-i)^2\frac{e^2}{2\pi} \int
dk_E\frac{(-k_E+2p_E)(-k_E+2p_E)}{(k_E^2-\lambda_0^2)((p_E-k_E)^2-\omega^2)}
\label{2.16b}
\end{equation}

 Then similarly we have the Feynman rules such that the one-loop vertex correction
is given by the following Feynman integral:
\begin{equation}
(-ie)\Lambda_0(p_E,q_E):=(i)^3(-i)^3\frac{e^3}{2\pi}\int dk_E\frac
{(2p_E-k_E)(2q_E-k_E)(p_E+q_E-2k_E)}{((p_E-k_E)^2-\omega^2)((q_E-k_E)^2-\omega^2)(k_E^2-\lambda_0^2)}
\label{2.15}
\end{equation}

Let us first compute the one-loop vertex correction and then
compute the photon self-energy and the electron self-energy.
As a statistics we extend the one dimensional integral $\int dk_E$
to  the $n$-dimensional integral $\int d^n k$ ($n\to 4$) where
$k=(k_E, {\bf k})$. This is similar to the dimensional
regularization in the conventional quantum field theories (However
here our aim is to increase the dimension for statistics which is
different from the dimensional regularization which is to reduce
the dimension from $4$ to $n$ to avoid the ultraviolet
divergence). With this statistics the factor $2\pi$ is replaced by
the statistical factor $(2\pi)^n$. From this statistics on
(\ref{2.15}) we have the following statistical one loop vertex
correction:
\begin{equation}
\begin{array}{rl}
 & \frac{e^3}{(2\pi)^n}\int_0^1 dx\int_0^1 2ydy\int d^n k
  \frac
 {4p_Eq_E(p_E+q_E)-2k_E((p_E+q_E)^2 +4p_Eq_E)+5k_E^2 (p_E+q_E)-2k_E^3}
 {[k^2 -2k(p xy+q(1-x)y)-p_E^2 xy-q_E^2 (1-x)y+m^2 y+\lambda^2 (1-y)]^3} \\
 &\\
 \end{array}
\label{2.15a}
\end{equation}
where $k^2=k_E^2 -{\bf k}^2$,  and ${\bf k}^2$ is from the free
parameters $\omega$  and $\lambda_0$  where we let
$\omega^2=m^2+{\bf k}^2$, $\lambda_0^2=\lambda^2+{\bf k}^2$ for
the electron mass $m$ and a mass-energy parameter $\lambda$ for
photon; and
\begin{equation}
k(p xy+q (1-x)y):= k_E(p_E xy+q_E(1-x)y) -{\bf k}\cdot {\bf
0}=k_E(p_E xy+q_E(1-x)y). \label{2.15a1}
\end{equation}
By using the formulae for computing Feynman integrals we have that
(\ref{2.15a}) is equal to (\cite{Zub}\cite{Tic}):
\begin{equation}
\begin{array}{rl}
& \frac{ie^3}{(2\pi)^n}\int_0^1 dx\int_0^1 2ydy
\frac{4p_Eq_E(p_E+q_E)\pi^{\frac{n}{2}}\Gamma(3-\frac{n}{2})}
{\Gamma(3)(\Delta -r^2)^{3-2}}
\frac{1}{(-\Delta +r^2)^{2-\frac{n}{2}}} \\
&\\
& -\frac{ie^3}{(2\pi)^n}\int_0^1 dx\int_0^1 2ydy
\frac{2((p_E+q_E)^2
+4p_Eq_E)\pi^{\frac{n}{2}}\Gamma(3-\frac{n}{2})r}
{\Gamma(3)(\Delta -r^2)^{3-2}}
\frac{1}{(-\Delta +r^2)^{2-\frac{n}{2}}} \\
&\\
& +\frac{ie^3}{(2\pi)^n}\int_0^1 dx\int_0^1 2ydy
\frac{5(p_E+q_E)\pi^{\frac{n}{2}}\Gamma(3-1-\frac{n}{2})\frac{n}{2}}
{\Gamma(3)(\Delta -r^2)^{3-2-1}}
\frac{1}{(-\Delta +r^2)^{2-\frac{n}{2}}} \\
&\\
& +\frac{ie^3}{(2\pi)^n}\int_0^1 dx\int_0^1 2ydy
\frac{5(p_E+q_E)\pi^{\frac{n}{2}}\Gamma(3-\frac{n}{2})r^2}
{\Gamma(3)(\Delta -r^2)^{3-2}}
\frac{1}{(-\Delta +r^2)^{2-\frac{n}{2}}} \\
&\\
& -\frac{ie^3}{(2\pi)^n}\int_0^1 dx\int_0^1 2ydy
\frac{\frac{(n+2)}{2}2\pi^{\frac{n}{2}}\Gamma(3-1-\frac{n}{2})r}
{\Gamma(3)(\Delta -r^2)^{3-2-1}}
\frac{1}{(-\Delta +r^2)^{2-\frac{n}{2}}} \\
&\\
& -\frac{ie^3}{(2\pi)^n}\int_0^1 dx\int_0^1 2ydy
\frac{2\pi^{\frac{n}{2}}\Gamma(3-\frac{n}{2})r^3}
{\Gamma(3)(\Delta -r^2)^{3-2}}
\frac{1}{(-\Delta +r^2)^{2-\frac{n}{2}}} \\
&\\
=:& (-ie)\Lambda(p_1,p_2)
\end{array}
\label{2.15aa}
\end{equation}
where we define:
\begin{equation}
\begin{array}{rl}
r=:& p_E xy+q_E (1-x)y \\
\Delta=:& p_E^2xy+q_E^2(1-x)y-m^2y-\lambda^2(1-y)
\end{array}
\label{2.15ab}
\end{equation}

We remark that in this statistics the $p_E$ and $q_E$ variables
are remained as the proper variables which is derived from the
proper time $s$.

Let us then introduce the Fermi-Dirac statistics on the electron
and we consider the on-mass-shell case as in the conventional QED.
We shall see this will lead to the theoretical results of the
conventional QED on the anomalous magnetic moment and the Lamb
shift.

As a Fermi-Dirac statistics we have shown in the above section
that the vertex term $-ie(p_E+q_E)$ is replaced with the vertex
term $-ie(p_E+q_E)\frac{\gamma^{\mu}}{2}$. Then as a Fermi-Dirac
statistics in the above section we have shown that the statistical
vertex is $-ie\gamma^{\mu}$ under the on-mass-shell condition. We
notice that this vertex agrees with the vertex term in the
conventional QED theory.

Let us then consider the Fermi-Dirac statistics on the one-loop
vertex correction (\ref{2.15aa}). Let us first consider the
following term in (\ref{2.15aa}):
\begin{equation}
\frac{ie^3}{(2\pi)^n}\int_0^1 dx\int_0^1
2ydy\frac{\pi^2(p_E+q_E)4p_Eq_E} {\Gamma(3)(\Delta
-r^2)^{3-2}}\frac{1}{(-\Delta +r^2)^{2-\frac{n}{2}}} \label{q1}
\end{equation}
where we can (as an approximation) let $n=4$. From Fermi-Dirac
statistics we have that this term gives the following statistics:
\begin{equation}
\frac{ie^3}{(2\pi)^4}\int_0^1 dx\int_0^1 2ydy
\frac{\pi^2(p_E+q_E)\frac12
\gamma^{\mu}4p_Eq_E}{\Gamma(3)(\Delta-r^2)^{3-2}} \label{q2}
\end{equation}
Then we consider the case of on-mass-shell. In this case we have
$p_E=m$ and $q_E=m$. Thus from (\ref{q2}) we have the following
term:
\begin{equation}
\frac{ie^3}{(2\pi)^4}\int_0^1 dx\int_0^1 2ydy\frac{\pi^2
\gamma^{\mu}4p_Eq_E} {\Gamma(3)(\Delta -r^2)^{3-2}} \label{q3}
\end{equation}
where a mass factor $m=\frac12(p_E+q_E)$ has been omitted and put
to the external spinor of the external electron as explained in
the above section on space-time statistics. In (\ref{q3}) we still
keep the expression $p_Eq_E$ even though in this case of
on-mass-shell because this factor will be important for giving the
observable Lamb shift, as we shall see. In (\ref{q3}) because of
on-mass-shell we have (As an approximation we let $n=4$):
\begin{equation}
(\Delta -r^2)^{3-2}=-\lambda^2(1-y)- r^2=-\lambda^2(1-y)-m^2 y^2
\label{q4}
\end{equation}
Thus in the on-mass-shell case (\ref{q3}) is of the following
form:
\begin{equation}
ie\gamma^{\mu}\frac{\alpha}{\pi}\int_0^1 dx\int_0^1 ydy\frac{\pi^2
p_Eq_E} {-\lambda^2(1-y)-m^2 y^2} \label{q5}
\end{equation}
where $\alpha=\frac{e^2}{ 4\pi}$ is the fine structure constant.
Carrying out the integrations on $y$ and on $x$ we have that as
$\lambda\to 0$ (\ref{q5}) is equal to:
\begin{equation}
(-ie)\gamma^{\mu}\frac{\alpha}{\pi}\frac{p_Eq_E}{ m^2}\log
\frac{m}{\lambda} \label{q6}
\end{equation}
where the proper factor $p_Eq_E$ will be for a linear space-time
statistics of summation. We remark that (\ref{q6}) corresponds to
a term in the vertex correction in the conventional QED theory
with the infra-divergence when $\lambda=0$ (\cite{Zub}). Here
since the parameter $\lambda$ has not been  determined we shall
later find other way to determine the effect of (\ref{q6}) and to
solve the infrared-divergence problem.

Let us first rewrite the form of  the proper value $p_Eq_E$. We
write $p_Eq_E$ in the following space-time statistical form:
\begin{equation}
p_Eq_E=-2p^{\prime}\cdot p \label{q7}
\end{equation}
where $p$ and $p^{\prime}$ denote two space-time four-vectors of
electron such that $p^2=m^2$ and $p^{\prime 2}=m^2$. Then we have
\begin{equation}
\begin{array}{rl}
& p_Eq_E\\
&\\
=&\frac13 (p_Eq_E+p_Eq_E+p_Eq_E)=\frac13(m^2-2p^{\prime}\cdot p+m^2)\\
&\\
=&\frac13(m^2-2p^{\prime}\cdot p +m^2)=\frac13(p^{\prime 2}-2p^{\prime}\cdot p +p^2)\\
&\\
=&\frac13(p^{\prime}-p)^2 \\
&\\
=:&\frac13 q^2
\end{array}
\label{q8}
\end{equation}
where following the convention of QED we define $q=p^{\prime}-p$.
Thus from (\ref{q6}) we have the following term:
\begin{equation}
(-ie)\gamma^{\mu}\frac{\alpha}{3\pi}\frac{q^2}{ m^2}\log
\frac{m}{\lambda} \label{q10}
\end{equation}
where the parameter $\lambda$ are to be determined. Again this
term (\ref{q10}) corresponds to a term in the vertex correction in
the conventional QED theory with the infrared-divergence when
$\lambda=0$ (\cite{Zub}).

Let us then consider the following term in (\ref{2.15aa}):
\begin{equation}
\frac{-ie^3}{(2\pi)^n}\int_0^1 dx\int_0^1 2ydy \frac{2((p_E+q_E)^2
+4p_Eq_E)\pi^{\frac{n}{2}}\Gamma(3-\frac{n}{2})r}
{\Gamma(3)(\Delta -r^2)^{3-2}}\frac{1}{(-\Delta
+r^2)^{2-\frac{n}{2}}} \label{q11}
\end{equation}
For this term we can (as an approximation) also let $n=4$ . As
similar to the conventional QED theory we want to show that this
term gives the anomalous magnetic moment and thus corresponds to a
similar term in the vertex correction term of the conventional QED
theory (\cite{Zub}).

By the Fermi-Dirac statistics the factor $(p_E+q_E)$ of
$(p_E+q_E)^2$ in (\ref{q11}) gives the statistical term
$(p_E+q_E)\frac12 \gamma^{\mu}$. Thus with the on-mass-shell
condition the factor $(p_E+q_E)$ gives the statistical term
$m\gamma^{\mu}$. Thus with the on-mass-shell condition the term
$(p_E+q_E)^2$ gives the statistical term $m\gamma^{\mu}(p_E+q_E)$.
Then we have that the factor $(p_E+q_E)$ in this statistical term
also give $2m$ by the on-mass-shell condition. Thus by the
Fermi-Dirac statistics and the on-mass-shell condition the factor
$(p_E+q_E)^2$ in (\ref{q11}) gives the statistical term $
\gamma^{\mu}2m^2$. Then since this is a (finite) constant term it
can be cancelled by the corresponding counter term of the vertex
giving the factor $-ie \gamma^{\mu}$ and having the factor $z_e-1$
in (\ref{2.1}). From this cancellation the renormalization
constant $z_e$ is determined. Since the constant term is depended
on the $\delta>0$ which is introduced for space-time statistics we
have that the renormalization constant $z_e$ is also depended on
the $\delta>0$. Thus the renormalization constant $z_e$ (and the
concept of renormalization) is related to the space-time
statistics.

At this point let us give a summary of this renormalization
method, as follows.

{\bf Renormalization}. 1). We use the renormalization method of
the conventional QED theory to obtain the renormalized physical
results. Here unlike the conventional QED theory the
renormalization method is not for the removing of ultraviolet
divergences since the QED theory in this paper is free of
ultraviolet divergences.

 2). We have mentioned in the above section on
photon propagator that the property of renormalizable is a
property of gauge invariance that it gives the physical results
independent of the chosen photon propagator.

3). The procedure of renormalization is as a part of the
space-time statistics to get the statistical results which is
independent of the chosen photon propagator. $\diamond$

Let us then consider again the above computation of the one-loop
vertex correction. We now have that the (finite) constant term of
the one-loop vertex correction is cancelled by the corresponding
counter term with the factor $z_e-1$ in (\ref{2.1}). Thus the
nonconstant term (\ref{q10}) is renormalized to be the following
renormalized form:
\begin{equation}
(-ie)\gamma^{\mu}\frac{\alpha}{3\pi}\frac{q^2}{ m^2}\log
\frac{m}{\lambda} \label{q10aa}
\end{equation}

Let us then consider the following term in (\ref{q11}):
\begin{equation}
\frac{-ie^3}{(2\pi)^n}\int_0^1 dx\int_0^1 2ydy
\frac{8p_Eq_E\pi^{\frac{n}{2}}\Gamma(3-\frac{n}{2})r}
{\Gamma(3)(\Delta -r^2)^{3-\frac{n}{2}}} \label{q12}
\end{equation}
where we can (as an approximation) let $n=4$. With the
on-mass-shell condition we have that $\Delta -r^2$ is again given
by (\ref{q4}). Then letting $\lambda=0$ we have that (\ref{q12})
is given by:
\begin{equation}
\frac{-ie\alpha}{4\pi}\int_0^1 dx\int_0^1 ydy \frac{8p_Eq_E}{-r}
\label{q13}
\end{equation}
With the on-mass-shell condition we have $r=my$. Thus (\ref{q13})
is equal to
\begin{equation}
(-ie)\frac{-\alpha}{4\pi m}8p_Eq_E \label{q14}
\end{equation}
Again the factor $p_Eq_E$ is for the exchange of energies for two
electrons with proper energies $p_E$ and $q_E$ respectively and
thus it is the vital factor. This factor is then for the
space-time statistics and later it will be for a linear statistics
of summation for the on-mass-shell condition. Let us introduce a
space-time statistics on the factor $p_Eq_E$, as follows. With the
on-mass-shell condition we write $p_Eq_E$ in the following form:
\begin{equation}
p_Eq_E=\frac12 (mp_E+q_E m)=\frac12 m(p_E+q_E) \label{q15}
\end{equation}
Then we introduce a space-time statistics on the proper energies
$p_E$ and $q_E$ respectively that $p_E$ gives a statistics $\beta
p$ and $q_E$ gives a statistics $\beta p^{\prime}$ where $p$ and
$p^{\prime}$ are space-time four vectors such that $p^2=m^2$;
$p^{\prime  2}=m^2$; and $\beta$ is a statistical factor to be
determined.

Then we have the following Gordan relation on the space-time four
vectors $p$ and $p^{\prime}$ respectively (\cite{Zub}\cite{Tic}):
\begin{equation}
\begin{array}{rl}
p^{\mu}& =\gamma^{\mu} (p\cdot \gamma) +i\sigma^{\mu\nu}p_{\nu} \\
p^{\mu\prime}& =(p^{\prime}\cdot \gamma) \gamma^{\mu} -i\sigma^{\mu\nu}p_{\nu}^{\prime} \\
\end{array}
\label{q16}
\end{equation}
where $p^{\mu}$ and $p^{\mu\prime}$ denote the four components of
$p$ and $p^{\prime}$ respectively. Thus from (\ref{q15}) and the
Gordan relation (\ref{q16}) we have the following space-time
statistics:
\begin{equation}
\frac12 (mp_E+q_E m)=\frac12 m\beta (\gamma^{\mu}( p\cdot \gamma)
+ (p^{\prime}\cdot \gamma) \gamma^{\mu}-i\sigma^{\mu\nu}q_{\nu})
\label{q17}
\end{equation}
where following the convention of QED we define $q=p^{\prime}-p$.

From (\ref{q17}) we see that the space-time statistics on $p_E$
for giving the four vector $p$ needs the product of two Dirac
$\gamma$-matrices. Then since the introducing of a Dirac
$\gamma$-matrix for space-time statistics requires a statistical
factor $\frac12$ we have that the statistical factor
$\beta=\frac14$.

Then as in the literature on QED when evaluated between
polarization spinors, the $p^{\prime}\cdot \gamma$ and
$\gamma\cdot p$ terms are deduced to the mass $m$ respectively.
Thus the term $\frac12 m\beta (\gamma^{\mu} p\cdot \gamma +
p^{\prime}\cdot \gamma \gamma^{\mu})$ as a constant term can be
cancelled by the corresponding counter term with the factor
$z_e-1$ in (\ref{2.1}).

Thus by space-time statistics on $p_Eq_E$ from the term
(\ref{q14}) we get the following vertex correction:
\begin{equation}
(-ie)\frac{i\alpha}{4\pi m}\sigma^{\mu\nu}q_{\nu} \label{q18}
\end{equation}
where $q=p-p^{\prime}$ and the factor $8$ in (\ref{q14}) is
cancelled by the statistical factor $\frac12\beta=\frac18$. We
remark that in the way of getting (\ref{q18}) a factor $m$ has
been absorbed by the  two polarization spinors $u$ to get the form
$\sqrt \frac{m}{E} u$ of the spinors of external electrons.

Then from (\ref{q18}) we get the following exact second order
magnetic moment:
\begin{equation}
\frac{\alpha }{2\pi}\mu_0 \label{q18a}
\end{equation}
where  $\mu_0=\frac{1}{2m}$ is the Dirac magnetic moment as in the
literature on QED (\cite{Zub}).

We see that this result is just the second order anomalous
magnetic moment obtained from the conventional QED
(\cite{Zub}\cite{Tic}-\cite{Yen}). Here we remark that we can
obtain this anomalous magnetic moment exactly while in the
conventional QED this anomalous magnetic moment is obtained only
by approximation under the condition that $|q^2|<<m^2$. The point
is that we do not need to carry out a complicate integration as in
the literature in QED when the on-mass-shell condition is applied
to the proper energies $p_E$ and $q_E$, and with the on-mass-shell
condition applied to the proper energies $p_E$ and $q_E$ the
computation is simple and the computed result is the exact result
of the anomalous magnetic moment.

Let us then consider the following terms in the one-loop vertex
correction (\ref{2.15aa}):
\begin{equation}
\begin{array}{rl}
 & \frac{ie^3}{(2\pi)^n}\int_0^1 dx\int_0^1 2ydy
\frac{5(p_E+q_E)\pi^{\frac{n}{2}}\Gamma(3-1-\frac{n}{2})\frac{n}{2}}
{\Gamma(3)(\Delta -r^2)^{3-2-1}}
\frac{1}{(-\Delta +r^2)^{2-\frac{n}{2}}}\\
&\\
& + \frac{ie^3}{(2\pi)^n}\int_0^1 dx\int_0^1 2ydy
\frac{5(p_E+q_E)\pi^{\frac{n}{2}}\Gamma(3-\frac{n}{2})r^2}
{\Gamma(3)(\Delta -r^2)^{3-2}}
\frac{1}{(-\Delta +r^2)^{2-\frac{n}{2}}}\\
&\\
& -\frac{ie^3}{(2\pi)^n}\int_0^1 dx\int_0^1 2ydy
\frac{\frac{(n+2)}{2}2\pi^{\frac{n}{2}}\Gamma(3-1-\frac{n}{2})r}
{\Gamma(3)(\Delta -r^2)^{3-2-1}}
\frac{1}{(-\Delta +r^2)^{2-\frac{n}{2}}}\\
&\\
&- \frac{ie^3}{(2\pi)^n}\int_0^1 dx\int_0^1 2ydy
\frac{2\pi^{\frac{n}{2}}\Gamma(3-\frac{n}{2})r^3}
{\Gamma(3)(\Delta -r^2)^{3-2}}
\frac{1}{(-\Delta +r^2)^{2-\frac{n}{2}}}\\
\end{array}
\label{q19}
\end{equation}
From the on-mass-shell condition we have shown that $\Delta
-r^2=-r^2$ where we have set $\lambda=0$. The first and the second
term are with the factor $(p_E+q_E)$ which by Fermi-Dirac
statistics gives the statistics $(p_E+q_E)\frac12\gamma^{\mu}$.
Then from the following integration:
\begin{equation}
\int_0^1 dx\int_0^1 2yrdy=\int_0^1 dx\int_0^1
2y^2(p_Ex+(1-x)q_E)dy \label{q20}
\end{equation}
we get a factor $(p_E+q_E)$ for the third and fourth terms. Thus
all these four terms by Fermi-Dirac statistics are with the
statistics $(p_E+q_E)\frac12 \gamma^{\mu}$. Then by the
on-mass-shell condition  we have that the statistics
$(p_E+q_E)\frac12 \gamma^{\mu}$ gives the statistics
$m\gamma^{\mu}$. Thus (\ref{q19}) gives a statistics which is of
the form $\gamma^{\mu}\cdot \mbox{constant}$. Thus this
statistical constant term can be cancelled by the corresponding
counter term with the factor $z_e-1$ in (\ref{2.1}).

Thus under the on-mass-shell condition the renormalized vertex
correction $(-ie)\Lambda_{R}(p^{\prime},p)$ from the one-loop
vertex correction is given by the sum of (\ref{q10}) and
(\ref{q18}):
\begin{equation}
(-ie)\Lambda_{R}(p^{\prime},p)
=(-ie)[\gamma^{\mu}\frac{\alpha}{3\pi}\frac{q^2}{ m^2}\log
\frac{m}{\lambda} + \frac{i\alpha }{4\pi m}\sigma^{\mu\nu}q_{\nu}]
\label{q21}
\end{equation}

\section{Computation of the Lamb shift: Part I}

The above computation of the vertex correction has not been
completed since the parameter $\lambda$ has not been determined.
This appearance of the nonzero $\lambda$ is due to the
on-mass-shell condition. Let us in this section complete the above
computation of the vertex correction by finding another way to get
the on-mass-shell condition. By this completion of the above
computation of the vertex correction we are then able to compute
the Lamb Shift.

As in the literature of QED we let $\omega_{\mbox{min}}$ denote
the minimum of the (virtual) photon energy in the scatting of
electron. Then as in the literature of QED we have the following
relation between $\omega_{\mbox{min}}$ and $\lambda$ when
$\frac{v}{c}<<1$ where $v$ denotes the velocity of electron and
$c$ denotes the speed of light (\cite{Zub}\cite{Tic}-\cite{Yen}):
\begin{equation}
\log 2\omega_{\mbox{min}} =\log\lambda+ \frac56 \label{q22}
\end{equation}
Thus from (\ref{q21}) we have the following form of the vertex
correction:
\begin{equation}
(-ie)\gamma^{\mu}\frac{\alpha}{3\pi}\frac{q^2}{ m^2} [\log
\frac{m}{2\omega_{\mbox{min}}} +\frac56] + (-ie)\gamma^{\mu}\cdot
\frac{ie\alpha i\sigma^{\mu\nu}q_{\nu}}{4\pi m} \label{q23}
\end{equation}
Let us then find a way to compute the following term in the vertex
correction (\ref{q23}):
\begin{equation}
(-ie)\gamma^{\mu}\frac{\alpha}{3\pi}\frac{q^2}{ m^2}\log
\frac{m}{2\omega_{\mbox{min}}} \label{q23a}
\end{equation}

The parameter $2\omega_{\mbox{min}}$ is for the exchanging (or
shifting) of the proper energies $p_E$ and $q_E$ of electrons.
Thus the magnitudes of $p_E$ and $q_E$ correspond to the magnitude
of $\omega_{\mbox{min}}$. When the $\omega_{\mbox{min}}$ is chosen
the corresponding $p_E$ and $q_E$ are also chosen and vise versa.

Since $\omega_{\mbox{min}}$ is chosen to be very small we have
that the corresponding proper energies $p_E$ and $q_E$ are very
small that they are no longer equal to the mass $m$ for the
on-mass-shell condition and they are for the virtual electrons.
Then to get the on-mass-shell condition we use a linear statistics
of summation on the vital factor $p_Eq_E$. This means that the
large amount of the effects $p_Eq_E$ of the exchange of the
virtual electrons are to be summed up to statistically getting the
on-mass-shell condition.

Thus let us consider again the one-loop vertex correction
(\ref{2.15aa}) where we choose $p_E$ and $q_E$ such that  $p_E<<m$
and $q_E<<m$. This chosen corresponds to the chosen of
$\omega_{\mbox{min}}$. We can choose $p_E$ and $q_E$ as small as
we want such that $p_E<<m$ and $q_E<<m$.  Thus we can let
$\lambda=0$ and set $p_E=q_E=0$ for the $p_E$ and $q_E$ in the
denominators $(\Delta -r^2)^{3-2}$ in (\ref{2.15aa}). Thus
(\ref{2.15aa}) is approximately equal to:
\begin{equation}
\begin{array}{rl}
 & \frac{ie^3}{(2\pi)^n}\int_0^1 dx\int_0^1 dy
\frac{4p_Eq_E(p_E+q_E)\pi^{\frac{n}{2}}\Gamma(3-2)} {-m^2} +
\frac{-ie^3}{(2\pi)^n}\int_0^1 dx\int_0^1 dy \frac{2((p_E+q_E)^2
+4p_Eq_E)\pi^{\frac{n}{2}}\Gamma(3-2)r}
{-m^2} \\
&\\
& +\frac{ie^3}{(2\pi)^n}\int_0^1 dx\int_0^1 y dy
\frac{5(p_E+q_E)\pi^{\frac{n}{2}}\Gamma(2-\frac{n}{2})\frac{n}{2}}
{(-\Delta +r^2)^{2-\frac{n}{2}}} + \frac{ie^3}{(2\pi)^n}\int_0^1
dx\int_0^1 dy \frac{5(p_E+q_E)\pi^{\frac{n}{2}}\Gamma(3-2)r^2}
{-m^2}\\
&\\
& +\frac{-ie^3}{(2\pi)^n}\int_0^1 dx\int_0^1 y dy
\frac{\frac{(n+2)}{2}2\pi^{\frac{n}{2}}\Gamma(2-\frac{n}{2})r}
{(\Delta -r^2)^{2-\frac{n}{2}}} + \frac{-ie^3}{(2\pi)^n}\int_0^1
dx\int_0^1 dy \frac{2\pi^{\frac{n}{2}}\Gamma(3-2)r^3}
{-m^2} \\
\end{array}
\label{q24}
\end{equation}
Let us then first consider the four terms in (\ref{q24}) without
the factor $\Gamma(2-\frac{n}{2})$. For these four terms we can
(as an approximation) let $n=4$. Carry out the integrations
$\int_0^1 dx\int_0^1 ydy$ of these four terms we have that the sum
of these four terms is given by:
\begin{equation}
\begin{array}{rl}
& (ie)\frac{\alpha\pi^2}{4\pi^3 m^2}
[4p_Eq_E(p_E+q_E)-\frac12((p_E+q_E)^2+4p_Eq_E)(p_E+q_E) \\
&\\
& \qquad +\frac{5}{9}(p_E+q_E)(p_E^2 +q_E^2 +p_Eq_E)
-\frac18(p_E^3 +q_E^3 +p_E^2q_E+p_Eq_E^2)] \\
&\\
=& (ie)\frac{\alpha\pi^2}{4\pi^3 m^2}
(p_E+q_E)[\frac{5}{72}p_E^2+\frac{5}{72}q_E^2-\frac{14}{9}p_Eq_E]
\end{array}
\label{q25}
\end{equation}
where the four terms of the sum are from the corresponding four
terms of (\ref{q24}) respectively.

Then we consider the two terms in (\ref{q24}) with the factor
$\Gamma(2-\frac{n}{2})$. Let $\delta :=2-\frac{n}{2}>0$. We have
\begin{equation}
\begin{array}{rl}
 \Gamma(\delta)\cdot (-\Delta +r^2)^{-\delta}
= (\frac{1}{\delta}+ \mbox{a finite limit term as} \, \delta\to
0) \cdot e^{-\delta \log (-\Delta+r^2)}
\end{array}
\label{q26}
\end{equation}
We have
\begin{equation}
\begin{array}{rl}
 \frac{1}{\delta}\cdot e^{-\delta \log (-\Delta+r^2)} 
= \frac{1}{\delta}\cdot [1-\delta \log (-\Delta+r^2)+
0(\delta^2)]
\end{array}
\label{q27}
\end{equation}
Then we have
\begin{equation}
\begin{array}{rl}
& -\frac{1}{\delta}\cdot \delta \log (-\Delta+r^2)
\\&\\=& 
-\log m^2y -\log \frac{1}{m^2}[m^2-p_E^2 x-q_E^2(1-x) +(p_E x+q_E(1-x))^2 y] \\
&\\
=& -\log m^2y -\log [1- \frac{p_E^2
x(1-xy)+q_E^2(1-x)(1-(1-x)y)}{m^2} +\frac{2p_Eq_Ex(1-x)y}{m^2}
+0(\frac{p_E^2+q_E^2}{m^2})]
\end{array}
\label{q28}
\end{equation}
Then the constant term $-\log m^2y$ in (\ref{q28}) can be
cancelled by the corresponding counter term with the factor
$z_e-1$ in (\ref{2.1}) and thus can be ignored. When $p_E^2<<m^2$
and $q_E^2<<m^2$ the second term in (\ref{q28}) is approximately
equal to:
\begin{equation}
\frac{p_E^2 x(1-xy)+q_E^2(1-x)(1-(1-x)y)}{m^2}
-\frac{2p_Eq_Ex(1-x)y}{m^2} \label{q29}
\end{equation}
Thus by (\ref{q29}) the sum of the two terms in (\ref{q24}) having
the factor $\Gamma(2-\frac{n}{2})$ is approximately equal to:
\begin{equation}
\begin{array}{rl}
& +\frac{ie^3}{(2\pi)^n}\int_0^1 dx\int_0^1 ydy
5(p_E+q_E)\pi^{\frac{n}{2}}\frac{n}{2}
[\frac{p_E^2 x(1-xy)+q_E^2(1-x)(1-(1-x)y)}{m^2} -\frac{2p_Eq_Ex(1-x)y}{m^2}] \\
&\\
& -\frac{ie^3}{(2\pi)^n}\int_0^1 dx\int_0^1 ydy
2\pi^{\frac{n}{2}}\frac{(n+2)}{2}r [\frac{p_E^2
x(1-xy)+q_E^2(1-x)(1-(1-x)y)}{m^2} -\frac{2p_Eq_Ex(1-x)y}{m^2}]
\end{array}
\label{q30}
\end{equation}
where we can (as an approximation) let $n=4$. Carrying out the
integration $\int_0^1 dx \int_0^1 ydy$ of the two terms in
(\ref{q30})  we have that (\ref{q30}) is equal to the following
result:
\begin{equation}
(ie)\frac{\alpha\pi^2}{4\pi^3 m^2} (p_E+q_E)[(-5\cdot
\frac{1}{9}\cdot 2p_Eq_E)
+(-\frac{7}{24}p_E^2-\frac{7}{24}q_E^2+\frac{3}{9}p_Eq_E)]
\label{q31}
\end{equation}
where the first term and the second term in the $[\cdot ]$ are
from the  first term and the second term in (\ref{q30})
respectively.
Combining (\ref{q25}) and (\ref{q30}) we have the following result
which approximately equal to (\ref{q24}) when $p_E^2<<m^2$ and
$q_E^2<<m^2$:
\begin{equation}
(-ie)\frac{\alpha\pi^2}{4\pi^3 m^2}
(p_E+q_E)[\frac{2}{9}p_E^2+\frac{2}{9}q_E^2+\frac{7}{3}p_Eq_E]
\label{q32}
\end{equation}
where the exchanging term $\frac{7}{3}p_Eq_E$ is of vital
importance.
Now to have the on-mass-shell condition let us consider a linear
statistics of summation on (\ref{q32}). Let there be a large
amount of virtual electrons $z_j, j\in J$ indexed by a set $J$
with the proper energies $p_{Ej}^2<<m^2$ and $q_{Ej}^2<<m^2$,
$j\in J$. Then from (\ref{q32}) we have the following linear
statistics of summation on (\ref{q32}):
\begin{equation}
(-ie)\frac{\alpha\pi^2}{4\pi^3 m^2}
(p_{Ej_0}+q_{Ej_0})[\frac{2}{9}\sum_{j}p_{Ej}^2+\frac{2}{9}\sum_{j}q_{Ej}^2+
\frac{7}{3}\sum_{j}p_{Ej}q_{Ej}] \label{q33}
\end{equation}
where for simplicity we let:
\begin{equation}
p_{Ej}+q_{Ej}=p_{Ej^{\prime}}+q_{Ej^{\prime}}
=p_{Ej_0}+q_{Ej_0}=2m_0 \label{aq33}
\end{equation}
for all $j,j^{\prime}\in J $ and for some (bare) mass $m_0<<m$ and
for some $j_0\in J$.

Then by applying a Fermi-Dirac statistics on the factor
$p_{Ej_0}+q_{Ej_0}$ in (\ref{q33}) we have the following
Fermi-Dirac statistics for (\ref{q33}):
\begin{equation}
\begin{array}{rl}
& (-ie)\frac{\alpha\pi^2}{4\pi^3 m^2} \frac12 \gamma^{\mu}
(p_{Ej_0}+q_{Ej_0})[\frac{2}{9}\sum_{j}p_{Ej}^2+\frac{2}{9}\sum_{j}q_{Ej}^2+
\frac{7}{3}\sum_{j}p_{Ej}q_{Ej}] \\
&\\
=&(-ie)\frac{\alpha\pi^2}{4\pi^3 m^2} \gamma^{\mu}
m_0[\frac{2}{9}\sum_{j}p_{Ej}^2+\frac{2}{9}\sum_{j}q_{Ej}^2+
\frac{7}{3}\sum_{j}p_{Ej}q_{Ej}] \\
\end{array}
\label{q33a}
\end{equation}

Then for the on-mass-shell condition we require that the linear
statistical sum $2m_0\frac{7}{3}\sum_{j}p_{Ej}q_{Ej}$ in
(\ref{q33a}) is of the following form:
\begin{equation}
2m_0\frac{7}{3}\sum_{j}p_{Ej}q_{Ej}=\beta_0 2m\frac{7}{3} q^2
\label{q34}
\end{equation}
where $q^2=(p^{\prime}-p^2)$ and the form
$2mq^2=2m(p^{\prime}-p^2)$ is the on-mass-shell condition; and
that $\beta_0$ is a statistical factor (to be determined) for this
linear statistics of summation and is similar to the statistical
factor $(2\pi)^n$ for the space-time statistics.

Then we notice that (\ref{q33a}) is for computing (\ref{q23a}) and
thus its exchanging term corresponding to $\sum_{j}p_{Ej}q_{Ej}$
must be equal to (\ref{q23a}). From (\ref{q33a}) we see that there
is a statistical factor $4$ which does not appear in (\ref{q23a}).
Since this exchanging term in (\ref{q33a}) must be equal to
(\ref{q23a}) we conclude that the statistical factor $\beta_0$
must be equal to $4$ so as to cancel the statistical factor $4$ in
(\ref{q33a}) (We also notice that there is a statistical factor
$\pi^2$ in the numerator of (\ref{q33a}) and thus it requires a
statistical factor $4$ to form the statistical factor $(2\pi)^2$
and thus $\beta_0=4$). Thus we have that for the on-mass-condition
we have that (\ref{q33a}) is of the following statistical form:
\begin{equation}
(-ie)\frac{\alpha\pi^2}{\pi^3 m^2} m\gamma^{\mu}[\beta_2
\frac{2}{9}m^2+\beta_2^{\prime}\frac{2}{9}m^2+\frac{7}{3}q^2]
\label{q35}
\end{equation}

 Then from (\ref{q35}) we have the following statistical
form:
\begin{equation}
(-ie)\frac{\alpha\pi^2}{\pi^3 m^2} \gamma^{\mu}[\beta_2
\frac{2}{9}m^2+\beta_2^{\prime}\frac{2}{9}m^2+\frac{7}{3}q^2]
\label{q36}
\end{equation}
where the factor $m$ of $m\gamma^{\mu}$ has been absorbed to the
two external spinors of electron.

Then we notice that the term corresponding to $\beta_2
\frac{2}{9}m^2+\beta_2^{\prime}\frac{2}{9}m^2$ in (\ref{q36}) is
as a constant term and thus can be cancelled by the corresponding
counter term with the factor $z_e-1$ in  (\ref{2.1}). Thus from
(\ref{q36}) we have the following statistical form of effect which
corresponds to (\ref{q23a}):
\begin{equation}
(-ie)\gamma^{\mu}\frac{\alpha}{\pi m^2}\frac{7}{3}q^2 \label{q36a}
\end{equation}

This effect (\ref{q36a}) is as the total effect of $q^2$ computed
from the one-loop vertex with the minimal energy
$\omega_{\mbox{min}}$ and thus includes the effect of $q^2$ from
the anomalous magnetic moment. Thus we have that (\ref{q23a}) is
computed and is given by the following statistical form:
\begin{equation}
(-ie)\gamma^{\mu}\frac{\alpha}{3\pi}\frac{q^2}{m^2}\log
\frac{m}{2\omega_{\mbox{min}}}
=(-ie)\gamma^{\mu}\frac{\alpha}{3\pi}\frac{q^2}{m^2}[7-\frac38]
\label{q36b}
\end{equation}
where the term corresponding to the factor $\frac38$ is from the
anomalous magnetic moment (\ref{q18}) as computed in the
literature of QED (\cite{Zub}). This completes our computation of
(\ref{q23a}). Thus under the on-mass-shell condition the
renormalized one-loop vertex $(-ie)\Lambda_{R}(p^{\prime},p)$ is
given by:
 \begin{equation}
(-ie)\Lambda_{R}(p^{\prime},p) =(-ie)[\gamma^{\mu}\frac{\alpha
q^2}{3\pi m^2}(7+\frac56-\frac38) + \frac{i\alpha}{4\pi
m}\sigma^{\mu\nu}q_{\nu}] \label{q21a}
\end{equation}
This completes our computation of the one-loop vertex correction.

\section{Computation of photon self-energy }

To compute the Lamb shift let us then consider the one-loop photon
self energy (\ref{2.16a}). As a statistics we extend the one
dimensional integral $\int dp_E$ to the $n$-dimensional integral
$\int d^n p$ ($n\to 4$) where $p=(p_E, {\bf p})$. This is similar
to the dimensional regularization in the existing quantum field
theories (However here our aim is to increase the dimension for
statistics which is different from the dimensional regularization
which is to reduce the dimension from $4$ to $n$ to avoid the
ultraviolet divergence). With this statistics the factor $2\pi$ is
replaced by the statistical factor $(2\pi)^n$. From this
statistics on  (\ref{2.16a})
  we have that the following statistical one-loop photon self-energy:
\begin{equation}
(-1)i^2(-i)^2\frac{e^2}{(2\pi)^n}\int_0^1 dx\int d^n p
\frac{4p_E^2+4p_E k_E +k_E^2} {(p^2+2pkx+k_E^2x-m^2)^{2}}
\label{p1}
\end{equation}
where $p^2=p_E^2 -{\bf p}^2$,  and ${\bf p}^2$ is from
$\omega^2=m^2+{\bf p}^2$; and:
\begin{equation}
 pk:=p_E k_E -{\bf p}\cdot {\bf 0}=p_E k_E
\label{ap1}
\end{equation}
As a Feynman rule for space-time statistics
 a statistical factor $(-1)$ has been
introduced for this photon self-energy since it has a loop of
electron particles.

By using the formulae for computing Feynman integrals we have that
(\ref{p1}) is equal to:
\begin{equation}
\begin{array}{rl}
\frac{(-1)ie^2}{(2\pi)^n}\int_0^1 dx[
\frac{k_E^2(4x^2-4x+1)\pi^{\frac{n}{2}}\Gamma(2-\frac{n}{2})}
{\Gamma(2)(m^2-k_E^2x(1-x))^{2-\frac{n}{2}}}
 +
\frac{\pi^{\frac{n}{2}}\Gamma(2-1-\frac{n}{2})\frac{n}{2}}
{\Gamma(2)(m^2-k_E^2x(1-x))^{2-1-\frac{n}{2}}}]
\end{array}
\label{p2}
\end{equation}

Let us first consider the first term in the $[\cdot]$ in
(\ref{p2}) . Let $\delta :=2-\frac{n}{2}>0$. As for the one-loop
vertex we have
\begin{equation}
\begin{array}{rl}
& \Gamma(\delta)\cdot (m^2-k_E^2x(1-x))^{-\delta}
=(\frac{1}{\delta}+ \mbox{a finite limit term as} \, \delta\to
0) \cdot e^{-\delta \log (m^2-k_E^2x(1-x))}
\end{array}
\label{p3}
\end{equation}
We have
\begin{equation}
\begin{array}{rl}
 \frac{1}{\delta}\cdot e^{-\delta \log (m^2-k_E^2x(1-x))} 
= \frac{1}{\delta}\cdot [1-\delta \log (m^2-k_E^2x(1-x))+
0(\delta^2)]
\end{array}
\label{p4}
\end{equation}
Then we have
\begin{equation}
\begin{array}{rl}
 -\frac{1}{\delta}\cdot \delta \log (m^2-k_E^2x(1-x))
= -\log m^2 -\log [1-\frac{k_E^2x(1-x) }{m^2} ]
\end{array}
\label{p5}
\end{equation}
Then the constant term $-\log m^2$ in (\ref{p5}) can be cancelled
by the corresponding counter term with the factor $z_A-1$ in
(\ref{2.1}) and thus can be ignored. When $k_E^2<<m^2$ the second
term in (\ref{p5}) is approximately equal to:
\begin{equation}
\frac{k_E^2x(1-x)}{m^2} \label{p6}
\end{equation}
Carrying out the integration $\int_0^1 dx $ in  (\ref{p1}) with
$-\log [1-\frac{k_E^2x(1-x) }{m^2} ]$ replaced by (\ref{p6}), we
have the following result:
\begin{equation}
\int_0^1 dx(4x^2-4x+1)\frac{k_E^2x(1-x)}{m^2}=\frac{k_E^2}{30m^2}
\label{p7}
\end{equation}
Thus  as in the literature in QED from the photon self-energy we
have the following term which gives contribution to the Lamb
shift:
\begin{equation}
\frac{k_E^2}{30m^2}=\frac{(p_{E}-q_{E})^2}{30m^2} \label{p8}
\end{equation}
where $k_E=p_{E}-q_{E}$ and $p_E$, $q_E$ denote the proper
energies of virtual electrons. Let us then consider statistics of
a large amount of photon self-energy (\ref{p5}). When there is a
large amount of photon self-energies we have the following linear
statistics of summation:
\begin{equation}
\frac{\sum _i k_{Ei}^2}{30m^2} \label{p9}
\end{equation}
where each $i$ represent a photon. Let us write
\begin{equation}
k_{Ei}^2=(p_{Ei}-q_{Ei})^2=p_{Ei}^2-2p_{Ei}q_{Ei}+q_{Ei}^2
\label{p10}
\end{equation}
Thus
\begin{equation}
\sum _ik_{Ei}^2=\sum _i(p_{Ei}-q_{Ei})^2 =\sum _i p_{Ei}^2-2\sum
_i p_{Ei}q_{Ei}+\sum _i q_{Ei}^2 \label{p11}
\end{equation}
Now as the statistics of the vertex correction we have the
following statistics:
\begin{equation}
\sum _i p_{Ei}q_{Ei}=4(p^{\prime}-p)^2=4q^2 \label{p12}
\end{equation}
where $4$ is a statistical factor which is the same statistical
factor of case of the vertex correction and $p$, $p^{\prime}$ are
on-mass-shell four vectors of electrons. As the the statistics of
the vertex correction this statistical factor cancels another
statistical factor $4$. On the other hand as the statistics of the
vertex correction we have the following statistics:
\begin{equation}
\sum _i p_{Ei}^2=\beta_3 m^2,  \qquad \sum _i q_{Ei}^2=\beta_4
m^2, \label{p13}
\end{equation}
where $\beta_3$ and $\beta_4$ are two statistical factors. As the
case of the vertex correction these two sums give constant terms
and thus can be cancelled by the corresponding counter term with
the factor $z_A-1$ in (\ref{2.1}). Thus from (\ref{p11}) we have
that the linear statistics of summation $\sum _ik_{Ei}^2$ gives
the following statistical renormalized photon self-energies
$\Pi_R$ and $\Pi_M$ (where we follow the notations in the
literature of QED for photon self-energies $\Pi_M$):
\begin{equation}
i\Pi_R(k_E)=ik_E^2
\Pi_M(k_E)=ik_E^2\frac{\alpha}{4\pi}\frac{8q^2}{30m^2}=ik_E^2\frac{\alpha}{3\pi}\frac{q^2}{5m^2}
\label{p14}
\end{equation}
where we let $k_{Ei}^2=k_E^2$ for all $i$.
Let us then consider the second term in the $[\cdot ]$ in
(\ref{p2}). This term can be written in the following form:
\begin{equation}
\begin{array}{rl}
 &\frac{\pi^{\frac{n}{2}}\Gamma(2-\frac{n}{2})\frac{n}{2}}
{(1-\frac{n}{2})\Gamma(2)(m^2-k_E^2x(1-x))^{2-1-\frac{n}{2}}} 
=\frac{\pi^{\frac{n}{2}}\Gamma(2-\frac{n}{2})\frac{n}{2}}
{(1-\frac{n}{2})\Gamma(2)}
[(m^2-k_E^2x(1-x))+ 0(\delta)] \\
&\\
= &k_E^2[\frac{1}{\delta}\cdot
\frac{(-1)\pi^{\frac{n}{2}}\Gamma(2-\frac{n}{2})\frac{n}{2}}
{(1-\frac{n}{2})\Gamma(2)} x(1-x)] + [\frac{1}{\delta}\cdot
\frac{\pi^{\frac{n}{2}}\Gamma(2-\frac{n}{2})\frac{n}{2}}
{(1-\frac{n}{2})\Gamma(2)} m^2+ 0(\delta))] \\
\end{array}
\label{pp1}
\end{equation}

Then  the first term in (\ref{pp1}) under the integration
$\int_0^1 dx$ is of the form $k_E^2\cdot \mbox{constant}$. Thus
this term can also be cancelled by the counter-term with the
factor $z_A-1$ in (\ref{2.1}). In summary the renormalization
constant $z_A$ is given by the following equation:
\begin{equation}
(-1)^3i(z_A-1)=(-i)\{ \frac{1}{\delta}
\cdot\frac{e^2\pi^{\frac{n}{2}}}{(2\pi)^n}\int_0^1
dx[(4x^2-4x+1)-\frac{nx(1-x)}{2-n}]+ c_A \}
 \label{ren}
\end{equation}
where $c_A$ is a finite constant when $\delta\to 0$. From this
equation we have that $z_A$ is a very large number when $\delta>0
$ is very small. Thus
$e_0=\frac{z_e}{z_Zz_A^{\frac12}}e=\frac{1}{n_e }e$ is a very
small constant when $\delta>0 $ is very small (and since
$\frac{e^2}{4\pi}=\alpha=\frac{1}{137}$ is small) where shall show
that we can let $z_e=z_Z$.
Then the second term in (\ref{pp1}) under the integration
$\frac{(-1)e^2}{(2\pi)^n}\int_0^1 dx$ gives a parameter $\lambda_3
>0$ for the photon self-energy since $\delta>0$ is as a parameter.
Combing the effects of the two terms in the $[\cdot ]$ in
(\ref{p2}) we have the following renormalized one-loop photon
self-energy:
\begin{equation}
i(\Pi_R(k_E)+\lambda_3) \label{pp2}
\end{equation}

Then we have the following Dyson series for photon propagator:
\begin{equation}
\begin{array}{rl}
& \frac{i}{k_E^2-\lambda_0} + \frac{i}{k_E^2-\lambda_0}
(i\Pi_R(k_E)+i\lambda_3) \frac{i}{k_E^2-\lambda_0}+... 
=\frac{i}{k_E^2(1+\Pi_M)-(\lambda_0-\lambda_3) }
=: \frac{i}{k_E^2(1+\Pi_M)-\lambda_R }\\
\end{array}
\label{ph2}
\end{equation}
where $\lambda_R$ is as a renormalized mass-energy parameter. This
is as the renormalized photon propagator. We have the following
approximation of this renormalized photon propagator:
\begin{equation}
\frac{i}{k_E^2(1+\Pi_M)-\lambda_R } \approx
\frac{i}{k_E^2-\lambda_R } (1-\Pi_M) \label{ph3}
\end{equation}

\section{Computation of the Lamb shift: Part II}

Combining the effect of vertex correction and photon self-energy
we can now compute the Lamb shift. Combining the effect of vertex
correction and photon self-energy $(-ie\gamma^{\mu})[-\Pi_M]$ we
have:
 \begin{equation}
(-ie)\Lambda_{R}(p^{\prime},p)+ (-ie\gamma^{\mu})[-\Pi_M]
=(-ie)[\gamma^{\mu}\frac{\alpha q^2}{3\pi
m^2}(7+\frac56-\frac38-\frac15) + \frac{i\alpha}{4\pi
m}\sigma^{\mu\nu}q_{\nu}] \label{p15}
\end{equation}
As in the literature of QED let us consider the states
$2S_{\frac12}$ and the $2P_{\frac12}$ in the hydrogen atom
\cite{Zub}\cite{Tic}-\cite{Yen}. Following the literature of QED
for the state $2S_{\frac12}$ an effect of $\frac{\alpha q^2}{3\pi
m^2}(\frac38)$ comes from the anomalous magnetic moment which
cancels the same term with negative sign in (\ref{p15}). Thus by
the method of computing the Lamb shift in the literature of QED we
have the following second order shift for the state
$2S_{\frac12}$:
 \begin{equation}
\Delta E_{2S_{\frac12}}=\frac{m\alpha^5}{6\pi }(7+\frac56-\frac15)
\label{p16}
\end{equation}
Similarly by the method of computing the Lamb shift in the
literature of QED from the anomalous magnetic moment we have the
following second order shift for the  state $2P_{\frac12}$:
 \begin{equation}
\Delta E_{2P_{\frac12}}=\frac{m\alpha^5}{6\pi }(-\frac18)
\label{p17}
\end{equation}
Thus the second order Lamb shift for the  states $2S_{\frac12}$
and $2P_{\frac12}$ is given by:
 \begin{equation}
\Delta E=\Delta E_{2S_{\frac12}}-\Delta E_{2P_{\frac12}}
=\frac{m\alpha^5}{6\pi }(7+\frac56-\frac15+\frac18) \label{p18}
\end{equation}
or in terms of frequencies for each of the terms in (\ref{p18}) we
have:
 \begin{equation}
\Delta \nu =952+ 113.03-27.13+16.96=1054.86 Mc/sec \label{p18a}
\end{equation}
This agrees with the experimental results given by
(\cite{Zub}\cite{Tic}-\cite{Yen}):
 \begin{equation}
\Delta \nu^{\mbox{exp}}=1057.86\pm 0.06 Mc/sec \quad \mbox{and}
\quad 1057.90\pm 0.06 Mc/sec \label{p19}
\end{equation}

\section{Computation of the electron self-energy}

Let us then consider the one-loop electron self-energy
(\ref{2.16a}). As a statistics we extend the one dimensional
integral $\int dk_E$ to the $n$-dimensional integral $\int d^n k$
($n\to 4$) where $k=(k_E, {\bf k})$. This is similar to
 the dimensional regularization in the existing quantum field theories (However here our aim is
 to increase the dimension for statistics which is different from the dimensional regularization
 which is to reduce the dimension from $4$ to $n$ to avoid the ultraviolet divergence). With
 this statistics the factor $2\pi$ is replaced by the statistical factor $(2\pi)^n$. From
 this statistics on  (\ref{2.16b})
  we have that the following statistical one-loop electron self-energy:
\begin{equation}
-i\Sigma(p_E):=i^2(-i)^2\frac{e^2}{(2\pi)^n}\int_0^1 dx\int d^n k
\frac{k_E^2-4p_E k_E +4p_E^2}
{(k^2-2kpx+p_E^2x-xm^2-(1-x)\lambda^2)^{2}} \label{p20}
\end{equation}
where $k^2=k_E^2 -{\bf k}^2$,  and ${\bf k}^2$ is from
$\omega^2=m^2+{\bf k}^2$ and $\lambda_0^2=\lambda^2+{\bf k}^2$;
and
 $kp:= k_Ep_E -{\bf k}\cdot {\bf 0}=k_Ep_E $.
By using the formulae for computing Feynman integrals we have that
(\ref{p20}) is equal to:
\begin{equation}
\begin{array}{rl}
& \frac{ie^2}{(2\pi)^n}\int_0^1 dx[
\frac{p_E^2(x^2-4x+4)\pi^{\frac{n}{2}}\Gamma(2-\frac{n}{2})}
{\Gamma(2)(xm^2+(1-x)\lambda^2-p_E^2x(1-x))^{2-\frac{n}{2}}} +
\frac{\pi^{\frac{n}{2}}\Gamma(2-1-\frac{n}{2})\frac{n}{2}}
{\Gamma(2)(xm^2+(1-x)\lambda^2-p_E^2x(1-x))^{2-1-\frac{n}{2}}}]\\
&\\
 =&\frac{ie^2}{(2\pi)^n}\int_0^1 dx\{
p_E^2(x^2-4x+4)\pi^{\frac{n}{2}} [(\frac{1}{\delta}+ \mbox{a
finite limit term as} \, \delta\to 0)
\cdot e^{-\delta \log (xm^2+(1-x)\lambda^2-p_E^2x(1-x))}]\\
&\\
&-\pi^{\frac{n}{2}}\frac{n}{2}
\frac{1}{\delta}[xm^2+(1-x)\lambda^2-p_E^2x(1-x) +0(\delta)]\} \\
&\\
 =&\frac{ie^2}{(2\pi)^n}\int_0^1 dx\{
p_E^2(x^2-4x+4)\pi^{\frac{n}{2}} [\frac{1}{\delta}-
\frac{1}{\delta}\cdot\delta \log
(xm^2+(1-x)\lambda^2-p_E^2x(1-x))+ 0(\delta)]\\
&\\
&- \pi^{\frac{n}{2}}\frac{n}{2}
\frac{1}{\delta}[xm^2+(1-x)\lambda^2-p_E^2x(1-x) +0(\delta)]\}\\
&\\
=&\frac{ie^2}{(2\pi)^n}\int_0^1 dx\{
p_E^2(x^2-4x+4)\pi^{\frac{n}{2}} [\frac{1}{\delta}-
\log(xm^2+(1-x)\lambda^2-p_E^2x(1-x))+ 0(\delta)]\\
&\\
&- \pi^{\frac{n}{2}}\frac{n}{2}
\frac{1}{\delta}[xm^2+(1-x)\lambda^2-p_E^2x(1-x) +0(\delta)]\}\\
&\\
=&\frac{ie^2}{(2\pi)^n}\int_0^1 dx\{
p_E^2(x^2-4x+4)\pi^{\frac{n}{2}} [\frac{1}{\delta}-
\log(xm^2+(1-x)\lambda^2)-\log(1-\frac{p_E^2x(1-x)}{xm^2+(1-x)\lambda^2})+ 0(\delta)]\\
&\\
 &- \pi^{\frac{n}{2}}\frac{n}{2}
\frac{1}{\delta}[xm^2+(1-x)\lambda^2-p_E^2x(1-x) +0(\delta)]\}\\
&\\
=:&\frac{ie^2}{(2\pi)^n}\int_0^1 dx\{
p_E^2(x^2-4x+4)\pi^{\frac{n}{2}} [\frac{1}{\delta}-
\log(xm^2+(1-x)\lambda^2)-\log(1-\frac{p_E^2x(1-x)}{xm^2+(1-x)\lambda^2})+ 0(\delta)]\\
&\\
 &+p_E^2\cdot \frac{1}{\delta}\pi^{\frac{n}{2}}\frac{n}{2}x(1-x)
\} +i\omega_3\\
\end{array}
\label{p21}
\end{equation}
where $\omega_3>0$ is as a mass-energy parameter.

Then we notice that from the expressions for $\Sigma_0(p_E)$ and
$\Lambda_0(p_E,q_E)$ in (\ref{2.16b}) and (\ref{2.15}) we have the
following identity:
\begin{equation}
\frac{\partial }{\partial p_E}\Sigma_0(p_E)=-\Lambda_0(p_E,p_E)+
\frac{i4e^2}{2\pi}\int dk
\frac{-k_E+2p_E}{(k_E^2-\lambda_0^2)((p_E-k_E)^2-\omega^2)}
\label{p23}
\end{equation}
This is as a Ward-Takahashi identity which is analogous to the
corresponding Ward-Takahashi identity in the conventional QED
theory \cite{Zub}.

From (\ref{2.16b}) and (\ref{2.15}) we get their statistical forms
by changing $\int dk$ to $\int d^nk$. From this summation form of
statistics and the identity (\ref{p23}) we then get the following
statistical Ward-Takahashi identity:
\begin{equation}
\frac{\partial }{\partial p_E}\Sigma(p_E)=-\Lambda(p_E,p_E)+
\frac{i4e^2}{(2\pi)^n}\int_0^1 dx \int d^n
k\frac{-k_E+2p_E}{(k^2-2kpx+p_E^2x-xm^2-(1-x)\lambda^2)^{2}}
\label{p23a}
\end{equation}
where $\Sigma(p_E)$ denotes the statistical form of
$\Sigma_0(p_E)$ and is given by (\ref{p20}) and $\Lambda(p_E,q_E)$
denotes the statistical form of $\Lambda_0(p_E,q_E)$ as in the
above sections.

After the differentiation of (\ref{p21}) with respect to $p_E$ the
remaining  factor $p_E$ of the factor $p_E^2$ of (\ref{p21}) is
absorbed to the external spinors as the mass $m$ and a factor
$\frac{\gamma^{\mu}}{2}$ is introduced by space-time statistics,
as the case of the statistics of the vertex correction
$\Lambda_0(p_E,q_E)$ in the above sections. From the absorbing of
a factor $p_E$ to the external spinors for both sides of this
statistical Ward-Takahashi identity we then get a statistical
Ward-Takahashi identity where the Taylor expansion (of the
variable $p_E$) of both sides of this statistical Ward-Takahashi
identity are with constant term as the beginning term. From this
Ward-Takahashi identity we have that these two constant terms must
be the same constant. Then the constant term, denoted by
$C(\delta)$, of the vertex correction of this Ward-Takahashi
identity is cancelled by the counter-term with the factor $z_e-1$
in (\ref{2.1}), as done in the above computation of the
renormalized vertex correction $A_R(p^{\prime},p)$ (At this point
we notice that in computing the constant term of the vertex
correction some terms with the factor $p_E$ has been changed to
constant terms under the on-mass-shell condition $p_E=m$. This
then modifies the definition of $C(\delta)$).

On the other hand let us denote the constant term for the electron
self-energy by $B(\delta)$. Then from the above statistical
Ward-Takahashi identity we have the following equality:
\begin{equation}
B(\delta)+ a_1\cdot\frac{1}{\delta} +b_1=C(\delta)\label{pw1}
\end{equation}
where $a_1$, $b_1$ are finite constants when $\delta\to 0$ and the
term $a_1\cdot\frac{1}{\delta}$ is from the second term in the
right hand side of (\ref{p23a}).

Let us then compute the constant term $B(\delta)$ for the electron
self-energy, as follows. As explained in the above the constant
term for the electron self-energy can be obtained by
differentiation of (\ref{p21}) with respect to $p_E$ and the
removing of the remaining factor $p_E$ of $p_E^2$. We have:
\begin{equation}
\begin{array}{rl}
& \frac{\partial}{\partial p_E}\{ \frac{ie^2}{(2\pi)^n}\int_0^1 dx
p_E^2(x^2-4x+4)\pi^{\frac{n}{2}}
[\frac{1}{\delta}- \log(xm^2+(1-x)\lambda^2-p_E^2x(1-x))]\\
&\\
 &+p_E^2\cdot
 \frac{1}{\delta}\pi^{\frac{n}{2}}\frac{n}{2}\frac{ie^2}{(2\pi)^n}\int_0^1x(1-x)dx
 +i\omega_3
\}\\
&\\
 =&\frac{ie^2}{(2\pi)^n}\int_0^1 dx
2p_E(x^2-4x+4)\pi^{\frac{n}{2}}
[\frac{1}{\delta}- \log(xm^2+(1-x)\lambda^2-p_E^2x(1-x))]\\
 &\\
& +\frac{ie^2}{(2\pi)^n}\int_0^1 dx
p_E^2(x^2-4x+4)\pi^{\frac{n}{2}} [\frac{2p_E
x(1-x)}{xm^2+(1-x)\lambda^2-p_E^2x(1-x)}]\\
&\\
 &+2p_E\cdot
\frac{1}{\delta}\pi^{\frac{n}{2}}\frac{n}{2}\frac{ie^2}{(2\pi)^n}\int_0^1
x(1-x)dx
\\
\end{array}
 \label{p23b}
\end{equation}

Then by Taylor expansion of (\ref{p23b}) and by removing a factor
$2p_E$ from (\ref{p23b}) the constant term for the electron
self-energy is given by:
\begin{equation}
B(\delta):=\frac{-e^2}{(2\pi)^n}\int_0^1 dx
(x^2-4x+4)\pi^{\frac{n}{2}} [\frac{1}{\delta}-
\log(xm^2+(1-x)\lambda^2)]
-\frac{1}{\delta}\pi^{\frac{n}{2}}\frac{n}{2}\frac{e^2}{(2\pi)^n}\int_0^1
x(1-x)dx
 \label{p23c}
\end{equation}

Then as a renormalization procedure for the electron self-energy
we choose a $\delta_1>0$ which is related to  the $\delta$ for the
renormalization of the vertex correction such that:
\begin{equation}
B(\delta_1)=B(\delta)+ a_1\cdot\frac{1}{\delta} +b_1 \label{pw2}
\end{equation}
This is possible since $B(\delta)$ has a term proportional to
$\frac{1}{\delta}$. From this renormalization procedure for the
electron self-energy we have:
\begin{equation}
B(\delta_1)=C(\delta) \label{pw3}
\end{equation}
This constant term $B(\delta_1)$  for the electron self-energy is
to be cancelled by the counter-term with the factor $z_Z-1$ in
(\ref{2.1}). We have the following equation to determine the
renormalization constant $z_Z$ for this cancellation:
\begin{equation}
(-1)^3i(z_Z-1)=(-i)B(\delta_1) \label{pw3a}
\end{equation}

Then from the equality (\ref{pw3}) we have $z_e=z_Z$ where $z_e$
is determined by the following equation:
\begin{equation}
(-1)^3i(z_e-1)=(-i)C(\delta) \label{pw3b}
\end{equation}

By cancelling the constant term $B(\delta_1)$ from the electron
self-energy (\ref{p21}) we then get the following renormalized
one-loop electron self-energy:
\begin{equation}
\begin{array}{rl}
& -ip_E^2 \Sigma_R(p_E) + i\omega_3^2 :=
 -ip_E^2\cdot \frac{\alpha}{4\pi}\int_0^1 dx (x^2-4x+4)
\log[1-\frac{p_E^2x(1-x)}{xm^2+(1-x)\lambda^2}]
+ i\omega_3^2\\
\end{array}
\label{p21a}
\end{equation}

 We notice that in (\ref{p21a}) we can let
$\lambda=0$ since there is no infrared divergence when
$\lambda=0$. This is better than the computed electron self-energy
in the conventional QED theory where the computed one-loop
electron self-energy is with infrared divergence when $\lambda=0$
\cite{Zub}.

From this renormalized electron self-energy we then have the
renormalized electron propagator obtained by the following Dyson
series:
\begin{equation}
\begin{array}{rl}
&\frac{i}{p_E^2-\omega^2}+\frac{i}{p_E^2-\omega^2}
(-ip_E^2\Sigma_R(p_E)+ i\omega_3^2)\frac{i}{p_E^2-\omega^2}+...\\
&\\
=&\frac{i}{p_E^2(1-\Sigma_R(p_E))-(\omega^2-\omega_3^2)}
=: \frac{i}{p_E^2(1-\Sigma_R(p_E))-\omega_R^2}
\end{array}
\label{p25}
\end{equation}
where $\omega_R^2:=\omega^2-\omega_3^2 $ is as a renormalized
electron mass-energy parameter. Then by space-time statistics from
the renormalized electron propagator (\ref{p25}) we can get the
renormalized electron propagator in the spin-$\frac12$ form, as
that the electron propagator $\frac{i}{\gamma_{\mu}p^{\mu}-m}$ in
the spin-$\frac12$ form can be obtained from the electron
propagator $\frac{i}{p_E^2-\omega^2}$.

\section{New effect of QED}\label{QED8}

Let us consider a new effect for electron scattering which is
formed by two seagull vertexes with one photon loop and four
electron lines. This is a new effect of QED because the
conventional spin $\frac12$ theory of QED does not have this
seagull vertex.
  The Feynman integral corresponding to the photon loop is given by
\begin{equation}
\begin{array}{rl}
& \frac{i^2(i)^2e^4}{2\pi}\int
\frac{dk_E}{(k_E^2-\lambda_0^2)((p_E-q_E-k_E)^2-\lambda_0^2)} \\
&\\
 =&\frac{e^4}{2\pi}\int_0^1 \int
\frac{dk_E}{(k_E^2-2k_E(p_E-q_E)x+(p_E-q_E)^2x -\lambda_0^2)^2} \\
&\\
 =&\frac{e^4}{2\pi}\int_0^1 \int
\frac{dk_E}{(k_E^2-2k_E(p_E-q_E)x+(p_E-q_E)^2x -\lambda_0^2)^2} \\
\end{array}
\label{8.1}
\end{equation}

Let us then introduce a space-time statistics. Since the photon
propagator of the (two joined) seagull vertex interactions is of
the form of a circle on a plane we have that the appropriate
space-time statistics of the photons is with the two dimensional
space for the circle of the photon propagator. From this two
dimensional space statistics we then get a three dimensional space
statistics by multiplying the statistical factor
$\frac{1}{(2\pi)^3}$ of the three dimensional space statistics and
by concentrating in a two dimensional subspace of the three
dimensional space statistics.

Thus as similar to the four dimensional space-time statistics with
the three dimensional space statistics in the above sections from
(\ref{8.1}) we have the following space-time statistics with the
two dimensional subspace:
\begin{equation}
\begin{array}{rl}
&\frac{e^4}{(2\pi)^4}\int_0^1 \int
\frac{d^3k}{(k_E^2-2k_E(p_E-q_E)x+(p_E-q_E)^2x -{\bf k}^2-\lambda_4^2)^2} \\
&\\
 =&\frac{e^4}{(2\pi)^4}\int_0^1 dx\int
\frac{d^3k}{(k^2-2k\cdot(p_E-q_E,{\bf 0})x+(p_E-q_E)^2x -\lambda_4^2)^2} \\
\end{array}
\label{8.2}
\end{equation}
where the statistical factor $\frac{1}{(2\pi)^3}$ of the three
dimensional space has been introduced to give the factor
$\frac{1}{(2\pi)^4}$ of the four dimensional space-time
statistics; and we let $k=(k_E, {\bf k})$, $k^2=k_E^2-{\bf k}^2$
and since the photon energy parameter $\lambda_0$ is a free
parameter we can write $\lambda_0^2={\bf k}^2+\lambda_4^2$ for
some $\lambda_4$.

Then  a delta function concentrating at $0$ of a one dimensional
momentum variable is multiplied to the integrand in (\ref{8.2})
and the three dimensional energy-momentum integral in (\ref{8.2})
is changed to a four dimensional energy-momentum integral by
taking the corresponding one more momentum integral.

From this we then get a four dimensional space-time statistics
with the usual four dimensional momentum integral and with the
statistical factor $\frac{1}{(2\pi)^4}$. After this additional
momentum integral we then get (\ref{8.2}) as a four dimensional
space-time statistics with the two dimensional momentum variable.

Then to get a four dimensional space-time statistics with the
three dimensional momentum variable a delta function concentrating
at $0$ of another one dimensional momentum variable is multiplied
to (\ref{8.2}) and the two dimensional momentum variable of
(\ref{8.2}) is extended to the corresponding three dimensional
momentum variable. From this we then get a four dimensional
space-time statistics with the three dimensional momentum
variable.

Then we have that (\ref{8.2}) is equal to:
\begin{equation}
\begin{array}{rl}
 \frac{e^4}{(2\pi)^4}\frac{i\pi^{\frac32}\Gamma(2-\frac32)}{\Gamma(2)}\int_0^1
\frac{dx}{((p_E-q_E)^2x(1-x) -\lambda_4^2)^{\frac12} }\\
\end{array}
\label{8.3}
\end{equation}

Then since the photon mass-energy parameter $\lambda_4$ is a free
parameter for space-time statistics we can write $\lambda_4$ in
the following form:
\begin{equation}
\lambda_4^2={\bf (p-q)}^2 x(1-x)
 \label{b6}
\end{equation}
where ${\bf p-q}$ denotes a two dimensional momentum vector.

Then we let $p-q=(p_E-q_E, {\bf p-q})$. Then we have:
\begin{equation}
(p_E-q_E)^2x(1-x)-\lambda_4^2=(p_E-q_E)^2x(1-x)- {\bf
(p-q)}^2x(1-x)=(p-q)^2x(1-x)
 \label{b7}
\end{equation}
 Then we have that (\ref{8.3}) is
equal to:
\begin{equation}
\begin{array}{rl}
&\frac{e^4}{(2\pi)^4}
\frac{i\pi^{\frac32}\Gamma(2-\frac32)}{\Gamma(2)}\int_0^1
\frac{dx}{((p-q)^2x(1-x))^{\frac12}} \\
&\\
=&\frac{e^4}{(2\pi)^4}\frac{i\pi\pi^{\frac32}\Gamma(2-\frac32)}{\Gamma(2)}
\frac{1}{((p-q)^2)^{\frac12}} \\
&\\
=& \frac{e^4i}{16\pi((p-q)^2)^{\frac12}} \\
&\\
=& \frac{e^2\alpha i}{4((p-q)^2)^{\frac12}} \\
\end{array}
\label{8.4a}
\end{equation}

Thus we have the following potential:
\begin{equation}
V_{seagull}(p-q)
 =\frac{e^2\alpha
i}{4((p-q)^2)^{\frac12}} \label{8.4}
\end{equation}
This potential (\ref{8.4}) is as the seagull vertex potential.

We notice that (\ref{8.4}) is a new effect for electron-electron
scattering or electron-positron scattering. Recent experiments on
the decay of positronium show that the experimental
orthopositronium decay rate  is significantly larger than that
computed from the conventional QED theory \cite{Wes}-\cite{Kni2}.
In the following section 24 to section 26 we show that this
discrepancy can be remedied with this new effect (\ref{8.4}).

\section{Reformulating the Bethe-Salpeter equation }\label{bs2}

To compute the orthopositronium decay rate let us first find out
the ground state wave function of the positronium. To this end we
shall use the Bethe-Salpeter equation. It is well known that the
conventional Bethe-Salpeter equation is with difficulties such as
the relative time and relative energy problem which leads to the
existence of nonphysical solutions in the conventional
Bethe-Salpeter equation \cite{Bet}-\cite{Nie}. From the above QED
theory let us reformulate the Bethe-Salpeter equation to get a new
form of the Bethe-Salpeter equation.  We shall see that this new
form of the Bethe-Salpeter equation resolves the basic
difficulties of the Bethe-Salpeter equation such as the relative
time and relative energy problem.

Let us first consider the propagator of electron. Since electron
is a spin-$\frac12$ particle its statistical propagator is of the
form $\frac{i}{\gamma_{\mu}p^{\mu} -m}$. Thus before the
space-time statistics the spin-$\frac12$ form of electron
propagator is of the form $\frac{i}{p_E-\omega}$ which can be
obtained from the electron propagator $\frac{i}{p_E^2-\omega^2}$
by the factorization: $p_E^2-\omega^2=(p_E-\omega)(p_E+\omega)$.
Then we consider the following product which is from two
propagators of two spin-$\frac12$ particles:
\begin{equation}
\begin{array}{rl}
& [p_{E1}-\omega_1][p_{E2}-\omega_2] \\
&\\
=& p_{E1}p_{E2}-\omega_1p_{E2}-\omega_2 p_{E1} +\omega_1\omega_2\\
&\\
 =:& p_E^2-\omega_b^2
\end{array}
 \label{bb01}
\end{equation}
where we define $p_E^2=p_{E1}p_{E2}$ and $\omega_b^2
:=\omega_1p_{E2}+\omega_2 p_{E1} -\omega_1\omega_2$. Then since
$\omega_1$ and $\omega_2$ are free mass-energy parameters we have
that $\omega_b$ is also a free mass-energy parameter with the
requirement that it is to be a positive parameter.

Then let us introduce the following reformulated relativistic
equation of Bethe-Salpeter type for two particles with
spin-$\frac12$:
\begin{equation}
\phi_0 (p_E, \omega_b)
=\frac{i^2}{[p_{E1}-\omega_1][p_{E2}-\omega_2]}
\int\frac{\lambda^{\prime}e^2 i}{((p_E-q_E)^2-\lambda_0^2)}
\phi_0(q_E, \omega_b)dq_E
 \label{bb1}
\end{equation}
where we use the photon propagator $\frac{i}{k_E^2-\lambda_0^2}$
(which is of the effect of Coulomb potential) for the interaction
of these two particles and we write the proper energy $k_E^2$ of
this potential in the form $k_E^2=(p_E-q_E)^2$; and
$\lambda^{\prime}$ is as the coupling parameter. We shall later
also introduce the seagull vertex term for the potential of
binding.

Let us then introduce the space-time statistics. Since we have the
seagull vertex term for the potential of binding which is of the
form of a circle in a two dimensional space from the above section
on the seagull vertex potential we see that the appropriate
space-time statistics is with the two dimensional space. Thus with
this space-time statistics from (\ref{bb1}) we have the following
reformulated relativistic Bethe-Salpeter equation:
\begin{equation}
\phi_0 (p) =\frac{-\lambda^{\prime}}{ p^2-\gamma_0^2}\int\frac{i
d^3q}{(p-q)^2} \phi_0(q)
 \label{bb2}
\end{equation}
where we let the free parameters $\omega_b$ and $\lambda_0$ be
such that $p^2=p_E^2-{\bf p}^2$ with $\omega_b^2={\bf
p}^2+\gamma_0^2$ for some constant $\gamma_0^2=\frac{1}{a^2}>0$
where $a$ is as the radius of the binding system; and
$(p-q)^2=(p_E-q_E)^2-{\bf (p-q)}^2$ with $\lambda_0^2={\bf
(p-q)}^2$. We notice that the potential $\frac{i\alpha}{(p-q)^2}$
of binding is now of the usual (relativistic) Coulomb potential
type. In (\ref{bb2}) the constant $e^2$ in (\ref{bb1}) has been
absorbed into the parameter $\lambda^{\prime}$ in (\ref{bb2}).

We see that in this reformulated relativistic  Bethe-Salpeter
equation the relative time and relative energy problem of the
conventional Bethe-Salpeter equations is resolved
\cite{Bet}-\cite{Nie}. Thus there will be no abnormal solutions of
the conventional Bethe-Salpeter equations for this reformulated
Bethe-Salpeter equation.

Let us then solve (\ref{bb2}) for the relativistic bound states of
particles.

We show that the ground state solution $\phi_0 (p)$ can be exactly
solved and is of the following form:
\begin{equation}
\phi_0 (p) = \frac{1}{ (p^2-\gamma_0^2)^2}
 \label{bb3}
\end{equation}

We have
\begin{equation}
\begin{array}{rl}
& \frac{1}{((p-q)^2)}\frac{1}{ (q^2-\gamma_0^2)^2} \\
&\\
 = & \frac{(2+1-1)!}{(2-1)!(1-1)!}\int_0^1 \frac{(1-x)dx}{[x(p-q)^2+(1-x) (q^2-\gamma_0^2)^2]^3}\\
 &\\
 =&\frac{(2+1-1)!}{(2-1)!(1-1)!}\int_0^1 \frac{(1-x)dx}{[q^2 + 2xpq
 +xp^2-(1-x)\gamma_0^2]^3}\\
 &\\
 =&2\int_0^1 \frac{(1-x)dx}{[q^2 + 2xpq +xp^2-(1-x)\gamma_0^2]^3}
 \end{array}
 \label{bb4}
\end{equation}

Thus we have
\begin{equation}
\begin{array}{rl}
& i\int \frac{d^3 q}{((p-q)^2)(q^2-\gamma_0^2)^2} \\
&\\
 = & i2\int_0^1 (1-x)dx \int \frac{d^3 q}{[q^2 + 2xpq
 +xp^2-(1-x)\gamma_0^2]^3}\\
 &\\
 = &i^2 \frac{2\pi^\frac32 \Gamma(3-\frac32)}{\Gamma(3)}\int_0^1
 \frac{(1-x)dx}{[+x(1-x)p^2-(1-x)\gamma_0^2]^{\frac{3}{2}}}\\
 &\\
 =&-\frac{2\pi^\frac32 \Gamma(3-\frac32)}{\Gamma(3)}\int_0^1
 \frac{dx}{[+xp^2-\gamma_0^2][(1-x)(xp^2-\gamma_0^2)]^{\frac{3}{2}}}\\
 &\\
 =& -\frac{2\pi^\frac32 \Gamma(3-\frac32)}{\Gamma(3)}
 \frac{\partial^2}{\partial(\gamma_0^2)^2}
 \int_0^1 dx
 [\frac{xp^2-\gamma_0^2}{1-x}]^{\frac{3}{2}}\\
 &\\
 =&-\frac{2\pi^\frac32 \Gamma(3-\frac32)}{\Gamma(3)}
 \frac{\partial^2}{\partial(\gamma_0^2)^2}
 \int_0^1 dx
 [\frac{p^2-\gamma_0^2}{1-x}-p^2]^{\frac{3}{2}}\\
 &\\
 =&-\frac{2\pi^\frac32 \Gamma(3-\frac32)}{\Gamma(3)}
 \frac{\partial^2}{\partial(\gamma_0^2)^2}
 \int_1^{\infty} \frac{dt}{t^2}
 [(p^2-\gamma_0^2)t-p^2]^{\frac{3}{2}}\\
 &\\
=& -\frac{2\pi^\frac32 \Gamma(3-\frac32)}{\Gamma(3)}
\int_1^{\infty} \frac{dt}{t^2}
 [(p^2-\gamma_0^2)t-p^2]^{\frac{-3}{2}}\\
 &\\
=& -\frac{2\pi^\frac32 \Gamma(3-\frac32)}{\Gamma(3)}
\frac{1}{(p^2-\gamma_0^2)}
\int_{\gamma_0^2}^{\infty}x^{\frac{-3}{2}}dx
 \\
 &\\
 =& -\frac{\pi^2}{2}
 \frac{1}{\gamma_0(p^2-\gamma_0^2)}
 \end{array}
 \label{bb5}
\end{equation}

Then let us choose $\lambda^{\prime}$ such that
$\lambda^{\prime}=\frac{2\gamma_0}{\pi^2}$. From this value of
$\lambda^{\prime}$ we see that the BS equation (\ref{bb2}) holds.
Thus the ground state solution is of the form (\ref{bb3}). We see
that when $p_E=0$ and $\omega_b^2={\bf p}^2+\gamma_0^2$ then this
ground state gives the well known nonrelativistic ground state of
the form $\frac{1}{({\bf p}^2+\gamma_0^2)^2}$ of  binding system
such as the hydrogen atom.

\section{Bethe-Salpeter equation with seagull vertex potential}\label{b2}

Let us then introduce the following reformulated relativistic
Bethe-Salpeter equation which is also with the seagull vertex
potential of binding:
\begin{equation}
\phi (p) = \frac{-\lambda^{\prime}}{ p^2-\gamma_0^2}
\int[\frac{i}{(p-q)^2}+
\frac{i\alpha}{4((p-q)^2)^{\frac12}}]\phi(q)d^3q
 \label{bb02}
\end{equation}
where a factor $e^2$ of both the Coulomb-type potential and the
seagull vertex potential is absorbed to the coupling constant
$\lambda^{\prime}$.

Let us then solve (\ref{bb02}) for the relativistic bound states
of particles.

Let us write the ground state solution in the following form:
\begin{equation}
\phi (p) = \phi_0 (p)+\alpha \phi_1 (p)
 \label{bb03}
\end{equation}
where $\phi_0(p) $ denotes the ground state of the BS equation
when the interaction potential only consists of the Coulomb-type
potential. Let us then determine the $\phi_1 (p)$.

From (\ref{bb02}) by comparing the coefficients of the $\alpha^j,
j=0,1$ on both sides of BS equation we have the following equation
for $\phi_1 (p)$:
\begin{equation}
\begin{array}{rl}
\phi_1(p) =& \frac{-\lambda^{\prime}}{ p^2-\gamma_0^2}
\int[\frac{i}{4((p-q)^2)^{\frac12}}]\phi_0(q)d^3q+
 \frac{-\lambda^{\prime}}{
p^2-\gamma_0^2}\int[\frac{i}{((p-q)^2)}+
\frac{i\alpha}{4((p-q)^2)^{\frac12}}]\phi_1(q)d^3q
\end{array}
 \label{b04}
\end{equation}

This is a nonhomogeneous linear Fredholm integral equation. We can
find its solution by perturbation. As a first order approximation
we have the following approximation of $\phi_1 (p)$:
\begin{equation}
\begin{array}{rl}
\phi_1(p) \approx &\frac{-\lambda^{\prime}}{ p^2-\gamma_0^2}\int
\frac{i}{ 4((p-q)^2)^{\frac12}}\phi_0(q)d^3q 
= \frac{-\lambda^{\prime}}{ p^2-\gamma_0^2}\int
\frac{i}{4((p-q)^2)^{\frac12}} \frac{1}{(q^2-\gamma_0^2)^2}
d^3q \\
&\\
=&\frac{-\lambda^{\prime}}{ p^2-\gamma_0^2}
\frac{i\Gamma(1+\frac12+2-1)}{4\Gamma(1+\frac12-1)\Gamma(1+2-1)}
\int_0^1 y^{\frac12}(1-y)dy \int
\frac{d^3q}{[q^2-2qpy+p^2y -(1-y)\gamma_0^2]^{2+\frac12}} \\
&\\
 =&\frac{-\lambda^{\prime}}{ p^2-\gamma_0^2}
\frac{i\Gamma(\frac12+2)}{4\Gamma(\frac12)\Gamma(2)} \int_0^1
y^{\frac12}(1-y)dy
\frac{i\pi^{\frac32}\Gamma(\frac52-\frac32)}{\Gamma(\frac52)(p^{2}y(1-y) -(1-y)\gamma_0^2)} \\
&\\
 =&\frac{\lambda^{\prime}}{ p^2-\gamma_0^2}
\frac{\pi^{\frac32}}{4\Gamma(\frac12)} \int_0^1 y^{\frac12}dy
\frac{1}{(p^2y -\gamma_0^2)}
=\frac{\lambda^{\prime}\pi}{ p^2-\gamma_0^2}
\frac{1}{4|p|\gamma_0}\log|\frac{|p|-\gamma_0}{|p|+\gamma_0}| \\
&\\
 =&\frac{\pi}{
p^2-\gamma_0^2}\frac{2\gamma_0}{\pi^2}
\frac{1}{4|p|\gamma_0}\log|\frac{|p|-\gamma_0}{|p|+\gamma_0}| 
=\frac{1}{2\pi(p^2-\gamma_0^2)|p|}\log|\frac{|p|-\gamma_0}{|p|+\gamma_0}| \\
\end{array}
 \label{b05}
\end{equation}
where $|p|=\sqrt{p^2}$.

Thus we have the ground state of the form $\phi(p)
=\phi_0(p)+\alpha\phi_1(p)$ where $p$ denotes an energy-momentum
vector with a two dimensional momentum. Thus this ground state is
for a two dimensional (momentum) subspace. We may extend it to the
ground state of the form $\phi(p) =\phi_0(p)+\alpha\bar\phi_1(p)$
where $p$ denotes a four dimensional energy-momentum vector with a
three dimensional momentum; and due to the special nature that
$\phi_1(p)$ is obtained by a two dimensional space statistics the
extension $\bar\phi_1(p)$ of $\phi_1(p)$ to with a three
dimensional momentum  is a wave function obtained by multiplying
$\phi_1(p)$ with a delta function concentrating at $0$ of a one
dimensional momentum variable and the variable $p$ of $\phi_1(p)$
is extended to be a four dimensional energy-momentum vector with a
three dimensional momentum.

Let us use this form of the ground state $\phi(p)
=\phi_0(p)+\alpha\bar\phi_1(p)$ to compute new QED effects in the
orthopositronium decay rate where there is a discrepancy between
theoretical result and the experimental result
\cite{Wes}-\cite{Kni2}.

\section{New QED effect of orthopositronium decay rate }\label{b02}

From the seagull vertex let us find new QED effect to the
orthopositronium decay rate where there is a discrepancy between
theory and experimental result \cite{Wes}-\cite{Kni2}. Let us
compute the new one-loop effect of orthopositronium decay rate
which is from the seagull vertex potential.

From the seagull vertex potential we have that the ground state of
positronium is modified from $\phi(p)=\phi_0(p)$ to $\phi(p)
=\phi_0(p)+\alpha\bar\phi_1(p)$. Let us apply this form of the
ground state of positronium to the computation of the
orthopositronium decay rate.

Let us consider the nonrelativistic case. In this case we have
$\phi_0({\bf p})= \frac{1}{({\bf p}^2+\gamma_0^2)^2}$ and:
\begin{equation}
\phi_1({\bf p})= \frac{-1}{2\pi({\bf p}^2+\gamma_0^2)|{\bf
p}|}\log\left|\frac{|{\bf p}|-\gamma_0}{|{\bf p}|+\gamma_0}\right|.
\label{adecay}
\end{equation}

Let $M$ denotes the decay amplitude. Let $M_0$ denotes the zero-loop
decay amplitude. Then following the approach in the computation of
the positronium decay rate \cite{Wes}-\cite{Kni2} the first order
decay rate $\Gamma$ is given by:
\begin{equation}
\Gamma =\int 8\pi^{\frac12}\gamma_0^{\frac52}[\phi_0({\bf
p})+\alpha\bar\phi_1({\bf p})]M_0({\bf p})d^3{\bf
p}=:\Gamma_0+\alpha\Gamma_{seagull} \label{decay}
\end{equation}
where $8\pi^{\frac12}\gamma_0^{\frac52}$ is the normalized
constant for the usual unnormalized ground state wave function
$\phi_0$ \cite{Wes}-\cite{Kni2}.

 We have that the first order decay rate $\Gamma_0$ is given
by \cite{Wes}-\cite{Kni2}:
\begin{equation}
\begin{array}{rl}
\Gamma_0& :=\frac{1}{(2\pi)^3}
\int\ 8\pi^{\frac12}\gamma_0^{\frac52}\phi_0({\bf p}) M_0({\bf p})d^3{\bf p}
=\frac{1}{(2\pi)^3}
\int \frac{8\pi^{\frac12}\gamma_0^{\frac52}}{({\bf p}^2+\gamma_0^2)^2} M_0({\bf p})d^3{\bf p} \\
&\\
 &\approx  \psi_0({\bf r}=0)M_0(0) 
=\frac{8\pi^{\frac12}\gamma_0^{\frac52}}{(2\pi)^3} \int
\frac{d^3{\bf p}}{({\bf p}^2+\gamma_0^2)^2} M_0(0) \\
&\\
&=\frac{8\pi^{\frac12}\gamma_0^{\frac52}}{(2\pi)^3} \frac{\pi^2}{\gamma_0}  M_0(0) 
= \frac{1}{(\pi a^3)^{\frac12}}  M_0(0) \\
 \label{b06}
 \end{array}
\end{equation}
where $\psi_0 ({\bf r})$ denotes the usual nonrelativistic ground
state wave function of  positronium;  and $a=\frac{1}{\gamma_0}$
is as the radius of the positronium. In the above equation the
step $\approx$ holds since $\phi_0({\bf p})\to 0$ rapidly as ${\bf
p}\to \infty$ such that the effect of $M_0({\bf p})$ is small for
${\bf p}\neq 0$; as explained in \cite{Wes}-\cite{Kni2}.

Then let us consider the new QED effect of decay rate from
$\bar\phi_1({\bf p})$. As the three dimensional space statistics
in the section on the seagull vertex potential we have the
following three dimensional space statistics of the decay rate
from $\bar\phi_1({\bf p})$:
\begin{equation}
\begin{array}{rl}
\Gamma_{seagull}& =\frac{1}{(2\pi)^3 }\int
8\pi^{\frac12}\gamma_0^{\frac52}\bar\phi_1({\bf p}) M_0({\bf
p})d^3{\bf
p} \\
&\\
& =\frac{1}{(2\pi)^3}\int
8\pi^{\frac12}\gamma_0^{\frac52}\phi_1({\bf p}) M_0({\bf
p})d^2{\bf
p} \\
&\\
&\approx \frac{8\pi^{\frac12}\gamma_0^{\frac52}}{(2\pi)^3}\int
\phi_1({\bf p})M_0(0) d^2{\bf
p}\\
&\\
&=\frac{-8\pi^{\frac12}\gamma_0^{\frac52}}{(2\pi)^3}\int
\frac{1}{2\pi({\bf p}^2+\gamma_0^2)|{\bf p}|}\log|\frac{|{\bf
p}|-\gamma_0}{|{\bf p}|+\gamma_0}|d^2{\bf p}
M_0(0)\\
&\\
 &= \frac{8\pi^{\frac12}\gamma_0^{\frac52}}{(2\pi)^3 2\pi}\frac{\pi^3}{2\gamma_0}
  M_0(0)
=\frac{1}{4(\pi a^3)^{\frac12}}  M_0(0) \\
 \label{b08}
 \end{array}
\end{equation}
where  the step $\approx$ holds as similar the equation
(\ref{b06}) since in the two dimensional integral of $\phi_1({\bf
p})$ we have that $\phi_1({\bf p})\to 0$ as ${\bf p}\to \infty$
such that it tends to zero as rapidly as the three dimensional
case of $\phi_0({\bf p})\to 0$.

Thus we have:
\begin{equation}
\alpha \Gamma_{seagull}=\frac{\alpha}{4} \Gamma_0
 \label{decay2}
\end{equation}

Then from the literature in the computation of the
orthopositronium decay rate we have that the computed
orthopositronium decay rate (up to the order $\alpha^2$) is given
by \cite{Wes}-\cite{Kni2}:
\begin{equation}
\Gamma_{\mbox{o-Ps}}
=\Gamma_0[1+A\frac{\alpha}{\pi}+\frac{\alpha^2}{3}\log
\alpha+B(\frac{\alpha}{\pi})^2-\frac{\alpha^3}{2\pi}\log^2
\alpha]=7.039934(10) \, \mu s^{-1}
 \label{decay3}
\end{equation}
where $A=-10.286 606(10)$, $B=44.52(26)$ and $\Gamma_0=\frac92
(\pi^2-9)m\alpha^6=7.211 169\,\mu s^{-1} $.

Then with the additional decay rate from the seagull vertex
potential (or from the modified ground state $\phi$ of the
positronium) we have the following computed orthopositronium decay
rate (up to the order $\alpha^2$):
\begin{equation}
\begin{array}{rl}
\bar\Gamma_{\mbox{o-Ps}}:=&
\Gamma_{\mbox{o-Ps}}+\alpha\Gamma_{seagull} \\
&\\
 =& \Gamma_0[1+(A+\frac{\pi}{4})\frac{\alpha}{\pi}+\frac{\alpha^2}{3}\log
\alpha +B(\frac{\alpha}{\pi})^2-\frac{\alpha^3}{2\pi}\log^2 \alpha]\\
&\\
 = &7.039934(10) +0.01315874 \,\,\mu s^{-1}=7.052092(84) \,\mu s^{-1}
\end{array}
 \label{decay4}
\end{equation}

This agrees with the two Ann Arbor experimental values where the
two Ann Arbor experimental values are given by:
$\Gamma_{\mbox{o-Ps}}(\mbox{Gas})=7.0514(14) \, \mu s^{-1}$ and
$\Gamma_{\mbox{o-Ps}}(\mbox{Vacuum})=7.0482(16) \, \mu s^{-1}$
\cite{Wes}-\cite{Nic}.

We remark that for the decay rate $\alpha\Gamma_{seagull}$ we have
only computed it up to the order $\alpha$. If we consider the
decay rate $\alpha\Gamma_{seagull}$ up to the order $\alpha^2$
then the decay rate (\ref{decay4}) will be reduced since the order
$\alpha$ of $\Gamma_{seagull}$ is of negative value.

If we consider only the computed orthopositronium decay rate up to
the order $\alpha$ with the term $B(\frac{\alpha}{\pi})^2$
omitted, then we have the following computed orthopositronium
decay rate \cite{Wes}-\cite{Kni2}:
\begin{equation}
\bar\Gamma_{\mbox{o-Ps}}:=
\Gamma_{\mbox{o-Ps}}+\alpha\Gamma_{seagull} =7.038202 +0.01315874
\,\,\mu s^{-1}=7.05136074 \,\mu s^{-1}
 \label{decay4a}
\end{equation}
This also agrees with the above two Ann Arbor experimental values
and is closer to these two experimental values.

On the other hand the Tokyo experimental value given by
$\Gamma_{\mbox{o-Ps}}(\mbox{Powder})=7.0398(29) \, \mu s^{-1}$
\cite{Asa}  may be interpreted by that in this experiment the QED
effect $\Gamma_{seagull}$ of the seagull vertex potential is
suppressed due to the special two dimensional statistical form of
$\Gamma_{seagull}$ (Thus the additional effect of the modified
ground state $\phi$ of the positronium is suppressed). Thus the
value of this experiment agrees with the computational result
$\Gamma_{\mbox{o-Ps}}$. Similarly the experimental result of
another Ann Arbor experiment given by $7.0404(8)\mu s^{-1}$
\cite{Val} may also be interpreted by that in this experiment the
QED effect $\Gamma_{seagull}$ of the seagull vertex potential is
suppressed due to the special two dimensional statistical form of
$\Gamma_{seagull}$.

\section{Graviton constructed from photon}\label{sec15a}

It is well known that Einstein tried to find a theory to unify
gravitation and electromagnetism \cite{Pai}\cite{Ein2}\cite{Wey}.
The search for such a theory has been one of the major research
topics in physics \cite{Wey}-\cite{Wit5}. Another major research
topic in physics is the search for a theory of quantum gravity
\cite{Ash}-\cite{Wu1}. In fact, these two topics are closely
related. In this section, we propose a theory of quantum gravity
that unifies gravitation and electromagnetism.

In the above sections the photon is as
the quantum Wilson loop with the $U(1)$ gauge group for electrodynamics. In this section, we show
that the corresponding quantum Wilson line
can be regarded as the photon propagator in analogy to the usual concept
of propagator. From this photon propagator, the graviton propagator and the
graviton are constructed. This construction forms the foundation of a theory
of quantum gravity that unifies gravitation and electromagnetism.

It is well known that Weyl introduced the concept of gauge to
unify gravitation and  electromagnetism \cite{Wey}. However this
gauge concept of unifying gravitation and electromagnetism was
abandoned because of the criticism of the path dependence of the
gauge (It is well known that this gauge concept later is important
for quantum physics as phase invariance) \cite{Pai}. In this paper
we shall use again Weyl's gauge concept to develop a  theory of
quantum gravity which unifies gravitation and electromagnetism. We
shall show that the difficulty of path dependence of the gauge can
be solved in this quantum theory of unifying gravitation and
electromagnetism.

Let us consider a differential of the form $g(s)ds$ where $g$ is a
field variable to be determined. Let us consider a symmetry of the
following form:
\begin{equation}
g(s)ds=g^{\prime}(s^{\prime})ds^{\prime}
\label{graviton}
\end{equation}
where $s$ is transformed to $s^{\prime}$ and $g^{\prime}$ is a field variable such
that (\ref{graviton}) holds.
From
(\ref{graviton}) we have a symmetry of the following form:
\begin{equation}
g(s)^{*}g(s)ds^2=g^{\prime *}(s^{\prime})g^{\prime }(s^{\prime})ds^{\prime 2}
\label{graviton2}
\end{equation}
where $g^{*}$ and $g^{\prime *}$ denote the complex conjugate of $g$ and $g^{\prime}$ respectively.
This symmetry can be considered as the symmetry for deriving the gravity since we
can write $g(s)^{*}g(s)ds^2$ into the following metric form for the four dimensional space-time
in general relativity:
\begin{equation}
g(s)^{*}g(s)ds^2=g_{\mu\nu}dx^{\mu}dx^{\nu}
\label{graviton3}
\end{equation}
where we write $ds^2=a_{\mu\nu}dx^{\mu}dx^{\nu}$ for some functions $a_{\mu\nu}$ by introducing
the space-time variable $x^{\mu}, \mu=0,1,2,3$ with $x^{0}$ as the time variable;
and $g_{\mu\nu}=g(s)^{*}g(s)a_{\mu\nu}$.  Thus from the symmetry (\ref{graviton})
we can derive general relativity.

Let us now determine the variable $g$. Let us consider
$g(s)=W(z_0,z(s))$, a quantum Wilson line with $U(1)$ group where
$z_0$ is fixed. When $W(z_0,z(s))$ is the classical Wilson line
then it is of path dependence and thus there is a difficulty to
use it to define $g(s)=W(z_0,z(s))$. This is also the difficulty
of Weyl's gauge theory of unifying gravitation and
electromagnetism. Then when  $W(z_0,z(s))$ is the quantum Wilson
line because of the quantum nature of unspecification of paths we
have that $g(s)=W(z_0,z(s))$ is well defined where the whole path
of connecting $z_0$ and $z(s)$ is unspecified (except the two end
points $z_0$ and $z(s)$).

Thus for a given transformation $s^{\prime}\to s$ and for any
(continuous and piecewise smooth) path connecting the two end
points $z_0$ and $z(s)$ the resulting quantum Wilson line
$W^{\prime}(z_0,z(s(s^{\prime})))$ under this transformation is
again of the form  $W(z_0,z(s))=W(z_0,z(s(s^{\prime})))$.
 Let $g^{\prime}(s^{\prime})=W^{\prime}(z_0,z(s(s^{\prime})))\frac{ds}{ds^{\prime}}$.
Then we have:
\begin{equation}
\begin{array}{rl}
 & g^{\prime *}(s^{\prime})g^{\prime }(s^{\prime})ds^{\prime 2}\\
 &\\
=&
W^{\prime *}(z_0,z(s(s^{\prime})))W^{\prime }(z_0,z(s(s^{\prime})))
(\frac{ds}{ds^{\prime}})^2 ds^{\prime 2}\\
&\\
 =&
W^{*}(z_0,z(s))W(z_0,z(s))(\frac{ds}{ds^{\prime}})^2 ds^{\prime 2}\\
&\\
 =& g(s)^{*}g(s)ds^2
\end{array}
\label{graviton4}
\end{equation}
This shows that the quantum Wilson line $W(z_0,z(s))$ can be the field variable for the  gravity
and thus can be the field variable for quantum gravity since $W(z_0,z(s))$ is
a quantum field variable.

Then we investigate the operator $W(z_0,z)W(z_0,z)$. From this
operator $W(z_0,z)W(z_0,z)$ we can compute the operator
$W^{*}(z_0,z)W(z_0,z)$ which is as the absolute value of this
operator $W(z_0,z)W(z_0,z)$. Thus this operator $W(z_0,z)W(z_0,z)$
can  be regarded as the graviton propagator while the quantum
Wilson line $W(z_0,z)$ is regarded as the quantum photon
propagator for the photon field propagating from $z_0$ to $z$.

Let us then compute the graviton propagator $W(z_0,z)W(z_0,z)$.
We have the following formula:
\begin{equation}
W(z,z_0)W(z_0,z)=e^{-\hat{t}\log [\pm (z-z_0)]}Ae^{\hat{t}\log [\pm (z_0-z)]}
\label{graviton5a}
\end{equation}
where $t=-\frac{e_0^2}{k_0}$ for the $U(1)$ group ($k_0>0$ is a
constant and we may for simplicity let $k_0=1$) where the term
$e^{-\hat{t}\log [\pm (z-z_0)]}$ is obtained by solving the first form
of the dual form of the KZ equation and the term $e^{\hat{t}\log [\pm
(z_0-z)]}$ is obtained by solving the second form of the dual form
of the KZ equation.

Then we change the  factor $W(z,z_0)$ of $W(z,z_0)W(z_0,z)$
in (\ref{graviton5a}) to the second factor $W(z_0,z)$ of
$W(z,z_0)W(z_0,z)$ by reversing the proper time direction of the
path of connecting $z$ and $z_0$ for $W(z,z_0)$. This gives the
graviton propagator $W(z_0,z)W(z_0,z)$. Then the reversing of the
proper time direction of the path of connecting $z$ and $z_0$ for
$W(z,z_0)$ also gives the reversing of the
 first form of the dual
form of the KZ equation to the
 second form of the dual form of the
KZ equation. Thus by solving the
 second form of dual form of the
KZ equation we have that $W(z_0,z)W(z_0,z)$ is given by:
\begin{equation}
W(z_0,z)W(z_0,z)=e^{\hat{t}\log [\pm (z-z_0)]}Ae^{\hat{t}\log [\pm (z-z_0)]}
=e^{2\hat{t}\log [\pm (z-z_0)]}A \label{graviton5}
\end{equation}
where we take the same sign of $\pm$ for the two factors $ e^{\hat{t}\log [\pm (z-z_0)]}$.  For this case the gravitational wave which is from the imaginary part of  $\log [\pm (z-z_0)]$  is remained.

In (\ref{graviton5}) let us define the following constant $G$:
\begin{equation}
G:=-2\hat{t}=2\frac{e_0^2}{k_0} \label{graviton5g}
\end{equation}
 We regard
this constant $G$ as the gravitational constant of the law of
Newton's  gravitation and  general relativity. We notice that from
the relation $e_0=\frac{1}{z_A^{\frac12}}e=\frac{1}{n_e} e$ where
the renormalization number $n_e=z_A^{\frac12}$ is a very large
number we have that the bare electric charge $e_0$ is a very small
number. Thus the gravitational constant $G$ given by
(\ref{graviton5g}) agrees with the fact that the gravitational
constant is a very small constant. This then gives a closed
relationship between electromagnetism and gravitation.

We remark that since in (\ref{graviton5}) the factor $-G\log
r_1=G\log\frac{1}{r_1}<0$ (where we define $r_1=|z-z_0|$  and
$r_1$ is restricted such that $r_1> 1$)  is the fundamental
solution of the two dimensional Laplace equation we have that this
factor (together with the factor $e^{-G\log
r_1}=e^{G\log\frac{1}{r_1}}$) is analogous to the fundamental
solution $-G\frac{1}{r}$ of the three dimensional Laplace equation
for the law of Newton's gravitation. Thus the operator
$W(z_0,z)W(z_0,z)$ in (\ref{graviton5}) can be regarded as the
graviton propagator which gives attractive effect when $r_1> 1$.
Thus the graviton propagator (\ref{graviton5}) gives the same
attractive effect of $-G\frac{1}{r}$ for the law of Newton's
gravitation.

On the other hand
when $r_1\leq 1$ we have that the factor $-G\log r_1=G\log\frac{1}{r_1}\geq 0$. In
this case we may consider that this graviton propagator gives repulsive effect.
This means that when two particles are very close to each other then
the gravitational force can be from attractive to become repulsive.
This repulsive effect is a modification of $-G\frac{1}{r}$ for
the law of Newton's gravitation for which the attractive force between two particles tends
to $\infty$ when the distance between the two particles tends to $0$.

Then by multiplying two masses $m_1$ and $m_2$ (obtained from the
winding numbers of Wilson loops in (\ref{closed2}) of two
particles to the graviton propagator (\ref{graviton5}) we have the
following formula:
\begin{equation}
Gm_1 m_2\log\frac{1}{r_1}
\label{graviton6c}
\end{equation}
From this formula (\ref{graviton6c}) by introducing the space variable $x$ as
a statistical variable via the Lorentz metric:
$ds^2=dt^2-dx^2$  we have the following statistical formula which is
the potential law of Newton's gravitation:
\begin{equation}
-GM_1 M_2\frac{1}{r}
\label{graviton6d}
\end{equation}
where $M_1$ and $M_2$ denotes the masses of two objects.

We remark that the graviton propagator (\ref{graviton5}) is for matters. We may by symmetry find
a propagator $f(z_0,z)$ of the following form:
\begin{equation}
f(z_0,z):=Ae^{-2\hat{t}\log [\pm (z-z_0)]} \label{graviton5c}
\end{equation}
When $|z-z_0|>1$ this propagator $f(z_0,z)$ gives repulsive effect between two particles
and thus is for anti-matter particles where by the term anti-matter we mean particles with
the repulsive effect (\ref{graviton5c}).  Then since $|f(z_0,z)| \to \infty$ as $|z-z_0|\to \infty$
we have that two such anti-matter particles can not physically exist.
However in the following section on dark energy we shall show
the possibility of another  repulsive effect among gravitons.

As similar to that the quantum Wilson loop $W(z_0,z_0)$ is as the
photon we have that the following double quantum Wilson loop can
be regarded as the graviton:
\begin{equation}
{\bf g} : =W(z_0,z)W(z_0,z)W(z,z_0)W(z,z_0)\,. \label{graviton8}
\end{equation}

\section{Graviton propagator and quantum graviton propagator}

In this section as similar to the photon case from the quantum graviton propagator we shall derive graviton propagators. We shall show that these graviton propagators can just be the parallel propagators for general relativity. From this identification we can then derive general relativity from the theory of quantum gravity in this paper.

Let us first consider a parallel propagator of general relativity:
\begin{equation}
P_{\nu}^{\mu}(\eta_0, \eta_1):=P\exp(-\int_{\eta_0}^{\eta_1}\Gamma_{\sigma\nu}^{\mu}\frac{dx^{\sigma}}{d\eta})
\label{graviton81}
\end{equation}
where $\Gamma_{\sigma\nu}^{\mu}$ denotes the Cristo symbol for a gravity fild.

Let $z_0=z(\eta_0)$ and $z=z(\eta_1)$. Let
\begin{equation}
\begin{array}{rl}
z-z_0= P_{\nu}^{\mu}(\eta_0, \eta_1)
\end{array}
\label{graviton82a}
\end{equation}
Then we have
\begin{equation}
\begin{array}{rl}
& W(z_0,z)W(z_0,z)\\
&\\
=& e^{2\hat{t}\log [\pm (z-z_0)]} A 
= [P_{\nu}^{\mu}(\eta_0, \eta_1)]^{2\hat{t}}A \\
&\\
=&P\exp(-\int_{\eta_0}^{\eta_1}2\hat{t}\Gamma_{\sigma\nu}^{\mu}\frac{dx^{\sigma}}{d\eta})A
=: P\exp(-\int_{\eta_0}^{\eta_1}\hat{\Gamma}_{\sigma\nu}^{\mu}\frac{dx^{\sigma}}{d\eta})A
\end{array}
\label{graviton82}
\end{equation}
Then we notice that $P\exp(-\int_{\eta_0}^{\eta_1}\hat{\Gamma}_{\sigma\nu}^{\mu}\frac{dx^{\sigma}}{d\eta})$
is a parellel propagator of general relativity where we define the Cristo symbol $\hat{\Gamma}_{\sigma\nu}^{\mu}:=2\hat{t}\Gamma_{\sigma\nu}^{\mu}$.
Thus a parellel propagator of general relativity is as a graviton propagator derived from the quantum graviton propagator. This shows that general relativity can be derived from the theory of quantum gravity in this paper.

\section{Dark energy, dark matter and the cosmological constant}\label{sec12}

We notice that a graviton  ${\bf g}$ in (\ref{graviton8})  consists of two loops.
By the method of  the
computation of the graviton propagator (\ref{graviton5}) we have
that (\ref{graviton8}) is given by:
\begin{equation}
\begin{array}{rl}
{\bf g}  = & W(z_0,z)W(z_0,z)W(z,z_0)W(z,z_0) \\
&\\
= & e^{\hat{t}\log[\pm(z-z_0)]}e^{\hat{t}\log[\pm(z-z_0)]}A_g e^{-\hat{t}\log[\pm (z-z_0)]}  e^{-\hat{t}\log[\pm (z-z_0)]} 
\end{array}
\label{graviton9a}
\end{equation}
where $A_g=AA$ denotes the initial operator for the graviton, and  we have:
\begin{equation}
\begin{array}{rl}
& W(z_0,z)W(z_0,z)  
=e^{\hat{t}\log[\pm(z-z_0)]}e^{\hat{t}\log[\pm(z-z_0)]} A 
\end{array}
\label{graviton111}
\end{equation}
and we have:
\begin{equation}
\begin{array}{rl}
& W(z,z_0)W(z,z_0) 
= A e^{-\hat{t}\log[\pm (z-z_0)]}  e^{-\hat{t}\log[\pm (z-z_0)]} 
\end{array}
\label{graviton112}
\end{equation}

When the two loops with the same winding direction and the same winding number we have:
\begin{equation}
\begin{array}{rl}
{\bf g}  = & W(z_0,z)W(z_0,z)W(z,z_0)W(z,z_0) \\
&\\
= & e^{2\hat{t}\log[\pm(z-z_0)]} A_g e^{-2\hat{t}\log[\pm (z-z_0)]}  \\
&\\
= &  e^{i2n\pi e_0^2}A_g,\, n= 0,\pm 1,\pm 2,\pm 3, ...
\end{array}
\label{graviton9}
\end{equation}

Thus as similar to
the quantization of energy of photons we have the following quantization of energy of gravitons:
\begin{equation}
h\nu =2\pi e_0^2 n,   \quad n= 0, \pm 1, \pm 2, \pm 3, ...
\label{graviton10}
\end{equation}

Since the quantum graviton propagator $ W(z_0,z)W(z_0,z)$ is  formed without the effect of forming closed loop which gives the renormalization of the bare charge $ e_0$ (This effect gives the renormalized charge $e=n_e e_0$), we have that the chrage of  the quantum graviton propagator is still the bare charge $ e_0$,  as shown in  (\ref{graviton10}). Thus the energies    (\ref{graviton10})  are very small when $n$ is not large since the bare charge $ e_0$ is very small. Thus the graviton energies  (\ref{graviton10}) are more difficult to be observed and thus can be  identified as the dark energy.

As similar to that a photon with a specific frequency can be as a
magnetic monopole because of its loop nature we have that the
graviton (\ref{graviton8}) with a specific frequency can also be
regarded as a magnetic monopole (which is similar to but different
from the magnetic monopole of the photon kind) because of its loop
nature. (This means that the loop nature gives magnetic property.)

Then let us consider the  case that the two loops of $ {\bf g}$  are with the counter winding  directions.
Then we have:
\begin{equation}
\begin{array}{rl}
& W(z_0,z)W(z_0,z)  
= e^{\hat{t}\log[(z-z_0)]}  e^{\hat{t}\log[- (z-z_0)]}  A 
\end{array}
\label{graviton11a}
\end{equation}
and we have:
\begin{equation}
\begin{array}{rl}
& W(z,z_0)W(z,z_0)  
= A e^{-\hat{t}\log[(z-z_0)]}  e^{-\hat{t}\log[- (z-z_0)]}   
\end{array}
\label{graviton12a}
\end{equation}
Thus we have:
\begin{equation}
\begin{array}{rl}
{\bf g}=& W(z_0,z)W(z_0,z)W(z,z_0)W(z,z_0) \\
&\\
=& e^{i\pi e_0^2n}A_g \quad  n= 0,\pm 1,\pm 2,\pm 3, \dots
\end{array}
\label{graviton13}
\end{equation}

In particular when the two loops are with counter winding numbers $m$ and $-m$, then  the graviton:
\begin{equation}
\begin{array}{rl}
{\bf g}= W(z_0,z)W(z_0,z)W(z,z_0)W(z,z_0) = A_g
\end{array}
\label{graviton14}
\end{equation}
is in the ground state  with zero energy ($n=0$) such that it is still with the two loops form where one loop is with winding number $m$ and the other loop is with winding number 
$-m$.


Then as similar to the
construction of electrons from photons  we construct matter from
gravitons
by the following formula:
\begin{equation}
Z^{\prime} {\bf g}Z={\bf g}ZZ^{\prime}
\label{graviton11}
\end{equation}
where $Z, Z^{\prime}$ are complex numbers as states acted by the graviton.
When the graviton is in  the ground state, this matter (\ref{graviton11}) can be identified as the dark matter.  Further  since the  energies   (\ref{graviton10})  of the excited states of graviton  are small  when the  level $n$ of the excited state is not large that   this matter (\ref{graviton11})  in such excited states can also be identified as the dark matter.

When a dark matter  (\ref{graviton11}) is in the ground atate this dark matter can be regarded as formed by a pair of particles with up and down spins respectively. In this case these two particles can be regarded as a pair of particle and antiparticle  such as electron and positron.

Similar to the mechanism of generating mass of electron we have
that the mechanism of generating the mass $m_d$ of these particles
is given by the following formula:
\begin{equation}
m_d c^2= 2 \pi  e_0^2 n_d=\pi  G  n_d=h \nu_d
\label{graviton12}
\end{equation}
for some integer $n_d$ and some frequency $\nu_d$.

Let us consider the quantum gravity effects on dark energy and dark matter. When a quantum graviton propagator connects  two gravitons or two dark matters this gives gravitational force on the  two gravitons or two dark matters, as the  gravitational force on the ordinary particles.
Let ${\bf g}_1$ and  ${\bf g}_2$ denote two gravitons. Then we have the follwing gravitational interaction between ${\bf g}_1$ and  ${\bf g}_2$:
\begin{equation}
{\bf g}_1 W(z_0,z)W(z_0,z) {\bf g}_2
\label{graviton11c}
\end{equation}
where $ W(z_0,z)W(z_0,z)$ denotes the quantum graviton propagator connecting the  two gravitons ${\bf g}_1$ and  ${\bf g}_2$.
Then we have the following gravitational interaction between  two dark matters $Z_1^{\prime}{\bf g}_1 Z_1$ and  $Z_2^{\prime}{\bf g}_2Z_2$:
\begin{equation}
Z_1^{\prime}{\bf g}_1  Z_1 W(z_0,z)W(z_0,z) Z_2^{\prime}{\bf g}_2 Z_2
\label{graviton11d}
\end{equation}
where $ W(z_0,z)W(z_0,z)$ denotes the graviton propagator connecting the two dark matters $Z_1^{\prime}{\bf g}_1  Z_1$ and  $Z_2^{\prime}{\bf g}_2 Z_2$.

In the case of the ordinary particles,  the two Wilson lines of the quantum graviton propagator connect to the single loop of the  ordinary particles to give the gravitational force between the ordinary particles. 
In the case of gravitons and dark matters  with two loops, the two Wilson lines of the quantum graviton propagator connect to the two loops of the gravitons and dark matters 
 to give the gravitational force between  the gravitons and dark matters. 

Let us then consider the case that the gravitons and dark matters  are in the ground state with the two loops having the winding numbers $m$ and $-m$ respectively.  In this case, since the two Wilson lines of the quantum graviton propagator connect to the two loops of  the gravitons and dark matters, 
the quantum graviton propagator still can give the gravitaional force to the  gravitons and dark matters when these gravitons and dark matters are in the ground state.

Then let us consider the factor $-\log r_1=-\log |z_0-z|$ of the  quantum graviton propagator in the above sections. We have shown that when  $\log r_1 >0$ the  quantum graviton propagator gives the usual attractive gravitational force between the interacting particles. On the other hand when  $\log r_1 <0$ the  quantum graviton propagator gives the unusual repulsive gravitational force between the interacting particles. Further we have that the  repulsive gravitational force tends to $\infty$ as  $\log r_1  \to -\infty$ (or $r_1\to 0$).

Then at the begining of the universe the gravitons and dark matters  in the ground state  (which is as the vacuum state) are grouped  to be very close together in each group that the distance $r_1$ among  the  gravitons and dark matters  in the ground state in each group   is very small that  the  quantum graviton propagators give the  repulsive gravitational force among the interacting gravitons (as dark energy) and dark matters. This gives the  accelerating expansion phenomena of the universe \cite{Rie}-\cite{Per2}.

In this expansion of  the universe, some of the gravitons and dark matters are separated into two photons and two  partices with up and down spins (such as electrons) respectively by the repulsive gravitational force among the gravitons and dark matters (When the  two  partices with up and down spins form the ground state of a dark matter before the separation, these two particles can be regarded as particle and antiparticle such as electron and positron).  In this separation the two loops of a graviton are separated such that the two loops of this graviton become two photons, and the  two loops of a dark matter are separated such that the two loops acting on the $Z, Z^{\prime}$ of this dark matter  become two  partices with up and down spins (such as electrons).
For these created  photons and  particles (such as electrons), when the distance $r_1$ among them become $r_1>1$ (or $\log r_1 >0$), the gravitational force among them then  becomes attractive. This then gives the usual observable part of the universe.

Let us then consider the  factor $-G\log r_1=-G\log |z_0-z|$ of the  quantum graviton propagator in the above sections.  When 
$-G\log r_1>0$, then as shown in the above we have that $0< r_1 <1$ gives the unusual repulsive  gravitational force between the interacting particles which gives the expansion of the universe. 
 On the other hand, if  $-G\log r_1<0$,
then by the same reason as shown in the above,  we have that  $ r_1 >1$ gives  the usual attractive  gravitational force between the interacting particles which gives the contraction of the universe. 
 Thus
 the  factor  $-G\log r_1$  has the property of the cosmological constant $\Lambda$ which is the constant with the property that $\Lambda>0$ gives the 
expansion of the universe and $\Lambda<0$ gives the contraction of the universe. 
Thus we determine that: 
\begin{equation}
\Lambda =-G\log r_1
\label{graviton16}
\end{equation}
is the cosmological constant  $\Lambda$.
 
We notice that  $ r_1$ is the distance between gravitons. Thus  the factor $-\log r_1$ determines the  energy density of vacuum.

Then when  $0< r_1<1$ and  $ r_1$ is near $1$ we have that $\Lambda =-G\log r_1$ is a very small positive constant (We notice that $G$ is already a very small positive constant). Then we can determine  $ r_1$ such that the value of $\Lambda =-G\log r_1$
is the experimental value of $\Lambda$.  Then in the expansion of the universe the value of  $r_1$ tends  to $1$ very slowly when $ r_1$ is near $1$.  In this case we have that $\Lambda =-G\log r_1$ is approximately a constant.  Thus 
$\Lambda =-G\log r_1$  ($0< r_1<1$ and  $ r_1$ is near $1$) is the  cosmological constant which  is a very small positive constant (and  tends to $0$ very slowly as $r_1$ tends  to $1$ very slowly)  in the expansion of the universe. 
This 
 then resolves the cosmological constant problem  \cite{Car5}.


\section{Conclusion}

In this paper a quantum loop model of photon is established. We
show that this loop model is exactly solvable and thus may be
considered as a quantum soliton. We show that this nonlinear model
of photon has properties of photon and magnetic monopole and thus
photon with some specific frequency may be identified with the
magnetic monopole. From the discrete winding numbers of this loop
model we can derive the quantization property of energy for the
Planck's formula of radiation and the quantization property of
electric charge. We show that the charge quantization is derived
from the energy quantization. On the other hand from the nonlinear
model of photon a nonlinear loop model of electron is established.
This model of electron has a mass mechanism which generates mass
to the electron where the mass of the electron is from the
photon-loop. With this mass mechanism for generating mass the
Higgs mechanism of the conventional QED theory for generating mass
is not necessary.

We derive a QED theory which is not based on the four dimensional
space-time but is based on the one dimensional proper time. This
QED theory is free of ultraviolet divergences. From this QED
theory the quantum loop model of photon is established. In this
QED theory the four dimensional space-time is derived for
statistics. Using the space-time statistics, we employ Feynman
diagrams and Feynman rules to compute the basic QED effects such
as the vertex correction, the photon self-energy and the electron
self-energy.  From these QED effects we compute the anomalous
magnetic moment and the Lamb shift. The computation is of
simplicity and accuracy and the computational result is better
than that of the conventional QED theory in that the computation
is simpler and it does not involve numerical approximation as that
in the conventional QED theory where the Lamb shift is
approximated by numerical means.

From the QED theory in this paper we can also derive a new QED
effect which is from the seagull vertex of this QED theory. By
this new QED effect and by a reformulated Bethe-Salpeter (BS)
equation which resolves the difficulties of the BS equation (such
as the existence of abnormal solutions) and gives a modified
ground state wave function of the positronium. Then from this
modified ground state wave function of the positronium a new QED
effect of the orthopositronium decay rate is derived such that the
computed orthopositronium decay rate agrees with the experimental
decay rate. Thus the {\it orthopositronium lifetime puzzle} is
completely resolved where we also show that the recent resolution
of this orthopositronium lifetime puzzle only partially resolves
this puzzle due to the special nature of two dimensional space
statistics of this new QED effect.

By this quantum loop model of photon a theory of quantum gravity
is also established where the graviton is constructed from the
photon. Thus this theory of quantum gravity unifies gravitation
and electromagnetism. In this unification of gravitation and
electromagnetism we show that the universal gravitation constant
$G$ is proportional to $e_0^2$ where $e_0$ is the bare electric
charge which is a very small constant and is related to the
renormalized charge $e$ by the formula $e_0=\frac{1}{n_e} e$ where
the renormalized number $n_e$ is a very large winding number of
the photon-loop. This relation of $G$ with $e_0$ (and thus with
$e$) gives a closed relationship between gravitation and
electromagnetism. 

Then  we show that gravitons are the dark energy and the
particles constructed by gravitons, as electrons  constructed by photons, are  dark matter. From this identification of  gravitons with  dark energy  and the quantum  graviton propagator we can compute the cosmology constant  and  show that the  cosmological constant is a very small positive constant  (and tends to 0 very slowly). This then  resolves the  cosmological constant problem.

\end{document}